\documentclass{article}
\usepackage{amssymb,amsmath,color,longtable}
\usepackage{epsfig}

\newcommand{\bnabla}{\mbox{\boldmath$\nabla$}}

\begin{document}


\title{ PINNING AND SLIDING OF DRIVEN
ELASTIC SYSTEMS:\\
\bigskip
FROM DOMAIN WALLS TO CHARGE DENSITY WAVES
\bigskip\bigskip}

\author{{\bf Serguei Brazovskii}
\footnote{Permanent address:
 LPTMS, B\^{a}t.100, Universit\'{e} Paris-Sud, 91406 Orsay, Cedex,
France}\\
\bigskip\\
Laboratoire de Physique Th\'{e}orique et des Mod\`{e}les
Statistiques, CNRS, Orsay,  France
\\
\bigskip
Yukawa Institute for Theoretical Physics, Kyoto University, Japan
\, \\
\\and\\
\, \\
 {\bf Thomas Nattermann}
 \footnote {Permanent address:
 \newline
  Institute for Theoretical Physics, University of Cologne, Z\"ulpicher Str.
77, D-50937 Cologne, Germany}\\
\bigskip\\
Laboratoire de Physique Th\'{e}orique et des Mod\`{e}les
Statistiques, CNRS, Orsay,  France\\
\bigskip
LPMMH, Ecole Superieure de Physique et de Chimie Industrielles, Paris, France}

\bigskip\bigskip
\date{17 March 2004}

\maketitle

\bigskip
 \centerline{\textbf{ADVANCES in PHYSICS, \emph{to be published.}}}
\bigskip

\newpage

\begin{abstract}
The review is devoted to  the  theory of  collective and
local pinning effects in various disordered non-linear
driven systems. Common feature of both approaches  is the
emergence of metastability. Although the emphasis is put on
charge and spin density waves  and magnetic domain walls, the
theory has also applications to flux lines and  lattices thereof,
dislocation lines, adsorbed mono-layers and related systems. In
the first part of the article we focus on the theory of the it
collective pinning which includes the {\it equilibrium
properties}  of elastic systems with frozen-in disorder as well as
the features close to the  dynamic depinning transition
enforced by an external driving force. At zero temperature and for
adiabatic changes of the force, the dynamic depinning transition
is continuous, the correlation length is large, the behavior is
dominated by scaling laws with non-trivial static and dynamical
critical indices. The application of functional renormalization
group methods allows for the detailed description of both
equilibrium as well as non-equilibrium properties.  The depinning
transition is also characterized by the appearance of it new
scaling laws. Thermal fluctuations smear out this transition and
allow for a  creep motion of the elastic objects even at
small forces. The application of an ac-driving force also destroys
the sharp transition which is replaced by a  velocity
hysteresis.

The second part of the review is devoted to the picture of the
local pinning and its applications. Local theories apply in the
region where correlation effects are less important, i.e. not too
close to the depinning transition, at low temperatures, at high
enough frequencies or velocities. The inclusion of  plastic
deformations results in a rich cross-over behavior of the
force-velocity relation as well as of the frequency dependence of
the dynamic response. Being easily affected at higher frequencies
or velocities, the local pinning becomes the easily accessed
source of dispersion, relaxation and dissipation. The picture of
the local pinning can be effectively used to explain experimental
data: qualitatively and even quantitatively. The advantages come
from the explicit treatment of metastable states, their creation
and relaxation, their relation to plasticity and topological
defects. The local pinning recovers and exploits new elements of
the energy landscape such as termination points of some branches
or irreversibility of other ones related to generation of
topological defects in the course of sliding. It provides also a
clue to quantum effects describing the quantum creep as a
tunnelling between retarded and advanced configurations.
\end{abstract}

\tableofcontents
\newpage

\section{Introduction}

Many ordering phenomena in solids are connected with the emergence
of a modulated structure. Examples are charge or spin density
waves
\cite{Monc-rev,Gruner88,Proc89,Dumas93,ECRYS93,ECRYS99,ECRYS02,Wigner},
Wigner crystals \cite{Wigner,OrigniacGiamarchi}, flux line
lattices \cite{Blatter+94,Nat+Scheidl00}, incommensurate phases of
adsorbed monolayers \cite{monolayers.review} etc. These structures
often interact with imperfections frozen in the solid leading to
pinning phenomena which change drastically the statics and
dynamics of the modulated structure.

In other cases topological defects like isolated flux lines in
superfluids or superconductors, dislocation lines in solids,
domain walls in magnets etc. appear as a result of competing
interactions, external fields, or in the process of fabricating
the material. Pinning of the motion of these objects is often
required if one wants to exploit physical properties of the
condensed structure. The pinning of  flux lines prevents
dissipation from their frictional motion in superconductors,
pinning of dislocations prevents plastic deformations of a solid
etc.\cite{review.magnetic.wall,Fisher98,Kadar98}. The generality
of this approach has been anticipated already long time ago
\cite{Haasen}, but strong similarities between different systems
mentioned above have been worked out in detail only later on.

The goal of the present article is to give a
presentation of unifying concepts in the theory of pinning
phenomena. To make the ideas more clear, we will not go too much
into the details of specific systems but we shall stress the
generality of the approach.
Some aspects of pinning have been considered in the past in great
detail for type-II superconductors and we refer the reader to
these articles for more details \cite{Blatter+94,Nat+Scheidl00}.

A perfectly rigid object of any dimension, e.g. a straight flux
line, a planar domain wall or an undistorted charge density wave
will never be pinned -- the pinning forces acting on different
parts of the object cancel each other. More accurately: the
resulting total pinning force is of the order of the square root
of the volume of the object. It is therefore necessary to consider
the distortions of these objects under the influence of pinning
centers which are often of elastic type. We will therefore speak
about ''elastic systems'' if we refer to arguments which are
essentially correct for all systems mentioned above.

The delicate question  of  pinning by randomness stays already for
almost three decades since the earliest  proposals of Larkin
\cite{larkin70}, through theories of vortex lattices
\cite{larkin.ovch79} and density waves
\cite{fukuyama76,fukuyama.lee78,lee.rice79}. The last decade has
brought new insights, largely provoked by the studies of vortices
in high-T$_c$ superconductors \cite{Blatter+94,Nat+Scheidl00}. In
particular new understanding was reached detecting the possibility
of a quasi long-range ordered (Bragg) glass phase with an
algebraic decay of structural correlations in a disordered system
\cite{natter90,korshunov93,giamarchi.doussal94}  and by the
understanding of the role of metastable states in non-stationary
effects. Essential progress was due to new advanced methods like
the functional renormalization group method
\cite{fisher85,fisher86} for the collective pinning problem. Still
full details are not completely accessible and applications are
not always straightforward. The picture of the local or weak
pinning was also revised and developed through the last decade in
conjunction to plasticity and a role of topological defects.

The undisputable domain of the collective pinning approach are the
effects occurring on large length and time scales. These are
recovered e.g. in studies of long time evolution, the low
frequency response to external forces, the creep below the zero
temperature depinning threshold, and the region around  the
threshold field. The \emph{collective} or \emph{weak} pinning
forces come from elastic interference of many impurities
\cite{larkin.ovch79,lee.rice79,fukuyama.lee78,fukuyama76}. Their
characteristic features are: large correlation volumes -- of the
order or beyond  the Larkin-length, high energy  barriers  between
metastable states, huge relaxation times
and small pinning forces.
The collective pinning determines the threshold $f_{c}$ of the
driving force to initiate the sliding  quite similar to
conventional rest friction \cite{friction}. The comprehensive
picture of the collective pinning will be given in the first part
of this review in Sections 2-5.

Complementary   insight can be obtained within a frame of the so
called \emph{strong or local pinning} which manipulates with only
a few of metastable states. This will be the topic of the second
part of this review (Sections 6-12). This simple but transparent
approach offers some effects which have not been noticed yet --
and still are not fully accessible -- within the very complicated
picture of the collective pinning. The {\it local} or {\it strong}
pinning comes from rare metastable pinning centers which provide
finite barriers, hence reachable relaxation times. Being easily
affected at higher frequencies or velocities, the local pinning
becomes the easily accessed source of dispersion, relaxation and
dissipation. This part will summarize the local pinning approach
to time dependent properties of sliding superstructures. We shall
follow the scheme of \cite{Larkin94,Larkin95,Brazov96,Brazov99}
for the theory of the pinning-induced \emph{metastable plastic
deformations} partly due to creation of dislocation loops or
lines. Depending on the pinning potential strength, we shall find
several regimes of the behavior of local deformations in the
course of a displacement with respect to the impurity site. The
key observation is that the local state at the pinning center --
the impurity -- may be either unique or bistable. The bi-stability
can be either restricted or unrestricted (i.e. preserved
throughout the whole period of sliding). In the latter case this
leads to the generation of diverging pairs of dislocation loops --
or $2\pi-$ solitons in a quasi 1-dimensional  picture. On this
basis we can obtain contributions to the pinning force  resulting
in the sliding velocity-driving force  characteristics and in the
frequency dependent response.

Special applications will be devoted to the best studied examples
of domain walls and CDWs where we shall particularly address two
commonly observed experimental features: the totally nonlinear
current-voltage curve and the anomalous low frequency  low
temperature behavior of the dielectric susceptibility.

\section{Equilibrium properties of elastic objects in random
environments}

In this section we summarize the equilibrium properties of elastic
objects in a random environment. There are two main classes of
systems: The first one refers to non-periodic objects like
isolated domain walls, flux or dislocation lines. These can be
generalized to the so-called {\it elastic manifolds models}. The
second class of models is periodic, like charge density waves,
flux line lattices or Wigner crystals. These are subsumed under
the expression {\it periodic media}.

To make the notation simple, we will describe the distortions of
the elastic systems from perfect order by a {\it scalar}
displacement field $u({\bf x})$, its generalization to $N$-vector
or more complicated fields is straightforward.  Correspondingly we
will mostly use the terminology of {\it domain walls} and {\it
charge density waves}, respectively. Where peculiarities in
systems with $N>1$ may occur we will mention them however.

\subsection{Models}

We consider a $D$-dimensional elastic object embedded into a host
medium of a dimension $d$.  Since the medium includes quenched
disorder, the energy can be written in the form
\begin{equation}
   {\cal H}=\int d^Dx\left\{\frac{1}{2}C(\bnabla u)^2+
   V_R({\bf x},u) -fu\right\}.
   \label{eq:Hamiltonian}
\end{equation}
   Here $D$ denotes the internal dimension of the object ($D=1$ for
   isolated flux or dislocation lines, $D=d-1$ for isolated
   domain walls, $D=d$
   for charge
   density waves, flux line lattices and  Wigner crystals etc.). Note,
   that $D$ is in general {\it different} from the space dimension $d$
   ($D\leq d$).   (The dimension $N$ denotes the number of
   components of the
displacement field $u$. For isolated flux lines or domain walls
$N=d-D$ whereas for flux line lattices $N=2$ and $N=1$ for CDWs.)
   $C$ is an elastic constant and in general a tensor, e.g., in
   flux line lattices at least three elastic constants are necessary
   for the description. In some cases the elasticity is non-local on
   intermediate length scales -- as for flux line lattices on scales
   smaller then the London penetration length --  or even at all length
   scales as for dislocation lines. The elastic constants
   will also show a temperature dependence since they have to vanish at the
   transition where the structure melts, e.g. at the Peierls
   transition for CDWs.  $f$ denotes
 an external force density which acts on the object.

The random potential $V_R({\bf x},u)$ results from the coupling of
the elastic object to the  impurity potential $ v_R({\bf x,z}),\,
$ which is generated by the frozen impurities, fluctuating
exchange constants etc.:

\begin{equation}
V_R({\bf x},u)=\int d^{d-D}z\, v_R({\bf x,z})\,\rho({\bf x,z},u)\,,
\label{eq:random_potential}
\end{equation}

$\rho({\bf x,z},u)$  represents the density of the elastic object
and will be specified below. ${\bf x}=(x_1,...x_D)$ and ${\bf
z}=(x_{D+1},...,x_d)$ denote a D- and $(d-D)$-dimensional position
vectors  parallel and perpendicular to the main orientation of the
object (e.g. the domain wall or the flux line), respectively. For
density waves and flux line lattices $D=d$ and hence there is no
perpendicular coordinate ${\bf z}$.

The average over disorder - which replaces the average over the
(infinite) sample - will be denoted $\langle...\rangle_R$.
Without restricting the generality we will
assume that $\langle v_R({\bf x,z})\rangle_R=0$ -- with other words,
we incorporate the  effect of the averaged disorder
potential into  the bare parameters of our
model and  consider only its fluctuations.
A less general but still reasonable choice is that ${v_R}({\bf
x,z})$ is Gaussian distributed and short range
correlated (with a correlation length $l$) such that it is
characterized by its second moment:
   \begin{equation}
   \langle v_R({\bf x}_1,{\bf z}_1)\,v_R({\bf x}_2,{\bf z}_2)\rangle_R=
   v_R^2\delta({\bf x}_1-{\bf x}_2)\delta({\bf z}_1-{\bf z}_2).
   \label{eq:V-correlations1}
   \end{equation}

If we rewrite the random potential $v_{R}(\mathbf{x},\mathbf{z})$
as a sum of impurity potentials
\begin{equation}\label{eq:n_imp}
    v_{R}(\mathbf{x},\mathbf{z})=
    \sum\limits_{i=1}^{N_{imp}}V_{i}\delta(\mathbf{x}-\mathbf{x}_{i})
    \delta(\mathbf{z}-\mathbf{z}_{i})-\bar{V_i}n_{imp}
\end{equation}
then the disorder average is defined as
\begin{equation}\label{eq:disav}
    \big<\ldots\big>_{R}=
    \prod\limits_{i=1}^{N_{imp}}\int\frac{d^{D}x_{i}\,d^{d-D}z_{i}}{V}
    \int\limits_{-\infty}^{\infty}dV_{i}p(V_{i})\ldots
\end{equation}
Here $N_{imp}$, $V$ and $n_{imp}$ denote the total number of
impurities, the volume of the sample and the impurity
concentration, respectively. $p(V_i)$ is the normalized
probability distribution of the potential strength $V_{i}$ with
$\overline{V_i^{k}}=\int dV_i\cdot V_i^{k}\cdot p(V_i)$ and
$v_{R}^{2}=\overline{V_i^{2}}\,n_{imp}$.

The correlations of the random potential $v_R$
are always short ranged provided the impurities are
short range correlated.
$\delta({\bf x})$, $\delta({\bf z})$  are $\delta$--functions which
may be smeared out over a length scale of the order $l$.
Since we will use it later we introduce here also the correlation
function
 \begin{equation}
   \langle V_R({\bf x}_1,u_1)\,V_R({\bf x}_2,u_2)\rangle_R=
   R({u_1,u_2})\delta({\bf x}_1-{\bf x}_2).
   \label{eq:V-correlations}
   \end{equation}
Because of the gaussian nature of the disorder correlations the
disorder averaged physical quantities like the free energy
depend only on the correlation function
$R({u_1,u_2})$.
The relevant contributions of $R({u_1,u_2})$ are those which
depend only on the difference $u_1-u_2$ on which we will
concentrate in the following.  For $|u|\ll l$, $R(u)$ is quadratic
in $u$ (with $R_{uu}(0)<0$), for larger $u$ different cases have
to be distinguished (see below).

\vspace{0.5cm}

In the following we consider some specific examples:

\vspace{0.5cm}
 (i) {\it Domain walls in magnets} : Here $u({\bf
x})$ describes the displacement of the domain wall from a planar
reference configuration, $D=d-1$ and $N=1$. The stiffness constant
$C$ is finite only above the roughening transition temperature
$T_R$, for $T<T_R$ the elastic description breaks down  in the
absence of disorder. But disorder leads always to roughening of
the interface, even at $T=0$ \cite{TN2000,Nattermann84,EmNa}.  The
driving force density is directly related to the magnetic field
$B$ by $f=2\mu_B B$.  If the random potential results from
fluctuations of  exchange coupling between the spins (this is the
so-called {\it random bond} case), which couple only to the domain
wall, then $\rho({\bf x},z, u)\sim \delta (z-u({\bf x}))$.  In
some cases the width of the domain wall may be large compared to
the lattice spacing and the $\delta$-function has to be replaced
correspondingly by a smeared out profile function. The
correlations of  $V_R({\bf x},u)$ are then  short range  in ${\bf
x}$ and have a correlation length corresponding to the maximum of $l$ and
the domain wall width in the  $u$-direction. To keep the notation
simple we will denote this maximum in the following also by $l$.
$R(u)$ is then a $\delta$-function of finite width of the order
$\sim l$. The extension of this model to arbitrary $D$ and $N$
(i.e. $D+N$ is not longer equal to $d$) is called the \emph{random
manifold} model.

If disorder comes from {\it random field} impurities then the
domain wall couples to the disorder in the domains. In this case
$\rho({\bf x},z,u) \sim \Theta (u({\bf x})-z)$ which gives the
Zeeman energy with a constant magnetization in each domain,
$v_R({\bf x}, z)$ represents now an uncorrelated random magnetic
field. In this case of a non-local coupling to the disorder it can
be shown that $R(u)\approx -|u|$
for $|u|\gg l$ \cite{Villain82}.

\vspace{0.5cm}
 (ii) {\it  Isolated magnetic flux lines or dislocation
lines}: $u({\bf x})\to{\bf u}({\bf x})$ denotes now a $N=d-1$
component displacement field and  $D=1$. The random potential
$V_R({\bf x,u})$ is short range correlated both in ${\bf x}$ and
${\bf u}$ and hence $R({\bf u})$ is again a smeared out
$\delta$-function. For flux lines, $f$ is given by the
Lorentz force ${\bf f}=\frac{1}{c}{\bf j\times {\hat b}}\Phi_0$,
where $\bf j$ denotes the transport current. $\Phi_0=hc/2e$ is the
flux quantum and $\bf\hat b$ the local direction of the magnetic
field. In the case of dislocation and vortex lines the elastic
energy is non-local \cite{Blatter+94}. In particular, for
dislocation lines one finds after Fourier transformation  a weakly
momentum dependent elastic modulus $C(k)=-C \log(a_0 k)$. $a_0$ is
of the order of the lattice constant. The force acting on the
dislocation line is the Peach-K\"ohler force \cite{LLVol.7}.
\vspace{0.5cm}

 (iii) {\it Charge density waves}: The condensed
charge density can be written as
\begin{equation}
\rho({\bf x},\varphi)=\rho_0 (1+{\bf Q}^{-1}{\bnabla}\varphi) +
\rho_1 \cos \left({\bf Qx}+\varphi({\bf x})\right)+...
\label{eq:rho.cdw}
\end{equation}
where ${\bf Q}$ denotes the wave vector of the charge density wave
modulation and the dots stand for higher harmonics. The first
factor describes the density change due to an applied strain. The
phase field $\varphi({\bf x})$ is related to a displacement
$u({\bf x})$ of the maximum of the density by
\begin{equation}
\varphi({\bf x})=-u({\bf x})Q,
\label{eq:u-phi-relation}
\end{equation}
which we will use from now on as the relation between $u$ and
$\varphi$.

Since charge density waves carry an electric charge, the stiffness
constant $C$ shows in general a strong dispersion due to the long
range Coulomb interaction
\begin{equation}\label{eq:elastic.dispersion}
C{\bf k}^2\rightarrow
C_{\parallel}k_{\parallel}^2+C_{\perp}k_{\perp}^2+
C_{dip}\frac{(k_{\parallel}\lambda/a_0)^2}{1+{\bf
k}^2\lambda^{2}}\,,
\end{equation}
which we will ignore except for special applications. Here
$\lambda$ denotes the screening length. If $\lambda $ diverges,
the system may be considered effectively as a 4-dimensional one
with $k_{\parallel}/k$ playing the role of a fourth dimension
\cite{efetov.larkin.77}.

Corresponding to the two $\varphi$-dependent contributions
proportional to $\rho_0$ and $\rho_1$  in (\ref{eq:rho.cdw}) there
are two contributions in (\ref{eq:random_potential}) which
sometimes are referred to as forward and backward scattering,
respectively. The resulting correlator $R(u)$ is \emph{periodic}
in $u=-\varphi/Q$ with the periodicity $2\pi/Q$. Indeed, for
charge density waves we obtain from equations
(\ref{eq:V-correlations}) and (\ref{eq:rho.cdw})
\begin{equation}
R(\varphi)\propto-\frac{1}{2}\rho_0^2 \big({\bf Q}^{-1}{\bf
\nabla}\varphi\big)^2 +\frac{1}{2} \rho_1^2\cos\varphi.
\label{eq:periodicR(u)}
\end {equation}
Here we have neglected strongly oscillating terms (which average
to zero) as well as terms which can be included into the elastic
energy. Finally, the external force $f$ is given by the applied
electric field, $f\sim E$.

For a more detailed discussion of other systems like flux-line
lattices, Wigner crystals etc. we refer the reader to the
appropriate literature
\cite{Blatter+94,Nat+Scheidl00,andrei88,willett89,giamarchi02}.

\subsection{Basic properties of disordered systems}

To get preliminary information about the influence of disorder on
the elastic object and the relevant length scales we consider
first small distortions $u({\bf x})$ around the state of perfect
order  $u({\bf x})\equiv 0$. The energy can then be written as a
series expansion in $u({\bf x})$:
   \begin{equation}
   {\cal H}=\int d^Dx\left\{\frac{1}{2}C(\bnabla  u)^2+
   V_R({\bf x},0)+V_{R,u}({\bf x},0){ u}+\cdots\right\},\quad
   V_{R,u}=\frac{\partial V_R}{\partial u}.
   \label{eq:energy_1order}
   \end{equation}
    Here we assumed that the distortions are small such that we can neglect for the
   moment higher order terms in the expansion of $V_R({\bf x},u)$.

   The {\it necessary} condition for the
   ground state of the Hamiltonian follows from the variation of
   ${\cal H}$ with respect to ${u}$
   \begin{equation}
   \frac{\delta{\cal H}}{\delta u}=-C\bnabla^2u+V_{R,u}({\bf x},0)=0\,,
   \label{eq:dH/du}
   \end{equation}
   which is the Poisson equation known from electrostatics ($u$ and
   $V_{R,u}({\bf x},0)/C$ playing the role of the electrostatic
   potential and the charge distribution, respectively).

Its solution is given by
\begin{equation}
u({\bf x})\sim \int d^Dx^{ \prime}\frac{V_{R,u}({\bf
x}^{\prime},0)}{C|{\bf x}-{\bf x}^{\prime}|^{D-2}}. \label{eq:u}
\end{equation}
From (\ref{eq:u}) we obtain with the help of
(\ref{eq:V-correlations})
   \begin{equation}
   w_R^2\big(|{\bf x}|\big)\Big|_{T=0}\equiv
   \left\langle\big(u({\bf x})-u({\bf 0})\big)^2\right\rangle_R\sim
   \frac{(-R_{uu}(0))}{(4-D)C^2}|{\bf x}|^{4-D}
   \label{eq:w^2}
   \end{equation}
where $R_{uu}(0)$ denotes the second derivative of the correlator
at $u=0$. At this point we want to mention already one problem
with the use of the approximation (\ref{eq:dH/du}): since the
random forces $V_{R,u}({\bf x},0)$  do not depend on $u$, any
constant can be added to the solution (\ref{eq:u}). In other
words, the object could be moved through the random environment
without any change in energy. Thus the linear approximation misses
barriers and metastable states.

As can be seen from (\ref{eq:w^2}), the relative displacement in two
different points ${\bf 0},{\bf x}$ increases with their
separation $|\mathbf{x}|$  below the critical dimension $D_c=4$,
the elastic object
is called to be {\it rough}. We can rewrite (\ref{eq:w^2}) in a more general
form as
   \begin{equation}
   w_R(L)\approx l\left(\frac{L}{L_p}\right)^{\zeta}\,,
   \label{eq:w(L)}
   \end{equation}
where we introduce the {\it roughness exponent} $\zeta$. In our
present calculation $\zeta\equiv\zeta_{\rm RF}=(4-D)/2$, where the
subscript RF stands for {\it random force} corresponding the
expansion of  $V_R({\bf x},u)$ up to a force term in equation
(\ref{eq:energy_1order}). The characteristic length scale $L_p$
   \begin{equation}
   L_p\approx\left(\frac{(4-D)l^2C^2}{|R_{uu}(0)|}\right)^{1/(4-D)}
   \label{eq:L_p}
   \end{equation}
   is called the {\it Larkin--length} \cite{larkin70}
in the context of flux line
   lattices, the {\it Fukuyama--Lee length}
\cite{fukuyama.lee78} in the context of charge
   density waves, the {\it Imry--Ma length}
\cite{imry.ma75} for random magnets etc..
   On this length scale the displacement is of the order of the
   correlation length $l$ of the random potential, i.e. the displacement
   field can choose between different energy minima and hence
  {\it meta-stability} appears. For weak disorder,
  $L_p$ is large compared
   with the mean impurity distance $n_{imp}^{-1/d}$ and
   hence pinning phenomena arise from
   the collective action of many impurities.
   In the situation of
   strong pinning, which we will discuss further in the second half
   of this article, individual
   pinning centers lead to strong distortions already on the scale
   shorter than the distance
   between impurities $n_{imp}^{-1/d}$. In the next chapters we will
   always assume
   $L_{p}\gg n_{imp}^{-1/d}$. The Larkin-length will then play
   the role of an effective
   small-scale cut-off for the phenomena considered in the following.

   Unfortunately, the result (\ref{eq:w^2}) is not applicable on
   length scales much larger than $L_p$ since equation (\ref{eq:dH/du}) is
   only a necessary condition for the ground state: it is the
   condition for a saddle point, not only for the absolute minimum. It
   is generally believed that the elastic object has a unique
(rough) ground state \cite{alava01}. But for
   distortions $u >l$ the system has in general many local minima and
   perturbative methods break down. A nice demonstration of the
   breakdown of perturbation theory in systems with several energy
   minima has been given by Villain and Semeria
   \cite{villain.semeria83}.  In this case more elaborate methods like
   the renormalization group approach have to be applied. This exceeds
   the scope of the present short review, but a schematic presentation
   of the method is given below.

   The result of this approach is that
   the {\it roughness} $w(L)$ can still be written in the form (\ref{eq:w(L)})
   but
   with $\zeta_{RF}$ replaced by a non-trivial {\it roughness exponent}
   $\zeta$ with $0\leq \zeta \leq 1$. (For $\zeta>1$
   $\mid\nabla u\mid \sim L^{\zeta-1}$ diverges on large length
   scales and the elastic approach adopted here breaks down.)

   For periodic media, in general, effects of disorder are weaker.
   In the  case of elastic manifolds it may pay off for the
   elastic object to make a large excursion to find a larger
   fluctuation in the impurity concentration, which lowers the
   free energy. On the contrary, for a periodic medium,
   the distortion has to be at most of the
   order of the period of the medium to reach a favorable
   interaction between the impurity and the medium.
Hence we expect that periodic systems belong to another
universality class with different, smaller exponents. The actual
result happened to be more drastic: the roughness exponent $\zeta$
for periodic media is zero corresponding to a \emph{ logarithmic}
increase of $w_R(L)$
\cite{natter90,korshunov93,giamarchi.doussal94,villain.fernandez84}
(for further details see Section \ref{sec:rgm-results}).

\vspace{0.5cm} The average $\langle...\rangle_T$
   over {\it thermal fluctuations} does not change the asymptotic
   behavior of the correlation function
\begin{equation}
w^2_R(|{\bf x}|)=\left\langle\langle u({\bf x})-u({\bf
0})\rangle^2_T\right\rangle_R \propto |{\bf x}|^{2\zeta}
\label{eq:w_R}
\end{equation}
which vanishes apparently in the absence of disorder. It is this
correlation function which dominates the {\it structural
properties} of the system. But the temperature will have a drastic
effect upon the time dependent properties.

Since we want to consider later on the motion of the elastic
object under an external force $f$ it is instructive to
characterize the energy landscape the elastic object is exposed
to. As a first step we consider changes in the free energy
$\mathcal{F}_R$ (which depends on a particular configuration of
the disorder) by going over to a different disorder configuration.
Such new disorder configuration may be created in some cases by
applying external forces (compare with
equation(\ref{eq:Hamiltonian2})) or changing the boundary
conditions. \emph{Sample to sample fluctuations} of the free
energy of a region of linear extension $L$ are then expected to
show also scaling behavior described by a new exponent $\chi$
 \begin{equation}\label{F-fluctuations1}
 \left \langle \left[ \mathcal{F}_R(L)-\langle
 \mathcal{F}_R(L)\rangle_R\right]^2\right\rangle_R^{1/2}
 \approx T_p\left(\frac{L}{L_p}\right)^{\chi}\equiv {F}(L),\qquad
 T_p=C l^2L_p^{D-2}, \qquad \chi=D-2+2\zeta.
 \end{equation}
where $T_p$ denotes a characteristic energy scale see e.g.
\cite{Huse+85,Natter8788,Mezard90,Fisher.Huse}.
The scaling relation between $\chi$ and $\zeta$ can be understood
from the fact that in the free energy the elastic and the random
part of the energy have to be of the same order and therefore it
is plausible that the scaling behavior of $\mathcal{F}_R(L)$ can
be read off from the scaling of the elastic energy, which scales
with the exponent $\chi$. An illustrative one--dimensional example
of the free energy fluctuations is considered in Appendix
\ref{sec:app1}.
\medskip

Further information about the system comes from a second
correlation function $w_T(|{\bf x}|)$ which describes the response
to a local force which couples to $\big(u({\bf x})-u({\bf
0})\big)$:
\begin{equation}
w^2_T(|{\bf x}|)\equiv \left\langle\langle[u({\bf x})-u({\bf
0})]^2\rangle_T-\langle u({\bf x})-u({\bf 0})\rangle_T^2
\right\rangle_R\sim  T|{\bf x}|^{2-D}/C +
\text{const.}\label{eq:w_T}
\end{equation}
Any pair correlation function of $u({\bf x})$ can be expressed by
a combination of $w_R(|{\bf x}|)$ and $w_T(|{\bf x}|)$. Equation
(\ref{eq:w_T}) is (exactly) the same result as in a non-random
system and a consequence of a statistical ``tilt-symmetry'' of the
system \cite{Schulz+88}. Apparently, $w_T(|{\bf x}|)$ vanishes at
$T=0$. The result (\ref{eq:w_T}) can be related to the static {\it
susceptibility}
\begin{equation}
\chi=\left \langle\frac{\partial}{\partial f}\langle
 u({\bf 0})\rangle_T \right\rangle_R = \frac{1}{T}\int d^Dx \left
\langle\langle u ({\bf x})u ({\bf 0})\rangle_T-\langle u ({\bf
x})\rangle_T \langle u ({\bf 0})\rangle_T\right\rangle_R
\label{eq:susceptibility}
\end{equation}
where $f$ is the force conjugate to $u$.

\begin{figure}[hbt]
   \centerline{\epsfxsize=5cm \epsfbox{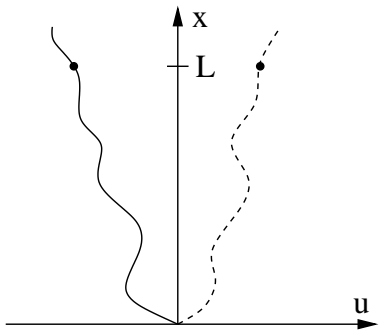}}
   \caption{Ground and low energy excited state of a $1D$ domain wall}
   \label{fig:rare_fluctuations}
   \end{figure}

To discuss the result for $w_T(|{\bf x}|)$ further we consider
the case when $u(0)$ is fixed to zero such that $ w^2_T(|{\bf
  x}|)=\left \langle\left\langle (u({\bf x}) -\langle u({\bf
    x})\rangle_T)^2\right\rangle_T \right\rangle_R\equiv
    \left \langle\left\langle \delta u^2({\bf x})\right\rangle_T \right\rangle_R$. The quantity $\delta
u({\bf x})=u({\bf x})-\langle u({\bf x})\rangle_T$ describes the
fluctuations of $u({\bf x})$ around its thermal average in a given
random environment. Naively one could assume that these
fluctuations are restricted to a narrow valley along the ground
state such that the fluctuations should not increase with $|{\bf
x}|=L$. The result (\ref{eq:w_T}) however suggests a different
picture: besides of the ground state there are {\it rare excited
states} which are very different in configuration from the ground
state, $\delta u({\bf x})\sim l(L/L_p)^{\zeta}$,
 the  energy of which  differs only by an amount of the order
$\Delta E \leq T$ from the ground state (compare with
Figure\ref{fig:rare_fluctuations}).

These excited states could become indeed true ground states if we would
change the random
potential locally (i.e. in the neighborhood of the initial ground state)
in an appropriate manner. Since at $T=0$ the free energy is given by the
ground state
energy we have to expect that rare excited states differ in energy as the
sample to sample variations of the free energy.

To calculate the fluctuations of $\delta u$ we make therefore for the
probability distribution of the energy of the excited states of an elastic
object of linear extension $L$ the scaling Ansatz
\begin{equation}
\label{eq:Delta-E}
 P(\Delta E,L)=F^{-1}(L)p(\Delta E/F(L))
\end{equation}
where  $F(L)$ denotes the sample to sample variations of the free
energy (compare with equation (\ref{F-fluctuations1}))  and $p(x)$
is an unknown normalized function with $p(0)=O(1)>0$. With this
Ansatz we find for the average fluctuations at  low temperatures
$T\ll F(L)$ \cite{Mezard90,Fisher.Huse}
\begin{equation}
\left\langle|\delta u({\bf x})|^n\right\rangle_{T,R} \approx
l^n\left(\frac{L}{L_p}\right)^{n\zeta}\int_0^Td(\Delta
E)\,P(\Delta E,L) \sim l^n
\frac{T}{T_p}\left(\frac{L}{L_p}\right)^{n\zeta-\chi}
\label{eq:thermal.fluc.u^n}
\end{equation}
which gives for $n=2$ the  exponent ${n\zeta-\chi}=2-D$ appearing
in (\ref{eq:w_T}).

The success of this approach gives us the possibility
to calculate also the {\it specific heat}
   \begin{equation}\label{eq:specific.heat}
   c(T)=L_0^{-D}\frac{\partial}{\partial T}\langle
   {\Delta E}\rangle_{R}\approx
   \frac{\partial}{\partial T}\int_{L_p}^{\infty}dL\,
   \nu(L)\int_0^Td(\Delta E)\cdot \Delta E\cdot P(\Delta E,L).
   \end{equation}
Here  $L_0$ and $\nu(L)$ denote the system size and the
\emph{size} distribution of the rare low energy excited states on
the scale $L$, respectively. In writing down
(\ref{eq:specific.heat}) we have decomposed the system in
$(L_0/L)^D$ blocks of linear extension $L$. Each of them gives a
contribution to the internal energy of the order of the integral
on the r.h.s. of (\ref{eq:specific.heat}). Then we have to sum
over the contributions from all length scales $L$, $L_p<L<L_0$.
The smallest scale is clearly given by the Larkin scale $L_p$. The
next independent contributions come from excitations on   larger
scales $bL_p, b^2L_p,...,b^nL_p$ etc., $b>1$. ($b$ has to be
chosen in such a way that excited states on scale $b^{k+1}L_{p}$
cannot be reached already on scale $b^{k}L_{p}$, which requires
$b^{\zeta}\gtrsim 2$.) The total number of scales is given by $\ln
(L_0/L_p)/\ln b$. If we replace this sum over $n$ by an integral
over $L$, we obtain a factor (the integration measure) $dL (L\ln
b)^{-1}(\frac{L_0}{L})^D$ and hence $\nu(L)\propto
\frac{1}{L^{1+D}}$ \cite{korshunov01}. (Roughly speaking we could
say that we integrate over  all momenta $d^dk$ with $k\sim
L^{-1}$.)

In principle, the distribution $P(\Delta E,L)$ may also depend on
the temperature. If we ignore this unknown dependence, we get
   \begin{equation}\label{eq:specific.heat.2}
   c(T)\sim \int_{L_p}^{\infty}dL
   L^{-D-1}\frac{T}{F(L)}p\left(\frac{T}{F(L)}\right)\approx
   \frac{T}{T_p}p(0)L_p^{\chi}\int_{L_p}^{\infty}\frac
   {dL}{L^{D+1+\chi}}\sim \frac{1}{D+\chi}\frac{T}{T_p}L_p^{-D}
   \end{equation}
since the integral is dominated by small $L$ and $p(T/T_p)\approx
p(0)$. This approach gives at $T\ll T_P$ a specific heat {\it
linear} in $T$ which is similar to physics of common two-level
systems in amorphous solids \cite{anderson.halperin.varma72}. This
analogy also builds a bridge to the local pinning picture which
thermodynamically is equivalent to the case of
\cite{anderson.halperin.varma72}. Within the collective picture,
similar results have been obtained, in another way, recently in
\cite{scheer03}. The precise value of $b$ remains unknown in this
approach but has no influence on the temperature dependence.
Measurements of the specific heat on a finite time scale $t$ will
lead to a reduced value of $c(T)$ since not all local energy
minima can be reached by thermally activated hopping. We will give
a time dependent correction factor (which takes this fact into
account) at the end of Section 3.2.

 So far we have considered mainly the case of  weak pinning. The large
scale properties ($L>L_p$) are dominated in this case by density
fluctuations of the disorder where many impurities are involved.
We want to stress that in the case of {\it strong pinning} {\it on
very large scales} pinning phenomena are again dominated by
density fluctuations of the impurities. Strong pinning on small
scales is however very different from weak pinning and will be
further discussed in the second part of this article. A
one--dimensional CDW--model with strong pinning \cite{fukuyama76,brazov79}
 is considered in Appendix \ref{sec:app2}.

\subsection{Results from the renormalization group method}
\label{sec:rgm-results}

In the rest of this section we give a short account of the
application of the renormalization group (RNG)
\cite{patashinsky,amit84} approach to elastic systems in random
environments \cite{balents_fisher,balents_et_al}.
The RNG method starts from a Fourier decomposition
of the displacement field $u({\bf x})=\sum_{\bf k}u_{\bf
k}e^{-i{\bf k}{\bf x}}$.
The {\it first} step of the procedure consists in the elimination
of the short wave length degrees of freedom $u_{\bf
k}^{\phantom{1}}\equiv u_{\bf k}^{>}$ with ${\bf k}$ in the
momentum shell $\Lambda_0/b\le |{\bf k}|\le\Lambda_0,\;b>1$, from
the partition function. $\Lambda_0={2\pi}/{a_0}$ denotes a
microscopic momentum cut--off. The result of taking the trace over
$u_{\bf k}^{>}$ can be written again in the form of a Boltzmann
factor with a new \emph{effective} Hamiltonian for the remaining
degrees of freedom. If thermal fluctuations are irrelevant, as for
$D\geq 2$-dimensional systems, this procedure reduces to the
problem of finding the values $u_{\bf k}^{>}=\tilde u_{\bf k}^{>}$
which minimize the energy, keeping all $u_{\bf
  k}^{\phantom{1}}\equiv u_{\bf k}^{<}$ with $|{\bf k}|\le\Lambda_0/b$
fixed.  Plugging these values $\tilde u_{\bf k}^>$ into the
Hamiltonian we get the new effective Hamiltonian which contains
less degrees of freedom \cite{balents_fisher}. Since the scale
$ba_0$ on which the $u_{\bf k}^>$ components describe
displacements is small, we can expect that there is only {\it one
minimum} as a solution for $\tilde u_{\bf k}^>$ such that the
application of perturbative methods is allowed. To avoid further
misunderstanding we stress here that we always assume in this
Section that the system may reach thermal equilibrium, even if the
time scales are huge. Besides of corrections to terms already
present in the initial Hamiltonian also {\it new terms} may be
generated in this procedure, their precise form depends on
$V_R({\bf x},u)$.  The concrete implementation of this procedure
is in general difficult and rests often on approximations valid
close to certain critical dimensions.

The {\it second} step of the renormalization group procedure consists
in a rescaling of length, time and fields according to
   \begin{equation}
   {\bf x}  =  {\bf x}^{\prime}b,\qquad t  = t^{\prime}b^z,\qquad
   u({\bf x}) = u^{\prime}({\bf x}^{\prime})b^{\zeta}
   \label{eq:x,t,u}
   \end{equation}
   with so far unspecified {\it dynamical exponent} $z$ ($\ge0$) and
   {\it roughness exponent} $\zeta$ ($\ge 0$).
   After the first step of the
   RG-transformation the remaining minimal length scale was $\Delta
   x_{\rm min}=a_0b$ which, after rescaling according to
   (\ref{eq:x,t,u}), goes over into the original minimal length
   $\Delta x^{\prime}_{\rm min}=\Delta x_{\rm min}/b=a_0$.

Using rescaling (\ref{eq:x,t,u}) in equation
(\ref{eq:Hamiltonian}), the elastic energy in the new coordinates
obtains a factor $b^{D-2+2\zeta}$.  Since in statistical physics
the Hamiltonian appears always in the combination ${\cal H}/T$, we
can absorb this factor in a rescaled temperature
   \begin{equation}
   T^{\prime}=Tb^{2-D-2\zeta}.
   \label{eq:rescaled_temp}
   \end{equation}
The rescaling has also to be applied to the second term in
(\ref{eq:Hamiltonian}).  Having introduced $T^{\prime}$, $V_R({\bf
x},u)$ is replaced by $ b^{2-2\zeta}V_R\big(b{\bf
x}^{\prime},u^{\prime}({\bf x}^{\prime})b^{\zeta} \big)\equiv
V_R^{\prime}({\bf x}^{\prime},u^{\prime})$. The correlator of
$V_R^{\prime}$ is then given by
  \begin{equation}
  \left\langle V_R^{\prime}({\bf x}^{\prime}_1,u_1^{\prime})
   V_R^{\prime}({\bf x}_2^{\prime},u_2^{\prime})\right\rangle_R=
   b^{4-4\zeta-D}\delta({\bf x}_1^{\prime}-
   {\bf x}_2^{\prime})R\left({b^{\zeta}}(u_1^{\prime}-
   u_2^{\prime})\right)\,.
   \label{eq:rescaling.corr_V}
   \end{equation}
As mentioned already, equations (\ref{eq:rescaled_temp}) and
(\ref{eq:rescaling.corr_V})  do {\it not} represent the whole
change of $T$, $R$  under the renormalization, since the new terms
produced in the first -- non--trivial -- step of the procedure
will  generate contributions to $R^{\prime}$ and $T^{\prime}$. In
general  the whole {\it function} $R(u)$ is transformed in a
non--trivial way. We will not discuss here the derivation of this
transformation   but only present the result. The interested
reader is referred to the original articles
\cite{fisher85,fisher86,balents_fisher,balents_et_al}.
For infinitesimal changes of $b=1+\delta
b$ one obtains a continuous flow of these quantities. To lowest
order in $4-D$ one finds  with $\delta b =\delta L/L$:
\begin{equation}
\begin{split}
\frac{\partial R(u)}{\partial\ln{L}}&=(4-D-4\zeta)R(u)+
   \zeta u R_u(u)+\frac{1}{2}R_{uu}(u)^2-
   R_{uu}(u)R_{uu}(0),\\
\frac{dT}{d\ln{L}}&=(2-D-2\zeta)T+\ldots
\end{split}
\label{eq:Delta_Rprim-lprim}
\end{equation}
where $R_u(u)=\frac{\partial}{\partial u }R(u)$ etc. To make the
notation easier we have adsorbed here the coefficients of the
quadratic terms in $R$ into the redefinition of $R(u)$.

 A word of caution seems to be indicated at this point: the rescaling
(\ref{eq:x,t,u}) is a matter of convenience, not a physical
necessity.  If one does not rescale at all, the first step of the
renormalization group still keeps the whole information about the
large scale behavior of the system. In this case one obtains the
{\it effective} physical quantities (those which one observes in
experiments) on the corresponding length scale $L$.

For systems with {\it uncorrelated disorder} ($V_{R}(\mathbf{x},u)$ is a random function
of both arguments) there is  an
important simplification since  there is no renormalization of
$T$. Indeed, the full stiffness constant $C$ can be measured by
replacing the periodic boundary conditions used so far by $
u(L,x_2,...,x_D)=u(0,x_2,...,x_D)+\delta u$, that is applying an
overall strain $\frac{\delta u}{L}$ in $x_1$--direction. The effective
elastic constant $C_{{\emph{eff}}}$ follows then from the free
energy $\mathcal{F}_R(L,u)$
  \begin{equation}\label{eq:Gamma_{eff}}
  C_{{\emph{eff}}}=L^{-D+2}\partial^2\mathcal{F}_R(L,u)/\partial \delta u^2\,.
  \end{equation}
The change of the boundary conditions can be compensated by
introducing a new variable $\tilde u({\bf x})=u({\bf x})-\delta
u\,x_1/L$. Changing from $u$ to $\tilde u$ in the Hamiltonian
(\ref{eq:Hamiltonian}) adds to the elastic energy a constant
contribution $\delta {\cal H}=\frac{C}{2}L^{D-2}{\delta u}^2$ and
changes the random potential into $\int d^dx V_R({\bf x},\tilde
u({\bf x})+\delta u\,x_1/L)$. Since $ V_R({\bf x},u)$ is by
definition a random function of its arguments, $V_R({\bf x},\tilde
u({\bf x})+\delta u\,x_1/L)$ can be replaced by a random potential
$\tilde V({\bf x},\tilde u({\bf x}))$ with the same statistical
properties. Thus, the only change in the Hamiltonian is given by
the constant $\delta {\cal H}=\delta \mathcal{F}_R(L,u)$ from
which one concludes with (\ref{eq:Gamma_{eff}})
$C_{\emph{eff}}=C$. Since $T$ appears only in the combination
$C/T$,  there is also no renormalization of $T$.

   \begin{figure}[hbt]
   \centerline{\epsfxsize=7cm \epsfbox{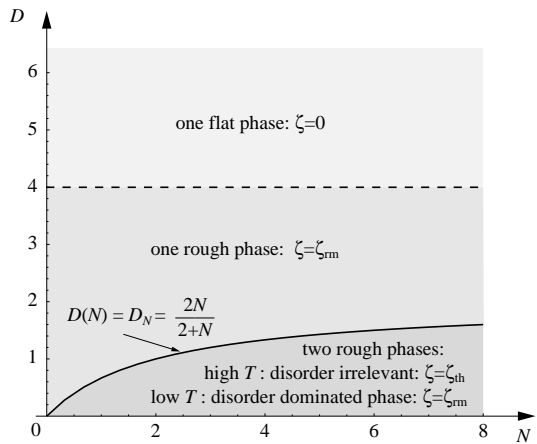}}
    \caption{The $D-N$ plane characterizing different elastic manifolds in
   a random medium ($\alpha=N$). For $D>4$  weak disorder and for $D>2$
   thermal fluctuations are irrelevant.
   For $4>D>D_N$ (i.e. $\lambda_R>0$) weak disorder is relevant leading to a non-zero roughness
   exponent. For $D_N>D$  a thermal depinning transition exists:
   at low temperature weak disorder leads to a non-trivial
   roughness exponent $\zeta$ whereas in the
   high temperature phase $\zeta_T=(2-D)/2$.}
   \label{fig:mani_phases.eps}
   \end{figure}
The  flow of $R(u)$ and $T$ as given by equation
(\ref{eq:Delta_Rprim-lprim}) will terminate in stable fixed points
(other forms of flow like limit cycles are excluded by general
reasons), which characterize physical phases. To find a {\it
finite fixed} point of the disorder (if it exists) we have to tune
the value of the exponent $\zeta$ in order to make elastic and
disorder energy scale in the same way. The fixed point condition
thus delivers the value of the exponent $\zeta$. Different fixed
points have in general different values of $\zeta$.  If several
fixed points exist -- this is the typical situation -- the domain
of attraction of a fixed point determines the {\it phase
boundary}. Clearly, the flow close to a fixed point depends on the
value of $\zeta$ at this fixed point.
   \begin{figure}[hbt]
   \centerline{\epsfxsize=10cm
   \epsfbox{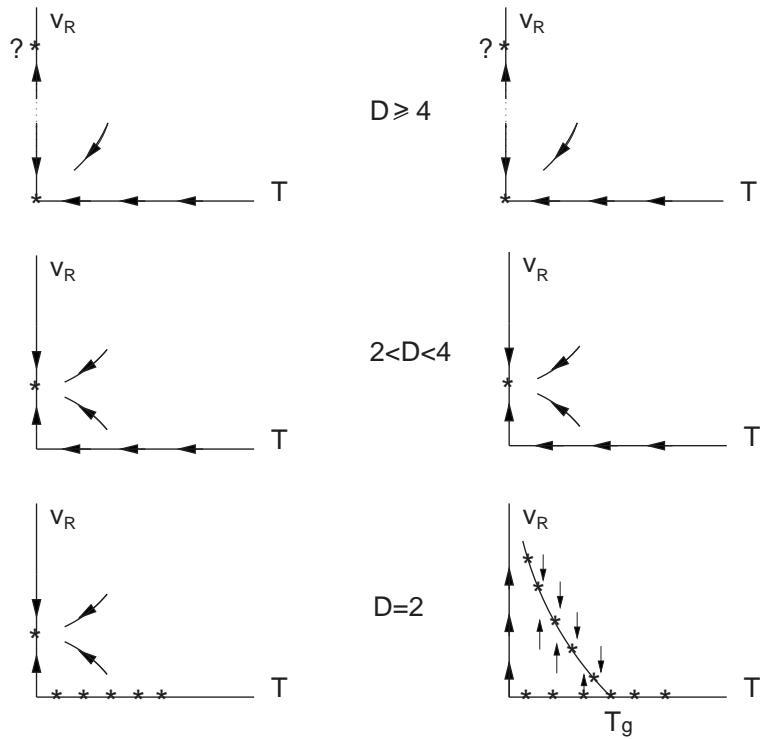}}
   \caption{ Schematic phase diagrams of impure elastic systems
    as a function
    of space dimension for (left) non-periodic media (single flux
    lines, domain walls) and (right) periodic media (flux line
    lattices, charge--density waves, Wigner crystals).
    $v_{R}$ and $T$ denote the
    strength of the disorder and temperature, respectively. Fixed points
    characterize the properties of a phase, their domain of
    attraction ends at the phase boundary. For realistic phase
    diagrams further degrees of freedom may become relevant
    (dislocations etc.). In $D\ge 4$ dimensions besides of the fixed point
    $v_{R}^{\ast}=0$
    there may be another strong disorder fixed point.}
   \label{fig:impure_elastic_systems}
   \end{figure}

\vspace{0.5cm}

 If there is no disorder, $R\equiv 0$, the roughness
exponent is $\zeta=\zeta_T\equiv\frac{2-D}{2}$ (for $D\le 2$)
according to (\ref{eq:Delta_Rprim-lprim}) and each value of $T$ is
a fixed point. Since we exclude negative roughness exponents
(there is always a constant contribution to the roughness), we put
$\zeta_T=0$ for $D>2$ and hence $T$ is transformed to zero.
Switching  on the disorder
 only the linear terms in $R$ in  equation
(\ref{eq:Delta_Rprim-lprim}) matter as long as the disorder
remains weak.
 If  the initial function $R(u)$ shows a simple power law
 behavior, $uR_u(u)\propto {-\alpha}R(u)$
 ($\alpha=-1$ for random field and $\alpha=N$
 for random manifold systems, respectively), we get
 $d\ln R/d\ln L=4-D-\zeta(4+\alpha)\equiv \lambda_R(\zeta)$.
Using as the initial value for $\zeta$ its value at the thermal
fixed point, $\zeta=\zeta_T$, we see that the disorder grows
provided $4-(4+\alpha)\zeta_T-D>0$.  To get a new stable fixed
point we have to choose then a new value for $\zeta$.
In this situation we have to take into account also the nonlinear
 contributions in equation (\ref{eq:Delta_Rprim-lprim}). The
 resulting fixed point function $R^*(u)$, which determines
 also the value of $\zeta$, depends on the initial function of
 $R(u)$ and can be  found often only numerically \cite{fisher85,fisher86}.
A characteristic feature of the $R^*(u)$ is the cusp singularity,
 $R_{uu}^*(u)-R_{uu}^*(0)\propto\mid u\mid$ for small $\mid u\mid$, which
 appears on length scales $L>L_p$
and is related to the appearance of metastability (see below).

The true roughness exponent for manifolds in \emph{random bond}
systems can then be written as an expansion in $\epsilon=4-D$. For
interfaces in random bond systems one finds
$\zeta=0.2083\epsilon+0.0069\epsilon^2+O(\epsilon^{3})$
\cite{fisher86,chauve+01}. It has been suggested that $\zeta_{RB}$ can
be written in closed form as $
\zeta_{RB}=\frac{4-D}{4+N\,\nu(D,N)}$ where $\nu(1,1)=0.5$,
$\nu(1,2)=\frac{2}{5}$ and $\nu(1,3)=\frac{8}{21}$
\cite{lassig98,Halpin-Healy.Zhang}.

For \emph{random field} system $\zeta_{RF}=\frac{4-D}{3}$
for $2<D<4$ \cite{fisher86}.

For $ \lambda_R(\zeta_T)>0$, $T/C$ is always transformed to zero,
thermal fluctuations are {\it irrelevant}. In the opposite case
$\lambda_R(\zeta_T)<0$ weak disorder is irrelevant. Increasing the
disorder strength, there is a phase transition between a disorder
dominated phase at low temperatures and a high temperature phase
where thermal fluctuations wipe out the disorder. The relation $
\lambda_R(\zeta_T)=0$ define the dimension $D(N)$ in
Figure\ref{fig:mani_phases.eps}.

\vspace{0.5cm}

The discussion applied so far to the so-called {\it random
manifold} models, which are given by a non-periodic disorder
correlators $R(u)$.
For {\it periodic media} the correlator $R(u)$
is periodic with period $b$
and this applies also to the fixed point function
$R^*(u)\sim\left(\frac{1}{36}-\Big(\frac{u}{b}\Big)^2
\Big(1-\frac{u}{b}\Big)^2\right)$. However the term $\zeta uR_u(u)$ in equation
(\ref{eq:Delta_Rprim-lprim}) violates the   periodicity, from
which one has to conclude that for periodic media $\zeta_{pm}=0$
corresponding  to a logarithmic increase of the roughness
\cite{natter90,korshunov93,giamarchi.doussal94,villain.fernandez84,emig+99}:
   \begin{equation}
   w^2_R(L)\sim (4-D)l^2\ln{\left(\frac{L}{L_p}\right)}\,.
   \label{eq:w(L)_sim_log(L/Lp)}
   \end{equation}
For some applications, e.g. for the calculation of the correlation
function $w_R(l)$, (but not for studying the effect of energy
barriers) one can derive the results for $L>L_p$ from the random
force model (\ref{eq:dH/du}) but with a modified correlator for
the random forces $V_{R,u}({\bf x})$
\begin{equation}\label{eq:eff.force.correlations}
\langle V_{R,u}({\bf x},0)V_{R,u}({\bf x'},0)\rangle_R = \int d^Dk
e^{i{\bf k(x-x')}}\frac{R_{uu}(0)}
{1+\frac{R_{uu}(0)}{R_{uu}^*}(kL_p)^{D-4}}
\end{equation}
where $R_{uu}^*\propto 4-D$. As can be seen from
(\ref{eq:eff.force.correlations}), the force correlations on
length scales $L\sim k^{-1}\ll L_p$ behaves as
$R_{uu}(0)\delta({\bf x}-{\bf x}^{\prime})$ as in the random force
approximation of Larkin \cite{larkin70}, whereas on length scales
$L\sim k^{-1}\gg L_p$ there is a long range contribution decaying
as $|{\bf x-x'}|^{-4}$.

Using the result equation  (\ref{eq:w(L)_sim_log(L/Lp)}) in the
structure factor $S({\bf k})$ one obtains a smeared (diffuse)
Bragg-peaks of finite width, $S({\bf k})\simeq |{\bf k}-{\bf
G}|^{-3+\eta_{\bf G}}$, despite of the fact that the system is
dominated by the influence of disorder, hence the name {\it
Bragg-glass} has been coined
\cite{natter90,korshunov93,giamarchi.doussal94,emig+99}.

In $D>4$ weak disorder flows to zero, i.e. it is irrelevant, but
for sufficiently strong disorder a separate strong disorder fixed
point for $R$ may exist. It should be noted that periodic media
allow the existence of topological defects like vortices or dislocations
which may destroy the Bragg-glass phase. However it can be shown
that for weak enough disorder this phase survives in $d=3$
dimensions
\cite{Kierfeld+,Carpentier+,Fisher97,Kierfeld98,kierfeld.vinokur00}.

\vspace{0.5cm}

 In $D=2$ the $T$--axis ($R=0$) is a line of fixed
points corresponding to $\zeta=0$. For random manifolds  disorder
is always relevant, i.e. $\zeta>0$ and there is a non--trivial
fixed point at $T^{\ast}=0$, $R^{\ast}>0$ (see the left side of
Figure \ref{fig:impure_elastic_systems}). For periodic media (like
flux line lattices in a thin film or Wigner crystals) the
situation is more complicated: there are two phases both with
$\zeta=0$ separated by a phase transition at $T=T_g$ (see the
right side of Figure \ref{fig:impure_elastic_systems}). In the
high--temperature phase
   \begin{equation}
   w^2(L)\sim\frac{T}{C}\ln{\left(\frac{L}{\xi}\right)}\,,
   \label{eq:w(L)_sim_log(L/xi)}
   \end{equation}
where as in the low--temperature phase ($T<T_g$)
   \begin{equation}
   w^2(L)\sim\left(\frac{T_g-T}{T_g}\right)^2\left[
   \ln{\left(\frac{L}{L_p}\right)}\right]^{2}\,.
   \label{eq:w(L)_sim_log^2(L/Lp)}
   \end{equation}
with the {\it glass temperature } $T_g\sim C$.
 The corresponding renormalization group flow diagrams for
 { random manifolds} and  periodic media in a random
matrix are shown in Figure \ref{fig:impure_elastic_systems}.

In $D=1$ dimension the roughness exponent for periodic media is
$\zeta=1/2$ both at zero and non-zero temperature
\cite{feigelman,q-slips} (compare Appendix \ref{sec:app2}).

\subsection{Metastability}

The most important feature of pinning is the appearance of
metastability. To demonstrate, how  metastability appears for weak
pinning on the Larkin scale, we consider here a $D=1$-dimensional
example. Let us assume that the renormalization group
transformation has been performed $n$ times until the Larkin
scale $L_p=a_0b^n$ is reached.
Below this scale   the perturbation theory is known to be valid
and there is no problem in deriving an effective Hamiltonian on
this length scale.

The latter can then be written as
   \begin{equation}
   {\cal H}^{(n)}=\sum_i^N\bigg\{\frac{C^{(n)}}{2}(u_{i+1}-u_i)^2
   + V_R^{(n)}(i,u_i)\bigg\}
   \label{eq:1DHamiltonian}
   \end{equation}
$C^{(n)}$ and $V_R^{(n)}(i,u_i)$ are the stiffness constant and
the random potential on scale $L_p$ and the $u_i$ are the
remaining degrees of freedom.
By definition of the Larkin scale the elastic and the random part
of the energy are of the same order of magnitude. In continuing
the real space RNG we eliminate now
half of the degrees of freedom
by minimizing the total Hamiltonian, keeping every second $u_i$
(say with $i$ even) fixed. This leads to a new effective
Hamiltonian
   \begin{equation}
   {\cal H}^{(n+1)}=\sum_{i=2m}
   \bigg\{\frac{C^{(n)}}{4}(u_i-u_{i+2})^2  +  V_R^{(n)}(i,u_i)
+ \delta V_R^{(n+1)}\left(i,\frac{u_i+u_{i+2}}{2}\right)\bigg\}\,,
   \label{eq:1DHamiltonian'}
   \end{equation}
where
   \begin{equation}\label{eq:1DHamiltonian''}
   \delta V_R^{(n+1)}(i,u)=
   \min_{u_{i+1}}\left[C^{(n)}(u-u_{i+1})^2+V_R^{(n)}(i+1,u_{i+1})\right].
   \end{equation}

\begin{figure}[hbt]
   \centerline{\epsfxsize=13cm
   \epsfbox{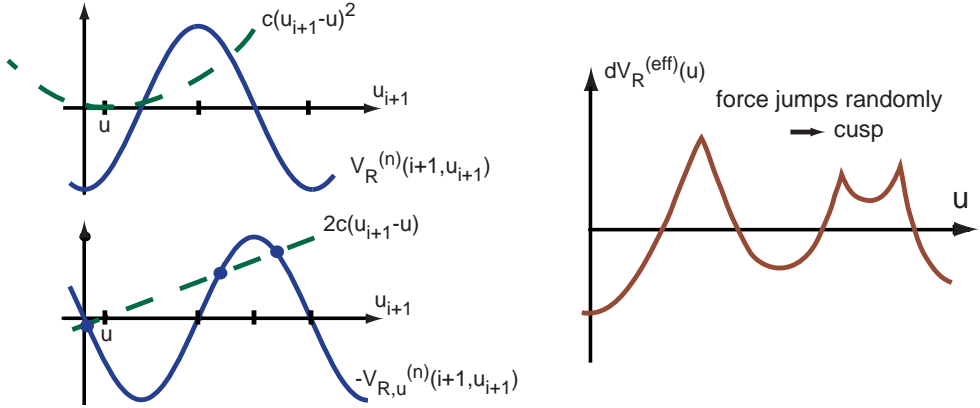}}
   \caption{Left:  The two contributions to the potential
   $\delta V_R^{(n+1)}(i,u)$ as well as their derivatives.
The points of intersection in the lower figure correspond to local
minima of the potential. Note that on scales $L\ll L_p$, where the
elastic energy is much larger than the random potential, there is
only one point of intersection and the derivative of the effective
potential is continuous. On the contrary, for $L\gg L_p$ more and
more meta stable states appear.
\newline
Right: The effective potential of $\delta V_R^{(n+1)}(i,u)$ as a
function of $u$. The potential exhibits random jumps of its
derivatives.} \label{fig:cusp}
   \end{figure}

The expression in $[...]$
consists of a parabolic potential with the minimum at $u_{i+1}=u$
plus a random potential. Since both are of the same order of
magnitude, there will be in general several local minima (compare
with Figure \ref{fig:cusp}). Let us start in a situation where
$u=0$ and $u_{i+1}=u_{(1)}$ is the true minimum of this
expression. Besides of this minimum in general several other local
minima at $u_{i+1}=u_{(m>1)}$ will exist. If we change the
external variable $u$ to values different from zero, the  minimum
$u_{(1)}$ may remain  for small values of $u$ still the global
minimum, but eventually for $u=u_c$ another minimum
$u_{i+1}=u_{(m)}$ will take over the role of the global minimum.
At this point the effective potential $\delta V_R^{(n+1)}(i,u)$ is
smooth, but the derivative jumps by an amount
   \begin{equation}
   2C^{(n)}(u_{(1)}-u_{(m)})+V_{R,u}^{(n)}(i+1,u_{(m)})-
   V_{R,u}^{(n)}(i+1,u_{(1)}).
   \end{equation}

As a result, the potential shows a {\it cusp} at $u=u_c$. Since
the magnitude and the positions $u_c$ of the jumps of the forces
are random, the effective potential acting on the degrees of
freedom on scales $L>L_p$ is scalloped as shown in Figure
\ref{fig:cusp}. Continuing this procedure by eliminating further
degrees of freedom we will obtain more of those cusps on larger
and larger length scales. Such a picture was advocated in
\cite{balents_fisher,balents_et_al}; we will demonstrate in the
Section \ref{sec:LP-CP} its derivation within the local pinning picture.
The forces generated by this effective potential
$V^{\emph{eff}}_{R}(i,u)$ change continuously between two cusps
where they jump from negative to positive values. The typical
distance between two consecutive cusps is of the order $l$ since
cusps first occur at the Larkin scales. On scales larger than $l$
the forces undergo a random walk such that
  \begin{equation}
  \big\langle\big(V^{\emph{eff}}_{R,u}(i,u)-
  V^{\emph{eff}}_{R,u}(i,u')\big)^2\big\rangle_R\propto \mid u-u'\mid
  \label{eq:potV}
  \end{equation}
for small $|u-u'|$. This structure of the force correlator is in
agreement with our RNG analysis. Indeed, since $-R_{uu}(u)$ is the
correlator of forces separated by a distance $u$, the difference
$2(R^*_{uu}(u)-R^*_{uu}(0))\propto -\mid u\mid$ denotes  the
square of the difference of these forces
averaged over the disorder. It has therefore the
same meaning and the same random walk property as the quantity in
equation (\ref{eq:potV}) \cite{balents_fisher,balents_et_al}.

So far we discussed the equilibrium, assuming an adiabatic
change of $u$. If we change however $u$ fast enough, the system
may not reach equilibrium and remain in the local minimum
$u_{(1)}$ until this minimum disappears completely or it may jump
to the new minimum with some delay. Such a situation will be
considered in the second part of this review when we consider the
dynamics of strong pinning.

\section{The close to equilibrium motion of elastic objects under an external
 dc- and ac-drive}\label{sec:creep_for_f<fp}

In this Section we want to consider the creep motion of the
elastic object in a random environment under the influence of a
{\it weak external driving force} density, $f\ll f_p$, where
$f_p=T_p/lL_p^D$. ($f_p$ is of the order of the zero temperature
depinning threshold $f_c$ discussed in the next Section.) We are
now in a {\it non--equilibrium} situation which requires its own
treatment. If $f$ is however small, as we will assume in this
Section, we are sufficiently close to the equilibrium such that we
can still use our findings of the previous Section
(for earlier descriptions of relaxation phenomena in CDWs
see e.g. reference \cite{Littlewood+88}).

\subsection{Constant driving force}

We first consider the case of a constant driving force $f$.
All changes in $f$ are assumed to be made {\it adiabatically}.
The equation of motion will be assumed to be overdamped
with a bare mobility $\gamma$
\begin{equation}
\frac{1}{\gamma}\frac{\partial u}{\partial t}=
-\frac{\delta {\cal H}}{\delta u}+\eta({\bf x},t)=
C\bnabla^2u-V_{R,u}({\bf x},u)+f+\eta({\bf x},t)
\label{eq:eq.motion}
\end{equation}
where $\eta({\bf x},t)$ denotes the thermal noise
\begin{equation}
\langle\eta({\bf x},t)\eta({\bf x}',t')\rangle_T=2\frac{T}{\gamma}
\delta({\bf x}-{\bf x}')\delta(t-t')\,. \label{eq:thermal noise}
\end{equation}
In the following we will repeatedly consider the elastic object on
a variable length scale $L$ which may vary from the microscopic
cut--off $a_0$ to the system size $L_0$.
The coupling between different length scales due to the anharmonic
random potential in (\ref{eq:Hamiltonian}) is at least partially
incorporated into these considerations by (i) the use of the
non--mean field exponents $\zeta$ and $\chi=D-2+2\zeta$ {\it and}
(ii) the condition, that in order to have a moving elastic object
on scale $L$, the system has to be able  to move on {\it all}
length scales below $L$.  From the dynamical point of view our
present analysis is a type of a mean-field treatment.

As follows from equations (\ref{F-fluctuations1}), the typical
free energy fluctuations on the scale $L$ are of the order $F(L)=
T_p(L/L_p)^{\chi}$.  The {\it energy barriers} between different
metastable states scale as $E_B(L)\approx T_p (L/L_p)^{\psi}$ with
an exponent ${\psi}$ which is in general different from $\chi$. In
the following we will however assume that ${\psi}=\chi$.  For some
systems this can be shown explicitly (see e.g.
\cite{Drossel+95,Mikheev+95}). In general this is not the case. A
counter example is  a system with an additional isolated potential
peak $V_p\delta({\bf x}-{\bf x}_0)\delta(u)$, which forms a
barrier which can never be overcome by thermally activated hopping
(unless we give up the elastic approximation and include
topological defects in the structure under consideration). In
equilibrium statistical mechanics, however, this potential peak
does not play a role, since it can easily be avoided by the
elastic object. This example directs us to the picture of the
local pinning considered in the Sec. \ref{sec:plastic} and later
on.

Next we include a small driving force density $f$. Taking into
account that the typical distance between different metastable
states is of the order $w(L)$ (see Figure
\ref{fig:rare_fluctuations}), we can write for the expression of
the total energy barrier
   \begin{equation}
   E_B(L,f)\approx F(L)-fL^Dw_R(L)= T_p\left(\frac{L}{L_p}\right)^{\chi}
   \left(1-\left(\frac{L}{L_f}\right)^{2-\zeta}\right)\,.
   \label{eq:total_E_B}
   \end{equation}

\begin{figure}[hbt]
   \centerline{\epsfxsize=8cm
   \epsfbox{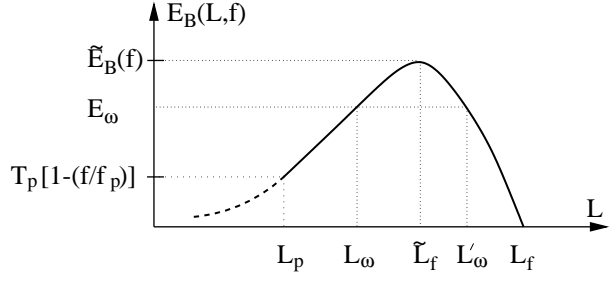}}
   \caption{ Energy barrier as a function of
   the length scale $L$ for a given driving force density $f$. The different lengths and
   energy scales are explained in the text.}
   \label{fig:E_B}
   \end{figure}

The second term on the rhs of equation (\ref{eq:total_E_B}),
$-fL^Dw_R(L)$, describes the reduction of the barrier due to the
tilt of the potential by the external force $fL^D$. In rewriting
this term we used (\ref{eq:w(L)}) and (\ref{eq:L_p}) and
introduced the force length scale $L_f$ associated with the
equilibrium length scale $L_p$:
   \begin{equation}
   L_f=L_p\left(\frac{f_p}{f}\right)^{1/(2-\zeta)}\,,\quad
   f_p=\frac{T_p}{lL_p^D}=C l L_p^{-2}\,.
   \label{eq:L_f}
 \end{equation}
 $E_B(L,f)$ is shown in
Figure \ref{fig:E_B}, it has a maximum at
$L=\tilde{L}_f=L_f(\frac{\chi}{\chi+2-\zeta})^{1/(2-\zeta)}<L_f$
and vanishes for $L=L_f$. Applying a small driving force $f$
corresponds to testing the system on a large length scale $L_f$.
Note, that $E_B(L=L_p,f)=T_p\big[1-(f/f_p)\big]$.

Assuming for the moment that (\ref{eq:total_E_B}), (\ref{eq:L_f})
are valid up to $f\sim f_p$, we see that $f\approx f_p$ determines
the \emph{depinning} threshold then there is no energy barrier
left in the system. Note however that we derived
(\ref{eq:total_E_B}) under the condition $f\ll f_p$, the vicinity
$f\sim f_p$ requires a special treatment which we will consider in
the following Section.

For  $f\ll f_p$ the elastic object is restricted in its motion by energy
barriers of maximal height
   \begin{equation}
   \tilde E_B(f)\equiv E_B(\tilde L_f,f)\approx
   T_p\left(\frac{f_p}{f}\right)^{\mu}\,,\quad \mu=\frac{\chi}{2-\zeta}\,,
   \label{eq:tilde_E_B}
   \end{equation}
which can be only overcome by thermally activated hopping. The
{\it creep velocity} of the elastic object  follows from
$v_{creep}\approx w(\tilde L_f)/\tau(\tilde L_f)$ where we use the
Arrhenius law for the hopping time $\tau\sim\omega_p^{-1}
e^{\tilde E_B(f)/T}$ and $\omega_p\approx C\gamma/L_p^2=\gamma
f_p/l$. This results in a creep velocity
   \begin{equation}
   v(f)\approx\frac{w(\tilde L_f)}{\tau(\tilde L_f)}\propto
   e^{-\frac{T_p}{T}\big(\frac{f_p}{f}\big)^{\mu}}\,.
   \label{eq:v_creep}
   \end{equation}
We omitted the prefactor on the r.h.s. which is beyond the
accuracy of the present considerations. This formula is valid for
$T\ll\tilde E_B(f)$ and was found first by Ioffe and Vinokur
\cite{ioffe.vinokur87} (see also \cite{Nattermann87}).
Equation (\ref{eq:v_creep}) was derived from a renormalization
group treatment in \cite{Radzihovsky98:mm,Chauve+98}.
In the opposite case $T\gg\tilde E_B(f)$ we expect a linear relation
between the driving force and the velocity:
   \begin{equation}
   v\simeq\gamma f.
   \label{eq:linear_v}
   \end{equation}
The border line
between the two cases, $T\approx\tilde E_B(f)$, defines a {\it temperature
     dependent force} $f_T$
   \begin{equation}
   f_T=f_p\left(\frac{T_p}{T}\right)^{1/\mu}\,.
   \label{eq:temp_dependent_force}
   \end{equation}
Note that the creep formula is valid only for $f\ll f_T$, i.e. for
$f\ll f_p$ and $T\ll T_p$.

\subsection{Periodically oscillating driving force}
\label{sec:Periodically_oscillating_driving_force}
\begin{figure}[hbt]
   \centerline{\epsfxsize=6cm
   \epsfbox{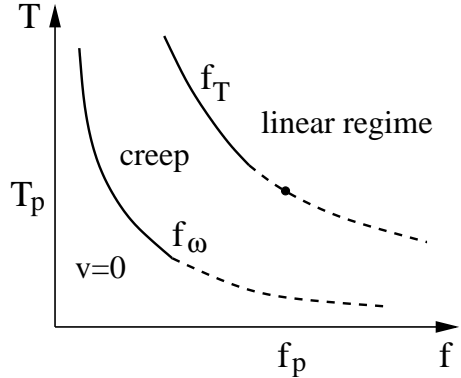}}
   \caption{The cross-over fields $f_T$ and $f_{\omega}(T)$ as a function
   of temperature. Note that for $\omega\to 0$ $f_{\omega}$ approaches the
   $T$-- and $f$--axis, respectively. For $f_{\omega}<f<f_T$
   the elastic object follows the external field in a creep like motion
   whereas for $f<f_{\omega}$ only segments of elastic object of size
   $L<L_{\omega}(T)$ (defined in (\ref{eq:max_height_barriers}))
   can follow the field and the average velocity of the whole object
   vanishes.}
   \label{fig:Tf_regime}
   \end{figure}

 Next we
consider the motion under influence of an {\it ac driving force}
with a finite frequency $\omega\ll \omega_p=\gamma f_p/l$
   \begin{equation}
   f(t)=f_0\sin{(\omega t)}\,.
   \label{eq:ac_driving_force}
   \end{equation}
From the Arrhenius law we conclude that in driving the system over
the period of time $\pi/\omega$ only barriers of maximal height
$E_{\omega}(T)$ on a corresponding length scale $L_{\omega}$ given
by (compare Figure \ref{fig:E_B}):
   \begin{equation}
   \frac{1}{\omega}\omega_p e^{-E_{\omega}(T)/T}
   \approx 1\qquad \text{i.e.}\qquad
   E_{\omega}(T)\equiv T_p\left(\frac{L_{\omega}}{L_p}\right)^\chi
   =T\ln{\left(\frac{\omega_p}{\omega}\right)}
   \label{eq:max_height_barriers}
   \end{equation}
can be overcome. If $E_{\omega}(T)>\tilde E_B(f)$, the oscillating force has enough time to
trigger jumps over all relevant barriers and hence the creep
formula (\ref{eq:v_creep}) is still valid (with $f$  replaced by
equation (\ref{eq:ac_driving_force})). In the opposite
limit this is not longer the case and hence there is no global motion, $v\equiv 0$. The relation
$E_{\omega}(T)= \tilde E_B(f)$ determines a temperature and
frequency dependent \emph{cross--over force} $f_{\omega}(T)$
   \begin{equation}
   f_{\omega}(T)\approx f_p\left(\frac{T_p}{T\ln{\left(\frac{\omega_p}{\omega}\right)}}
   \right)^{1/\mu}
   \approx
   f_T\left(\ln{\left(\frac{\omega_p}{\omega}\right)}\right)^{-1/\mu},
   \label{eq:frequency_force}
   \end{equation}
which separates the creep region $f_{\omega}<f\ll f_T$ from the region
$f<f_{\omega}$ where $v\equiv 0$.
 The crossover lines $f_T$ and $f_{\omega}$ are depicted in Figure
\ref{fig:Tf_regime}. For a discussion of the region $f \approx f_p
$ see Section \ref{sec:critical-region-f}.

For $f<f_{\omega}$ the elastic object as a whole cannot follow the
rapidly oscillating external driving field.
However there is still a local motion of parts of the elastic object
corresponding to length scales
\begin{equation}\label{L.omega}
  L<L_{\omega}=L_p\left(\frac{T}{T_p} \ln
{\frac{\omega_p}{\omega}}\right)^{1/\chi}
\end{equation}
 (see Figure
\ref{fig:E_B}). We consider to this point in the next Section
where we will treat these fluctuating parts as {\it two--level
systems}. As a side remark we note that for similar reasons
the specific heat obtains, if measured over a time scale $t$,
an extra factor $\left[1-\left(\frac{T}{T_p}
\ln\omega_pt\right)^{-(D+\chi)/\chi}\right]>0$ on the r.h.s. of equation
(\ref{eq:specific.heat.2}).

   \bigskip

\subsection{Dynamic response of the pinned elastic object:
Two-level systems }\label{sec:dyn-response}

In this subsection we consider  the influence of an external
time--dependent field $f(t)=f\sin \omega t$ on a pinned elastic
object in the region $f\ll f_{\omega}$. Thermal motion over energy
barriers $E_B(L<L_{\omega})$ mediates transitions between
configurations which have an energy difference $\Delta E \lesssim
T$. As we discussed already in Section 2.3, the distribution
$P(\Delta E,L)$ of $\Delta E$ is smooth and has a width of order
$F(L)=T_p(L/L_p)^{\chi}$. Hence  there are only rare pairs of
metastable configurations with $\Delta E\leq T$. We therefore
model the object as an ensemble of noninteracting two--level
systems \cite{ioffe.vinokur87,korshunov01}.

 In the following we discuss the dissipation in a two-level system
 due to an applied ac-field. We begin with a discussion of a given
 two-level systems on length scale $L$. The separation between the
  two minima
is $w\sim w(L)$ and the energy difference is $\Delta E$. Then the
probability that in thermal equilibrium the system is in the
higher--energy minimum is
   \begin{equation}
   n_0(\Delta E)=\frac{e^{-\Delta E/T}}{1+e^{-\Delta E/T}}=
   \big[\exp{(\Delta E/T)}+1\big]^{-1}\,.
   \label{eq:occupation.number}
   \end{equation}

The external field $f(t)$ disturbs the energy difference $\Delta
E$ by $\delta E(t,L)\approx f(t)L^{D}w_R(L)$. Therefore the system
relaxes to the new time--dependent equilibrium configuration
$n(\Delta E+\delta E)\approx n_0(\Delta E)+\delta n(t)$. The time
dependence of $\delta n$ is controlled by the relaxation time
   \begin{equation}
   \label{eq:lifetime}
   \tau(L)\approx
   \omega_p^{-1}\exp{\big(E_B(L)/T\big)},\qquad \omega_p\approx
   C\gamma /L_p^2=\gamma f_p/l
   \end{equation}
 of the two-level system and by the time dependence of $\delta E(t,L)$. In a linear
approximation the time dependence of $\delta n(t)$ is therefore
described by the equation
   \begin{equation}
   \left[\frac{\partial}{\partial t}+\frac{1}{\tau}\right]\delta n+
   \frac{\partial n_0}{\partial \Delta E}\frac{\partial\delta E}{\partial t}
   =0\,.
   \label{eq:lin.app}
   \end{equation}
The power dissipated in this way by the two-level system
of linear size $L$ is given by
   \begin{equation}
   {\cal W}(L,\omega)\sim
   -Re\left\langle\delta n^{\ast}L^Dw_R(L)\frac{df}{dt}\right\rangle_{\omega}\,,
   \label{eq:dissipated-power}
   \end{equation}
where $Re\langle...\rangle_{\omega}$ denotes the real part of the
Fourier transform. With this and the Fourier transform of equation
(\ref{eq:lin.app}), we get the power absorbed by the given
two-level system:
   \begin{equation}
   {\cal W}(L,\omega)\sim
   \frac{1}{4T}\left[\cosh{\left(\frac{\Delta E}{2T}\right)}
   \right]^{-2}(\delta E(t,L))^2\frac{\omega^2\tau}{1+\omega^2\tau^2}\,,
   \label{eq:absorbed-power}
   \end{equation}
where $\delta E=fL^{D}w_{R}(L)$.
To get the  power  dissipated by all two-level systems on scale
$L...L+\delta L$ we have to average this expression with $P(\Delta E,L)$,
equation(\ref{eq:Delta-E}), and multiply it with density
$\nu(L)dL\approx\frac{dL}{L^{D+1}}$ (compare with equation
(\ref{eq:specific.heat})) of the two-level systems on scale $L$.
Since the distribution function for the $\Delta E$ is smooth and
has a width of the order $F(L)\gg T$, only the fraction $T/F(L)$
of them contributes to the average. Hence we obtain for the total
power density dissipated by all two-level systems
   \begin{eqnarray}
   {\cal W}_{total}(\omega) & \sim &
   \int_{L_p}^{\infty}dL\,\nu(L)\int_0^{\infty}d(\Delta E)
   P(\Delta E,L){\cal W}(L,\omega)\nonumber\\
   & \approx & \int_{L_p}^{\infty}\frac{dL}{L}
   \left(\frac{1}{L}\right)^{D}\frac{\delta E^2(L)}{F(L)}
   \frac{\omega^2\tau(L)}{1+\omega^2\tau^2(L)}
   \label{eq:total-power}
   \end{eqnarray}
The energy dissipation $ {\cal W}(L,\omega)$ is related to the
imaginary part of the dynamic susceptibility
\begin{equation}
\chi(L=2\pi/\mid{\bf k}\mid,T,\omega)\sim \int d(t-t^{\prime})
e^{i\omega (t-t^{\prime})} \langle\langle\partial u_{\bf
k}(t)/\partial f_{-\bf k}(t^{\prime})\rangle_T\rangle_R
\label{eq:complex_suscept}
   \end{equation}
 \begin{equation}
   {\cal W}(L,\omega)\sim\frac{1}{2}\omega\chi^{\prime\prime}(L,T,\omega)
   f^2\,.
   \label{eq:power-suscept}
   \end{equation}
The total susceptibility is given by the integral over $L$ with
the probability distribution $\nu(L)dL$ similarly to
(\ref{eq:total-power}). The main contribution to the real par
$\chi^{\prime}(\omega)$ comes from the length scale $L$ which
fulfills  the condition $\partial\chi^{\prime}(L,\omega)/\partial
L=0$. This yields $(\omega\tau)^{-2}\approx\big[\chi
E_B(L)/T-1\big]$. For low frequencies and temperatures this gives
$L\approx L_{\omega}$ (compare with equation (\ref{L.omega}))  and
hence
   \begin{equation}
   \chi^{\prime}(T,\omega)\sim\int dL\nu(L)P\big(\Delta E(L),L\big)\chi^{\prime}(L,T,\omega)
   \approx \frac{L_p^2}{C}\left[\frac{T}{T_p}
   \ln{(\frac{\omega_p}{\omega})}\right]^{2/\chi}.
   \label{eq:L-low.temp}
   \end{equation}
Decreasing the frequency leads to an increase of the
susceptibility which is a typical experimental trend, as well as the
logarithmic dependence on the frequency $\omega$.

\section{Critical Depinning at $T=0$}\label{sec:critical-region-f}

\subsection{Constant driving force, $f\geq f_c$}
\label{subsec:const_driving_f}
\begin{figure}[hbt]
   \centerline{\epsfxsize=6cm
   \epsfbox{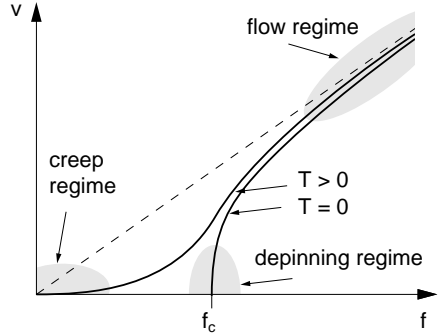}}
   \caption{The velocity of the driven elastic object as a function
   of the driving force $f$ both at zero and non-zero temperatures.}
   \label{fig:vf}
   \end{figure}

 So far we considered the region
$f\ll f_p$. Increasing $f$ we expect to reach a {\it critical
force density} $f_c$ (as it turns out it is of the order of $f_p$)
at which the elastic object is depinned. Above the depinning
transition the elastic object moves even without the help of
thermal activation with a finite velocity $v$ which reaches the
non-critical regime $v\approx \gamma f$ at large driving forces
$f$. Qualitatively the $v-f$ diagram is depicted in Figure
\ref{fig:vf}.
 All changes in $f$
are again to be performed adiabatically. The equation of motion in
the absence of thermal fluctuations reads
\begin{equation}
\frac{1}{\gamma}\frac{\partial u}{\partial t}=C {\bf
\nabla}^2u+f+g({\bf x},u)=-\frac{\delta{\cal H}}{\delta u}\,.
\label{eq:motion}
\end{equation}

Here we introduced the pinning force density
\begin{equation}
g({\bf x},u)\equiv -V_{R,u}({\bf x},u)\equiv -\frac{\partial
V_R}{\partial u} \label{eq:g(u)}
\end{equation}
 resulting from the random potential.
It is also Gaussian distributed with the correlator
   \begin{equation}
   \langle g({\bf x},u) g({\bf x'},u')\rangle_R=\delta({\bf x}-{\bf x'})
   \Delta(u-u'),\quad \Delta(u)=- R_{uu}(u)\equiv
   -\frac{\partial^2R}{\partial u^2}
   \label{g_correlator}
   \end{equation}
for the bare (unrenormalized) correlators. If the object is
completely stiff, i.e. $u({\bf x},t)\equiv u(t)$, -- this is the
situation if $L_{0}\ll L_p$, --, then the  average
pinning force density $\langle g({\bf x},u)\rangle_R$ vanishes and
its fluctuations are of the order $(L_0^{-D}\Delta(0))^{1/2}$ and
hence arbitrarily small for a macroscopic object.
  Thus a rigid object would never be pinned.

Next we shall use perturbation theory for weakly distorted elastic
objects \cite{larkin.ovch79,lee.rice79}.
To this aim it is convenient to go over to a {\it
co-moving frame} by rewriting $u({\bf x},t)=vt+\tilde u({\bf
x},t)$ with $\langle\tilde u\rangle_R=0$ and look for the lowest
non-zero correction to the velocity. Indeed such an approach works
well at high velocities where the displacements $\tilde u$
are small. This gives
   \begin{equation}
   \gamma^{-1}v-f=\langle g({\bf x},vt+\tilde u({\bf x},t))\rangle_R\approx
   \langle g_{u}({\bf x},vt)\tilde u({\bf x},t)\rangle_R
   \label{eq:v1}
   \end{equation}
where $ g_{u}({\bf x},u)=\partial g({\bf x},u)/\partial u$. In
the lowest order perturbation theory
   \begin{eqnarray}\label{eq:perturbation.theory.u.tilde}
   \tilde u ({\bf x},t)&=& \int d^Dx^{\prime} \int_{-\infty}^t dt'
   G_0({\bf x}-{\bf x}',t-t')g({\bf x}',vt'), \\
   G_0({\bf x},t)&=& \gamma
   \int_{\bf k}e^{i \bf {kx}}e^{-C\gamma {\bf k^2}t}\nonumber
   \end{eqnarray}
and with the help of (\ref{g_correlator}), we get from the r.h.s.
of equation (\ref{eq:v1})
   \begin{equation}
   \gamma^{-1}v-f=(4\pi C)^{-D/2}\int_{(C\Lambda^2)^{-1}}^{\infty}d\tau
   \tau^{-D/2}\Delta_u(v\tau/\gamma) \label{eq:v2}
   \end{equation}
The large scale momentum cut--off $\Lambda\sim a_0^{-1}$ appearing
in (\ref{eq:perturbation.theory.u.tilde}) was taken into account
by adding a factor
$e^{-k^2/\Lambda^2}$. Moreover, we introduced $\tau =t\gamma$ as a new variable.
We imply here that $D>2$ and hence the integral (\ref{eq:v2})
is convergent at large $\tau$. Below we will see that this restriction
is unnecessary.
Decreasing $f$ and hence $v$,  one finds for $v\rightarrow 0+$ a
non-zero threshold  $f_c$ if and only if $\Delta_u(0+)$ is
non-zero:
   \begin{equation}
   f_c\sim -\frac{\Delta_u(0+)\Lambda^{D-2}}{(D-2)C}
   \label{f_c}
   \end{equation}
The result (\ref{f_c}) looks at a first glance different from
$f_p\approx C l L_p^{-2}$ which we found in the previous Section.
Most importantly, the correlator $\Delta(u)=-\Delta(-u)$ of the
random forces has to have a {\it cusp-like singularity} at the
origin since $-\Delta_u(0+)=\Delta_u(0-)$ has to be positive. This
is not what one gets naively from a weak random potential
$V_R({\bf x},u)$ which shows an analytic behavior of $\Delta(u)$
for small $u$ and hence $\Delta_u(0+)=0$.

This long standing problem has been overcome by the
renormalization group theory of critical depinning at $T=0$
\cite{narayan.fisher92,natter+92,narayan.fisher93,Ertas+94,chauve+01}.
It was shown that the force-force correlator $\Delta(u)$ develops
indeed a cusp-like singularity on scales $L>L_p$ after the degrees
of freedom on scales $L<L_p$ have been integrated out.
To lowest order in $\epsilon=4-D$ the renormalization group equation
for $\Delta(u)$ in this non-equilibrium situation is identical to that
for $-R_{uu}(u)$ following from equation (\ref{eq:Delta_Rprim-lprim})
(calculated under
equilibrium conditions) by differentiating $R(u)$ twice with
respect to $u$. Note that this simple relation breaks down to order
$\epsilon^2$ \cite{chauve+01}.
Hence
$\Delta_u(u\to 0)\neq 0$. The force correlator on these scales
becomes scale dependent and reads
   \begin{equation}\label{eq:Delta.replacement}
   \Delta (u)\rightarrow \Delta(u;L)\propto
   (Cl/L_p^{\tilde\zeta})^2L^{-4+D+2\tilde\zeta}
  \Delta^*\left(u(L/L_p)^{-\tilde\zeta}/l\right),
   \end{equation}
where the function $\Delta^*(y)$ has a cusp  for small $y$.
Similarly, the mobility $\gamma$ is replaced by an effective scale
dependent expression
   \begin{equation}\label{eq:effective.mobility}
   \gamma\rightarrow \gamma(L)\approx
   \gamma\left(\frac{L}{L_p}\right)^{2-\tilde z}\,.
   \end{equation}
$\tilde\zeta$ and $\tilde z$ are two new {\it non-equilibrium
critical exponents} which can be calculated by an expansion in
$\epsilon=4-D$. Note that these exponents are in general {\it
different} from the equilibrium exponents introduced in the
earlier Sections. If we replace $\Lambda^{-1}$ by $L_p$ and
$\Delta_u(0)$ by $\Delta(u,L_p)$ in equation (\ref{f_c}), and put
$\tilde\zeta=0$ (since the corresponding integral is dominated by
small scales $L\approx L_p$), we indeed arrive at $f_c\approx
f_p$. A detailed calculation gives \cite{natter+92,chauve+01}
\begin{equation}\label{eq:threshold}
  f_c=\frac{-1}{2-\tilde\zeta}\Delta^{*}_y(0+)f_p.
\end{equation}
Qualitatively, the cusp singularity can be understood as follows:
In order to obtain a non-zero depinning threshold the average
value of $\lim_{v\rightarrow 0}\langle g({\bf x},vt+\tilde u({\bf
x},t))\rangle_R$ in equation (\ref{eq:v1}) has to be negative. In
other words, the elastic object ``sees'' in a pinned configuration
more increasing than decreasing potential hills, even at $v\to 0$!
\footnote{As mentioned already, within the collective pinning
regime this is only possible on scales larger than the
Larkin-length, since for $L<L_p$ the elastic object is essentially
undistorted. More generally, the cusp appears together with
metastable states; it follows from the possibility to switch
between ascending and descending branches (in terminology of the
Section \ref{ss:ms}, preferably selecting the lowest one (see the
derivation of the force correlator in the Section
\ref{sec:LP-CP}).}
 The next step in (\ref{eq:v1}) is to expand $g({\bf
x},vt+\tilde u)$ with respect to $\tilde u$. This leads to a
product $\langle g_{u}({\bf x},vt)g({\bf x},vt')\rangle_R$ with
$t>t'$, which, in the limit $v\rightarrow 0$ (hence $vt\rightarrow
u_0, vt'\rightarrow u_0-\epsilon_v$) has to be negative as well:
   \begin{equation}
   \langle  g_{u}({\bf x},u_0) g({\bf x},u_0-\epsilon_v)
   \rangle_R \approx
   \langle  g_{u}({\bf x},u_0)[g({\bf x},u_0)-\epsilon_v
   g_{u}({\bf x},u_0)]\rangle_R<0
   \label{curvature}
\end{equation}
For a typical potential dominating the correlations of the random
forces two  cases are possible:

(i) If $g_{u}({\bf x},u_0)<0$, i.e. the force is locally
decreasing with increasing $u$, then the force at $u_0-\epsilon_v$
has to be positive, i.e. accelerating. This is the situation
shortly before one reaches a potential minimum.  For $g<0$ one
concludes from the r.h.s. of (\ref{curvature})
$|g_u(u_0)|>|g(u_0)|/\epsilon_v$, i.e. in the limit $v\sim
\epsilon_v \rightarrow 0$ the curvature of the potential of pieces with $g<0$
becomes arbitrarily large and correspondingly these  pieces of the
potential shrink to zero.

(ii) If on the contrary $ g_u({\bf x},u_0)>0$, i.e. the force is
locally increasing with increasing $u$, then the force at
$u_0-\epsilon_v$ has to be negative, i.e. retarding.  This is the
situation shortly before one reaches a potential maximum. For
$g>0$ one concludes from the r.h.s of (\ref{curvature})
 $g_u(u_0)>g(u_0)/\epsilon_v$, the curvature in these pieces of the
potential diverges for  $v\sim \epsilon_v \rightarrow 0$ and hence
these pieces of the potential disappear as well.

\begin{figure}[hbt]
   \centerline{\epsfxsize=8cm
   \epsfbox{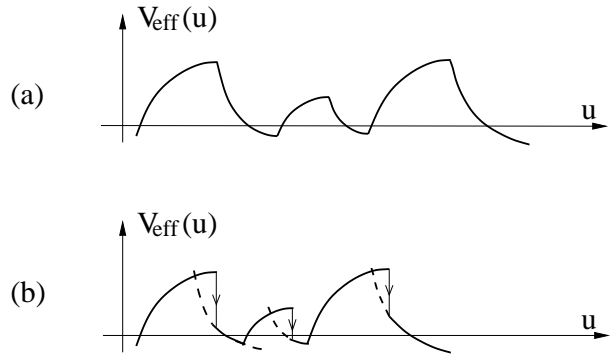}}
   \caption{(a) The effective potential as it emerges on scale $L>L_p$
   for an elastic object moving with a positive velocity. The shape results
   from the requirement that $\Delta^{\prime}_u(0+)$ is negative {\it}and that at
   the boundaries of each segment $g_u(u)g(u)=0$. (b) The effective
   potential resulting only from the condition that $g_u(u)g(u)$ is
   negative everywhere. We consider the case (b) to be the generic
   one.
Note, that orientation of the potential is reversed for a motion
with
   a negative velocity.}
   \label{fig:V_eff_new}
   \end{figure}

Therefore for $v\sim \epsilon_v \rightarrow 0$ the effective
potential consists mainly of pieces where $g$ and $g_u$ have
different signs. Regions with the same sign of $g$ and $g_u$
disappear gradually from the effective potential emerging on
scales $L>L_p$ (or at least give only a small contribution to it).
If we assume that only pieces with $gg_u\le 0$ indeed remain in
the effective potential and assume that these pieces extend up to
$gg_u= 0$, then a continuous and piecewise differentiable
potential can be constructed from alternating segments where $g$
and $-g_{u}$ are both negative or positive, respectively. Such
potential pieces are given by $V_+(u)$ for $ 0<u<u_+$ and $V_-(u)$
for $ u_-< u<0$ which we can model as e.g. (compare Figure
\ref{fig:V_eff_new}):
    \begin{equation}
    V_+(u)=f_+u\left(1-\frac{1}{3}\frac{u^2}{u_+^2}\right),\qquad
    V_-(u)=f_-\frac{u^2}{u_-}\left(1-\frac{1}{3}\frac{u}{u_-}\right).
    \label{V_eff}
\end{equation}
The forces $f_{\pm}$ and positions $u_{\pm}$ may change from
segment to segment but have to guarantee the continuity of the
potential. According to what we said above, the average value of
$f_+=O(f_p)$ has to be identified with the depinning threshold.
Such a shark-fin potential is schematically drawn in Figure
\ref{fig:V_eff_new}a. Its appearance in a driven situation is
rather obvious since the elastic object will pass the regions
where $g_ug>0$ very quickly. Together with the property
$\Delta(u)=\Delta(-u)$ this explains the physical origin of the
cusp. (See also the above footnote; the precise meaning of the
"shark-fin" potentials is given by "termination points" of
metastable branches of the Section \ref{ss:ms} and corresponding
figures.)

For a negative driving force a corresponding discussion leads to a
potential in which $g$ and $g_u$ have the same signs, regions with
the different signs of $g$ and $g_u$ disappear. This corresponds
to a potential which differs from that drawn in Figure
\ref{fig:V_eff_new} by changing the direction of the $u$- axis. If
we do not impose the additional condition that $gg_u=0$ at the
boundaries of the intervals where $g_u>0$ or $g_u<0$,
respectively, then the potential may be discontinuous as depicted
in Figure \ref{fig:V_eff_new}b.

Another problem following from perturbation theory is the fact
that for $D\le 2$ the r.h.s. of equation (\ref{eq:v2}) could
diverges. This also has been overcome by the renormalization group
theory of the critical depinning
\cite{narayan.fisher92,natter+92,narayan.fisher93,chauve+01}:
$D=2$ is not longer the lower critical dimension due to the
appearance of non-classical critical exponents in the renormalized
perturbation theory.

  Close to the
depinning transition the velocity - which can be considered as an
order parameter of the transition - vanishes as a power law
   \begin{equation}
   v\approx v_p\left(\frac {f-f_c}{f_c}\right)^{\tilde\beta}, \quad  f>f_c.
   \label{eq:v}
   \end{equation}
Here we introduced
the characteristic velocity scale $v_p=\gamma f_p$.
Approaching the depinning transition
there is a diverging correlation length
   \begin{equation}
   \xi\approx L_p\left|\frac{f-f_c}{f_c}\right|^{-\tilde\nu}.
   \label{eq:xi}
   \end{equation}
The appearance of a diverging correlation length on both sides of the
depinning transition has to be expected for the following reason: if we
approach $f_{c}$ from values $f<f_{c}$ by increasing $f$ adiabatically,
larger and larger avalanches of local motion of the elastic object will occur until
we reach a critical state at $f=f_{c}$. A further increase of $f$ will then
lead to a macroscopic motion of the elastic  object. At $f=f_{c}$ a local
perturbation will hence trigger a global response, corresponding to an infinite
correlation length. With increasing velocity spatial fluctuations in the
local velocity will be reduced and the correlation length shrinks again.
We will come back to this point in the following.

On length scales $L_p\ll L\ll\xi$ the non-linearities of the
pinning potential dominate and
distortions of the elastic object obey dynamical scaling
with non trivial exponents
$\tilde\zeta$ and $\tilde z$:
   \begin{equation}
   \langle\left(\tilde u({\bf x},t)-\tilde u({\bf x'},t')\right)^2
   \rangle_R^{1/2}=
   l\left(\frac{|{\bf x}-{\bf x}^{\prime}|}{L_p}\right)^{\tilde\zeta}
   \tilde\Phi\left(\Big(\frac{|{\bf x}-{\bf x}^{\prime}|}{L_p}\Big)^{\tilde z}
   \Big/\omega_p(t-t^{\prime})\right)\,.
   \label{dyn.scaling}
   \end{equation}
Here $\tilde \zeta$ is the non-equilibrium roughness exponent and
$\omega_p=v_p/l$. It turns out that $1<\tilde z <2$, i.e. the
dynamics close to the depinning transition is {\it
super-diffusive}, reflecting the rapid motion of the object after
the maximum of the shark fin potential has been overcome. The
scaling function $\tilde\Phi(y)$ behaves as
$y^{-\tilde\zeta/\tilde z}$ for $y\to 0$ and approaches a constant
for $y\to\infty$. The critical exponents satisfy the new scaling
relations \cite{natter+92}
   \begin{equation}
   \tilde\nu=\frac{1}{2-\tilde\zeta}=
   \frac{\tilde\beta}{\tilde z-\tilde\zeta}\geq \frac{2}{D+\tilde\zeta}.
   \label{exponent.relations}
   \end{equation}
These exponents were calculated first to order $\epsilon=(4-D)$ in
\cite{natter+92} and recently to order $\epsilon^2$
\cite{chauve+01}. For charge density waves $\tilde\zeta=0$
\cite{narayan.fisher92}(i.e. the roughness increases
logarithmically with $L$) and ${\tilde
  z}=2-\frac{\epsilon}{3}-\frac{\epsilon^2}{9}$
\cite{narayan.fisher92,chauve+01}, whereas for domain
walls $\tilde\zeta=\frac{\epsilon}{3}(1+0.14331\epsilon)$ and ${\tilde
  z}=2-\frac{2\epsilon}{9}-0.04321\epsilon^2$ \cite{chauve+01}.

In the opposite regime $L\gg\xi$ the problem is essentially linear
and $u$ can be replaced by $vt$ in the argument of $g({\bf x},u)$.
This can be seen qualitatively as follows: On the time scale $t$
the elastic object advances on average by an amount $vt$. Randomly
distributed pinning centers will lead to a local distortion which,
according to (\ref{dyn.scaling}), spreads over a region
$L(t)\approx L_p\left(\omega_pt\right)^{1/\tilde z}$. The local
retardation or advancement of the object due to the fluctuation in
the density of the pinning centers scales as $\tilde u(t)\approx
l(L(t)/L_p)^{\tilde\zeta}\approx l(\omega_pt)^{\tilde\zeta/\tilde
z}$. Since $\tilde\zeta<\tilde z$, $\tilde u(t)$ grows more slowly
than $vt$. Thus on time scales
$t>t_{v}=\omega_p^{-1}\left(v_p/v\right)^{\tilde
  z/(\tilde z-\tilde\zeta)}$ and length scales $L>\xi\equiv L(t_{v})$
the non-linearities in the argument of $V_{R,u}({\bf x},vt+\tilde
u)$ can be neglected and the linearized theory applies. In this
case $u({\bf x},t)$ is replaced by $vt$ in the argument of the
random forces. Random forces act then as a thermal noise with the
temperature $\sim v^{-1}$.

So far we considered the elastic theory of critical depinning. If
we include topological defects in the theory, the transition may
become hysteretic, as was shown in \cite{vinokur.natter97}.
We will come back to the influence of topological
 defects on pinning
phenomena in the second part of this article.

As a side remark we mention here that an alternative
characterization of the depinning transition can be reached if we
pin the elastic object at the boundary of the system  by an
infinitely strong surface barrier such that the  displacement at
certain surfaces vanish. In a charge density wave this can be
reached by applying an external electric field but preventing a
current flow by the absence of external leads. In a non-random
elastic system an external force $f$ then generates a parabolic
displacement profile as a solution of (\ref{eq:motion}). The
situation is different in a system with random pinning forces: as
long as $f<f_c$ the elastic object cannot move and the pinning by
surface barriers does not matter. For $f>f_c$ on the other hand
the surface barriers prevent the elastic object from moving and a
parabolic profile  will emerge. A detailed investigation shows
that this is indeed the case \cite{natter03}. Using the
decomposition $u({\bf x})=u_0({\bf x})+\tilde u({\bf x})$ with the
Ansatz
   \begin{equation}
   u_0({\bf x})=-\frac{1}{2}\sum\limits_{i=1}^{D}
   \tilde{\mathcal{C}}_i(x_i-x_{i,0})^2+u_0
   \label{eq:ansatz_u}
   \end{equation}
and $\langle\tilde u({\bf x})\rangle_R=0$, one can determine $f_c$
from the vanishing of $\tilde{\mathcal{C}}=\sum_i \tilde{\mathcal{C}}_i$. Which of
the curvatures $\tilde{\mathcal{C}}_i$ are non-zero depends on the
specific pinning conditions on the surface. Note, that this
decomposition is similar to the description of the dynamics of the
depinning transition in a co--moving frame. From equation (\ref{eq:motion})
we obtain
   \begin{equation}
   C\tilde{\mathcal{C}}-f=\left\langle g\big({\bf x},u_0({\bf x})+
   \tilde u({\bf x})\big)\right\rangle\approx
   \left\langle g_u\big({\bf x}, u_0({\bf x})\big)\tilde u({\bf x})
   \right\rangle\,,
   \end{equation}
which replaces equation (\ref{eq:v1}) of the case of a moving elastic object.
In lowest order of perturbation theory we obtain then (similar to
the derivation of equation (\ref{eq:v2})) its RG--counterpart
\cite{natter03}
   \begin{equation}
   \tilde{\mathcal{C}}=\frac{f-f_c}{C}\,.\label{eq:a}
   \end{equation}
It is to be expected that this relation is true to all orders in
$\epsilon=4-D$. Indeed, on the scale of the correlation length $\xi$
the height $a\xi^2$ of the parabola is expected to scale like the roughness
$l(\xi/L_p)^{\tilde\zeta}$. Thus
   \begin{equation}
   a\xi^2\approx\frac{f_c}{C}\left(\frac{f-f_c}{f_c}\right)^{1-2\tilde\nu}
   L_p^2\approx l\left(\xi\big/L_p\right)^{2-\frac{1}{\tilde\nu}}
   \end{equation}
and with (\ref{eq:xi}), (\ref{eq:a}) and the scaling relation
equation (\ref{exponent.relations})
$\tilde\nu=\frac{1}{2-\tilde\zeta}$ we get indeed the expected
result. Thus we may also characterize the depinning transition by
the vanishing of the parabolic displacement profile. If one
decreases the forces again the curvature shows a pronounced
rhombic hysteresis profile \cite{natter03}. The problems of
inhomogeneous profiles are related to contemporary space resolved
studies of sliding CDWs, see the Section \ref{sec:plastic} for
discussion and references.

\subsection{The depinning transition at finite temperatures}
\label{subsec:depinning_finite_temp}

At $f\le f_c$ and $T=0$ the velocity is zero, but one has to
expect that as soon as thermal fluctuations are switched on, the
velocity will become finite.
Scaling theory predicts in this case an Ansatz
\cite{Fisher83,Middleton91} (generalizing (\ref{eq:v}))
   \begin{equation}
   v(f,T)\sim T^{\tilde\beta/\tau}\Phi\left(\frac{f-f_c}{T^{1/\tau}}\right)
   \label{eq:v(f,T)}
   \end{equation}
with $\Phi(x)\to const.$ for $x\to 0$ and $\Phi(x)\sim
|x|^{\tilde\beta}$ for $|x|\gg 1$, such that $v(f_c,T)\sim
T^{\tilde\beta/\tau}$. Here $\tau>0$ is a new exponent which still
has to be determined.

 This prediction
seems to be in contradiction with simple scaling considerations
applied directly to the equation of motion (\ref{eq:eq.motion}).
Indeed, after renormalization of this equation at $T=0$ up to
length scale $\xi$ and time scale $\omega_p^{-1}(\xi/L_p)^{\tilde
z}$ all terms in this equation scale as $(f-f_c)$. If we consider
now the thermal noise as a small perturbation (at the fixed point
describing the depinning transition) then its contribution to the
equation of motion is of the order
$(f-f_c)\left(\frac{T}{T_p}\right)^{1/2}
\left(\frac{f-f_c}{f_c}\right)^{\tilde\nu\tilde\chi/2}$, where
$\tilde\chi=D-2+2\tilde\zeta>0$ and hence
$\tau=-\tilde\nu\tilde\chi <0$. Thus thermal fluctuations seem to
be irrelevant at this transition.

\begin{figure}[hbt]
   \centerline{\epsfxsize=12cm
   \epsfbox{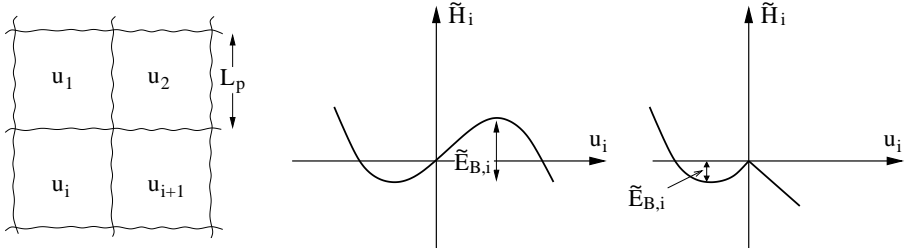}}
   \caption{The  decomposition of the system
   into different Larkin domains (left), the effective potential
   for the coordinate
   $u_i$ in the case of a smooth (middle) and a ratcheted potential (right)}
   \label{fig:middleton}
   \end{figure}

However, this is not true. The previous argument considers the
influence of thermal fluctuations on length scales of the order
$L\approx \xi$. The relevant thermal fluctuations which depin the
elastic object act however on much smaller scales of the order of
$L_p\ll \xi $ as we will see now, following an earlier argument by
A. Middleton \cite{Middleton91}. At the critical point $f=f_c$
essentially only barriers on the scale $L\approx
L_p$ are left as we saw in the previous Sections. It is therefore
sufficient to consider only this length scale. To this aim we
coarse--grain the system into regions of linear size $L_p$ and
denote the corresponding Larkin--domain by the subscript
$i=1,...,N=(L/L_p)^D$ (compare with Figure \ref{fig:middleton}).

Each domain is then essentially characterized by a single local
coordinate $u_i$ related to $u({\bf x})$ by $u_i\approx
\int_{i}d^Dx u({\bf x})$, where $\int_{i}d^Dx$ denote the
integration over the $i$--th domain.  Treating the interaction
between different domains within a mean--field approximation one
can write (in spirit of the Kim-Anderson approach) a local energy
expression for the domain $i$
\begin{equation}
{\cal H}_i=V_i(u_i)-fu_i-\kappa_i(f-f_c)u_i
\label{eq:H_i}
\end{equation}
$V_i(u_i)$ denotes the effective potential for the coordinate
$u_i$ at the depinning threshold $f=f_c$.  The term
$-\kappa_i(f-f_c)u_i$, $\kappa_i>0$ describes the mean--field type
coupling to neighboring domains. The full energy is then given by
$ {\cal H}=\sum_{i=1}^N{\cal H}_i$.

Two types of effective potentials were considered
\cite{Middleton91}: (i) a {\it smooth potential} $V_{s}(u)=\tilde
f_p{u}(1- \frac{u^2}{3u_0^2})$ and a {\it ratcheted kick
potential} with $V_{rk}(u)=\tilde f_pu(1+ \frac{u}{2u_0})$  ($u\le
0$, compare with Figure \ref{fig:middleton}). $u_0$ denotes the
position of the potential minimum. In general the values of $u_0$
($>0$) and $\tilde f_p$ ($\ge f_c>0$) will be different in
different domains. For simplicity we assume in the following that
$u_0\sim l L_p^D$ is the same for all domains.

At the threshold $f=f_c$ there is a metastable    state
corresponding to the left minimum of the potential. The most
unstable domain is then characterized by a value of $\tilde f_p$
which slightly exceeds $f_{c}$ and $\kappa_i\approx 1$.  If we
increase now $f$ from $f=f_c$ to $f=f_c+\delta f$, then the minima
in all domains with $\tilde f_p -f_c <(1+\kappa_i)\delta f $ will
disappear. These unstable domains will trigger transitions in
neighboring domains which destabilize further domains and so on
until the whole object is depinned.  The height of the energy
barrier for the most unstable domains in the region $\delta f<0$
is given by $ E_B\sim
u_0f_c\left((1+\kappa)\frac{f_c-f}{f_c}\right)^{3/2}$ for the
smooth and $E_B\sim
u_0f_c\left((1+\kappa)\frac{f_c-f}{f_c}\right)^{2}$ for the
ratcheted kick potential, respectively. These barriers become
irrelevant at temperatures $T\ge E_{B}$. An increase of $T$ from
$T=0$ has the same effect as increasing $f$ by a value $\sim
T^{2/3}$  from which we conclude $\tau=3/2$ for the smooth
potential. For the ratcheted kick potential the increase of $T$
has the same effect as an increase of $f$ by a value $\sim
T^{1/2}$ and hence $\tau=2$.  The different exponents reflect the
different non-linearities of the effective potential.  Our
analysis of the effective potential depicted in Figure
\ref{fig:V_eff_new}.a (with the analytical form given by equation
(\ref{V_eff})) is different from \cite{Middleton91}, but the
exponent $\tau=3/2$ is identical with that of the smooth
potential, since the barriers in both cases have the same
dependence on $(f_c-f)$, as one can check easily. However this
remark is only true for a potential of the type depicted in Figure
\ref{fig:V_eff_new}.a, for a potentials of the type depicted in
Figure \ref{fig:V_eff_new}.b the exponent $\tau$ is probably
non-universal. The exponent $\tau$ found  from simulations for
domain walls in random field systems \cite{roters+cond-mat} varies
indeed in the range $1.5<\tau<2$. Naturally the exponent
$\tau=3/2$ appears in treatment of termination points (Sections
\ref{ss:ms},\ref{app:ms}) which specify the meaning of the suggested
above instabilities.

\subsection{Depinning due to an ac-field}

In this subsection we want to discuss the effect of an ac-field of
a finite frequency $\omega$ on the pinning of the elastic object
at zero temperature.  The equation of motion is still given by
(\ref{eq:motion}) with $f\rightarrow f(t)=f\sin{(\omega t)}$.
   \begin{figure}[htbp]
   \centerline{\epsfxsize=8cm
   \epsfbox{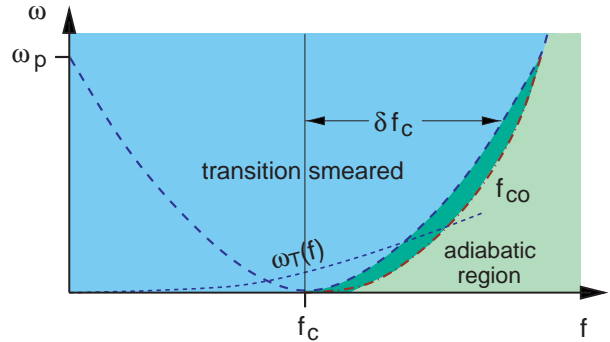}}
   \caption{Schematic frequency-field diagram for the depinning in an
   ac external field (with $f>f_c$): For $0<\omega\ll \omega_p$ the
   depinning transition is smeared but traces of the $\omega=0$
   transition are seen in the frequency dependency of the velocity at
   $f=f_c$. This feature disappears for $\omega\gg \omega_p$.}
   \label{fig:fig2_GNP}
   \end{figure}
A finite frequency $\omega$  of the driving force acts as an
infrared cutoff for the propagation of perturbations, resulting
from the local action of pinning centers on the object. As follows
from (\ref{eq:motion}) with the renormalization
(\ref{eq:effective.mobility}) these perturbations can propagate
during one cycle of the external force up to the (renormalized)
diffusion  length

\begin{equation}\label{eq:diffusion  length}
  \tilde L_{\omega}=L_p(\gamma C/\omega L_p^2)^{1/\tilde z}\equiv
L_p(\omega_p/\omega)^{1/\tilde z}\,,\quad \omega_{p}=\frac{C\gamma}{L_{p}^{2}}\,.
\end{equation}
If $\tilde L_{\omega}<L_p$, i.e. $\omega>\omega_p$, then there is
no renormalization and $\tilde z$ has to be replaced by 2. During
one cycle of the ac-drive, perturbations resulting from local
pinning centers affect the configuration of the elastic object
only up to scale $\tilde L_{\omega}$, such that the resulting
curvature force $Cl\tilde L_{\omega}^{-2}$ is always larger than
the pinning force -- there is no pinning anymore.

In the opposite case $\tilde L_{\omega}>L_p$, i.e.
$\omega<\omega_p$, the pinning forces can compensate the curvature
forces at length scales larger than $L_p$. As a result of the
adaption of the elastic object to the disorder, pinning forces are
renormalized. This renormalization is truncated at $\tilde
L_{\omega}$. Contrary to the adiabatic limit $\omega\rightarrow
0$, there is no sharp depinning transition if $\omega>0$. Indeed,
a necessary condition for the existence of a sharp transition in
the adiabatic case was the requirement that the fluctuations of
the depinning threshold in a correlated volume of linear size
$\xi$, $\delta f_c\approx f_c (L_p/\xi)^{(D+\tilde\zeta)/2}$, are
smaller than $(f-f_c)$, i.e., $(D+\tilde\zeta)\tilde\nu\geq 2$
(compare equation (\ref{exponent.relations})) \cite{natter+92}.
For $\omega>0$ the correlated volume has a maximal size
$L_{\omega}$ and hence the fluctuations $\delta f_c$ are given by
\begin{equation}
   \frac{\delta f_c}{f_c}\approx
   \left(\frac{L_p}{\tilde L_{\omega}}\right)^{(D+\tilde\zeta)/2}=
   \left(\frac{\omega}{\omega_p}\right)^{(D+\tilde\zeta)/(2\tilde z)}.
   \label{eq:Delta-f_p}
\end{equation}
   \begin{figure}[htb]
   \centerline{\epsfxsize=8cm
   \epsfbox{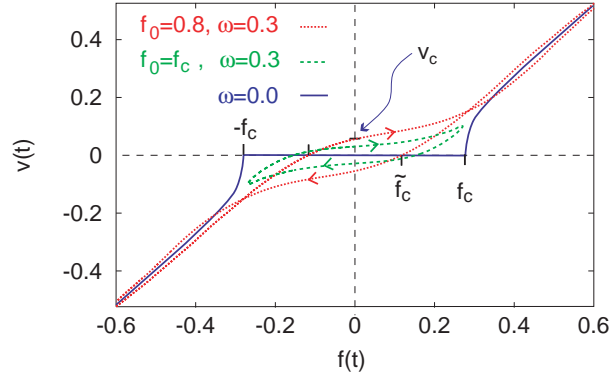}}
   \caption{Velocity hysteresis of a $D=1$ dimensional interface in
   a random environment.}
   \label{fig:fig1_GNP}
   \end{figure}
Thus, different parts of the elastic object see different
depinning thresholds -- the depinning transition is {\it smeared}.
$\delta f_c$ has to be considered as a lower bound for this
smearing. A full understanding of the velocity hysteresis requires
the consideration of the coupling between the different $\tilde
L_{\omega}$--segments of the elastic object. Approaching the
depinning transition from sufficiently large fields, $f(t)\gg f_c$
(and $\omega\ll\omega_p$), one first observes the critical
behavior of the adiabatic case as long as $\xi \ll L_{\omega}$.
The equality $\xi\approx \tilde L_{\omega}$ defines a field
$f_{co}$ signaling a cross-over to an {\it inner} critical region
where singularities are truncated by $\tilde L_{\omega}$.

Note that $f_{c0}-f_c=f_c(\omega/\omega_p)^{1/(\tilde\nu \tilde
z)}\geq\delta f_c$ (cf. Figure \ref{fig:fig2_GNP}). It is then
obvious to make the following scaling Ansatz for the mean
interface velocity ($f_0>f_c$, $v_p=\omega_p l$)
   \begin{equation}
   v\left(f(t)\right)\approx v_p
   \left(\frac{\omega}{\omega_p}\right)^{\frac{\tilde\beta}{\tilde\nu
   \tilde z}}\varphi_{\pm}\left[\left(\frac{f(t)}{f_c}-1\right)
   \left(\frac{\omega_p}{\omega}\right)^{\frac{1}{\tilde \nu \tilde z}}\right].
   \label{eq:interface_velocity}
   \end{equation}
Here the subscript $\pm$ refers to the cases of $\dot f >(<) 0$,
respectively, and $\varphi_{\pm}[x\to \infty]\sim x^{\tilde
\beta}$. For $f(t)-f_c\gg f_c$ the classical exponent $\tilde
\beta=1$ applies. For $|x|\ll 1$, $\varphi_{\pm}$ approaches a
constant $c_{\pm}$. Function $\varphi_{-}$ changes sign at a
critical value $\tilde f_c(\omega)\approx f_c (1-c_{-}
(\omega/\omega_p)^{1/(\tilde\nu \tilde z)})$. The velocity shows a
typical {\it double hysteresis} (Figure \ref{fig:fig1_GNP}).
Qualitatively, the hysteresis loop can be understood to result
from the motion in the ratchet like potential, Figure
\ref{fig:V_eff_new}. The reader is referred for details can to
reference \cite{glatz.natter.pokr02}. There is interesting related
work on the influence of an alternating current on pinned vortex
lattices by Kohandel and Kardar \cite{kohandel99,kohandel00}.

\section{Macroscopic perturbations and external constraints. }\label{sec:macro}

 Now we shall discuss briefly the influence of
external forces or constraints on the statistical properties of
our model equation (\ref{eq:Hamiltonian}). The topic includes the
important case of topologically non-trivial distortions which can
be enforced by appropriate boundary conditions or applied external
forces. We will assume that these constraints are kept constant or
are changed only adiabatically such that we can apply equilibrium
statistical mechanics. We will assume that a field $C{\bf
A(\textbf{x})}$  couples in the Hamiltonian linearly to
${\bnabla}u $, i.e. there is an extra piece $\delta{\cal H}_A$ in
the Hamiltonian
\begin{equation}\label{eq:A-perturbation}
  \delta {\cal H_A}=\int d^Dx C {\bf A(\textbf{x})} {\bnabla} u\,.
\end{equation}
Examples for
$\textbf{A}$ are given e.g. by

(i)  a constant external force $f$ coupling to $u$ for which $
C\textbf{A}=f \textbf{x}/D $ or

(ii) by a field which enforces a dislocation line into the system
(if we consider a periodic systems - see the following Sections).
In the latter case $\textbf{A}$ obeys the relation $\oint_{\cal C}
\textbf{A}d\textbf{x}=-b=2\pi nQ$ where the curve $\cal C$
encloses a dislocation line and $n$ is integer.

In general, extra pieces of the Hamiltonian of the form
(\ref{eq:A-perturbation}) lead to an instability: the system feels
a constant driving force or a torque. In order to prevent an
unlimited response we have to assume the existence of additional
surface forces which keep the system in equilibrium. This will be
done in this Section (for more details see \cite{natter03}).

It is convenient to go over to the new field $\tilde u$ by
\begin{equation}
\label{eq:tilde:u} \tilde u(\textbf{x}) = u(\textbf{x}) +
\int_0^{\bf x}\textbf{A}(\textbf{y})d \textbf{y}.
\end{equation}
In the case (ii) $\tilde u({\bf x})$ may depend on the path
along which the integration is performed.
Different pathes may lead to changes of $\tilde u({\bf x})$
by $mQ$ where $m$ is integer. Since the random potential is periodic
in $u$ with periodicity $2\pi/Q$ such an ambiguity is however
irrelevant.
After this transformation  the Hamiltonian is rewritten as
\begin{equation}
   {\cal H+\delta \cal H_A}=\int d^Dx\left\{\frac{1}{2}C(\bnabla \tilde u)^2+
   V_R\left({\bf x},\tilde u(\textbf{x})-\int_0^{\bf x}
   \textbf{A}(\textbf{y})d\textbf{y}\right)- \frac{1}{2}C\textbf{A}^2\right\}
   \label{eq:Hamiltonian2}
\end{equation}
In both cases (i) and (ii) the new Hamiltonian
(\ref{eq:Hamiltonian2}) has the same statistical properties as the
original one, equation (\ref{eq:Hamiltonian}), since $ V_R({\bf
x},u(\textbf{x}))$ is a {\it random }function of both arguments.
This can most easily seen by using the replica method, in which
the disorder averaged free enthalpie
\begin{equation}
\label{Eq:enthalpy} \left\langle G\{\textbf{A}\}\right\rangle_R
=-\left\langle \ln Tr e^{-(\cal{ H+\delta
H_A)}/{T}}\right\rangle\equiv-\lim_{n\rightarrow
0}\frac{T}{n}\left[Tr e^{-\frac{{\cal H}_n}{T}} -1\right]
\end{equation}
follows from the {\it replica Hamiltonian}
\begin{equation}
\label{eq:replica.Hamiltonian} {\cal H}_n=\int d^{D}x
\sum_{a,b=1}^n\frac{C}{2}\left\{(\bnabla \tilde
u_a)^2\delta_{a,b}- \frac{C}{T} R(\tilde u_a-\tilde
u_b)-{n}\textbf{A}^2\right\}\,.
\end{equation}
 Apparently, the replica
Hamiltonian is unchanged, apart from the additional term
 $-n\int d^Dx\frac{C}{2}\textbf{A}^2$.
It is worth to mention that this is true only if the random
potential $V_R(\bf x,u)$ is strictly uncorrelated in $\bf x$. If
the correlations are given by a smeared out $\delta$-function of
width $a_0$, $R(\tilde u_a({\bf x})-\tilde u_b({\bf x}))$ in
equation (\ref{eq:replica.Hamiltonian}) has to be replaced by
$\int d^{D}x^{\prime}R[\tilde u_a({\bf x})-\tilde u_b({\bf x'})-\int_{\bf
x'}^{\bf x}{\bf A}({\bf y})d{\bf y}]\delta_{a_0}({\bf x}-{\bf
x'})$.

 The disorder averaged free
 energy follows then as
\begin{equation}
\label{EQ:LEGENDRE.TRANSFORM.F-G} \left\langle F\{\langle\bnabla
u\rangle\} \right\rangle_R=\left\langle
G\{\textbf{A}\}\right\rangle_R-\int d^Dx C\textbf{A}\langle\bnabla
u\rangle=\langle F\{0\}\rangle+\int d^Dx\frac{C}{2}\langle\bnabla
u\rangle_{T,R}^2,
\end{equation}
where
\begin{equation}\label{}
  \langle\bnabla u\rangle_{T,R}=C^{-1}\delta
G\{\textbf{A}\}/\delta\textbf{A}=-\textbf{A}.
\end{equation}
 If $\textbf{A}$ represents a dislocation then also the mean
displacement $\langle u\rangle_{T,R}$ shows a dislocation
structure. Correlation functions of $u(\bf{x})$ in the presence of
external forces can now  easily be calculated by using the
decomposition equation (\ref{eq:tilde:u}), since $\tilde u$ is not
affected by the presence of $\bf A$.
\medskip

The glassy phases discussed previously  have been found under the
assumption that topological defects have been excluded. We may now
consider the stability of these phases with respect to topological
defects. In particular, we will briefly consider the stability of
the Bragg glass in charge density waves phase with respect to
dislocations.

Adding a dislocation increases the disorder averaged free energy
according to equation(\ref{EQ:LEGENDRE.TRANSFORM.F-G}) by
 $\int d^Dx\frac{C}{2}{\bf A}^2=\frac{b^2}{2\pi}L_0^{D-2}\ln\frac{L_0}{a_0}$,
 $L_0$
denotes the size of the system. This expression is the energy of
the dislocation line  (D=3) in a pure system and hence
dislocations seem to be always disfavored. However the dislocation
may take advantage from {\it fluctuations} in the disorder
distribution and choose a position where its energy is lowered
with respect to the average value. To this aim one has to consider
the sample to sample fluctuations of the free energy $\langle
F^2\{\langle\bnabla u\rangle\}\rangle_R-\langle F\{\langle\bnabla
u\rangle\}\rangle_R^2$.
This  is a difficult problem and only preliminary results exists,
which support the existence of a quasi-long range ordered phase in
$d=3$ dimensions \cite{Kierfeld+,Carpentier+,Fisher97,Kierfeld98}
provided the disorder is sufficiently weak. In $d=2$ dimensions,
where the disorder in the forward scattering term growth under
renormalization as $\ln L$, dislocation always
appear \cite{Zeng+99}.

\section{Plastic deformations and topological defects.\label{sec:plastic}}

The starting point of the collective pinning picture considered so
far was that the displacements $u$ grow unlimitedly at large
distances. At the same time, local deformations (i.e. strains -
gradients ${\bf\nabla}u$) were assumed to remain small thus
allowing for the universal elastic media description, expansions
of bare energies in terms of ${\bf\nabla}u$ or forces in terms of
$u$, etc. Nevertheless the elasticity can be broken already at the
local level in which case we refer to plastic deformations
\cite{LLVol.7}. The effect is not only quantitative which would
simply affect basic parameters. It happens that plastic
deformations related to impurities can originate  metastable
states which is the principle ingredient of the pinning picture.
These plastic metastable  states create a set of pinning effects
of their own nature, but also they clarify, or even challenge
sometimes, the complex picture of the collective pinning (see more
in the Sec. \ref{sec:LP-CP}).

In general, plastic deformations invoke displacements which are
not small at a microscopic scale, e.g. the domain wall width or
the crystal periodicity. The deformations may be topologically
trivial like large curvatures of domain walls, or as vacancies and
interstitials in Wigner or vortex crystals; then plasticity comes
from the strong pinning potential itself. The plastic deformations
can be topologically nontrivial, and these are locally stable even
without impurities. Among our cases they appear only in periodic
systems because of their ground state degeneracy
$\bf{x}\Rightarrow\bf{x}+\bf{b}$, where $\bf{b}$ is any of $d$
primitive periods of the sliding crystal. Here the topological
defects acquire forms of dislocation lines, or dislocation loops
and their particular limits of solitons in quasi 1d systems.
(concerning dislocations, see \cite{friedel} for a general review,
\cite{LLVol.7,nabarro} for the theory, \cite{Dumas86,Feinb88} for
the special case of CDWs and for helpful illustrations.).

\begin{figure}[hbt]
\centerline{\epsfxsize=6cm \epsfbox{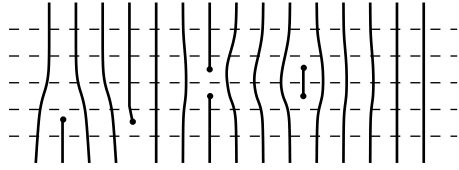}}
\caption{Topological defects in a CDW. The solid lines describe
the maxima of the charge density. The dashed lines represent
chains of the host crystal. From left to right: dislocations of
opposite signs and their pairs of opposite polarities. Embracing
only one chain of atoms, the pairs become a vacancy and an
interstitial or $\mp2\protect\pi$ solitons. Bypassing each of
these defects, the phase changes by $2\protect\pi$ thus leaving
the lattice far from the defect unperturbed.}
\label{fig:dislocations}
\end{figure}

The basic object is the dislocation line crossing the whole sample
(reduced to the dislocation point in $d=2$ systems), see the
Figure \ref{fig:dislocations}. It is a kind of a vortex of
displacements $\bf{u}$ (which are now the $D=d$ dimensional
vectors $u\Rightarrow\bf{u}$) such that going around the
dislocation line $\bf{u}$ acquires a finite increment
$\delta\bf{u}$ called the Burgers vector. In principle, it can be
any allowed translation of the regular lattice, but only
dislocation lines with minimal values of $\delta\bf{u}$,
coinciding with one of the primitive translations $\delta\bf{u=b}$
are stable which we shall imply below. In our perspective, finite
displacements only along the sliding direction $x$ are important,
so in our studies we shall assume that ${\bf b}=(b,0,0)$ is chosen
while dislocation lines and dislocation loops lie in the
perpendicular plane, these are the so called edge dislocations. In
$d=3$, the dislocation line either must cross the whole sample or
it must be closed to the dislocation loop (two dislocation points D,D* in
d=2). All paths across the dislocation loop acquire the
displacement $\bf{b}$ in compare to outer paths. The smallest
dislocation loop embraces just one line of atoms with one unit
cell missed or acquired along this selected line. This limit is
the $\pm2\pi$ soliton in quasi 1d systems (for short reviews on
theory see \cite{Brazov96,Brazov89} and also \cite{Brazov-Matv},
for experimental aspects see \cite{Nad89}). In isotropic crystals
(Wigner crystals, vortex lattices) the elementary dislocation loop
is the symmetry broken state of the vacancy or the addatom. Going
along two paths parallel to $\bf{b}$, one above and another below
the dislocation line,  the difference $\bf{b}$ of lattice
displacements will be accumulated. Then the dislocation line can
be viewed as the leading edge for an additional atomic plane being
introduced to (or withdrawn from) the crystal (from the side
boundary, or from another dislocation line - the counterpart D*
with the opposite circulation $\bf{b}^{\ast}=-\bf{b}$).

Here we already arrive at the first general significance of
dislocation lines for sliding crystals: their necessity to bring
in or modify the sliding regime providing the so called phase slip
processes\footnote{ Phase slippage is a common phenomenon in
condensed matter systems with complex order parameters. It has
been intensively studied in narrow superconducting channels
\cite{Langer,Ivlev}, in superfluid helium \cite{Varoquaux} and in
quasi one-dimensional CDW systems. Phase slips have been
incorporated to the picture of the collective pinning only
recently \cite{q-slips}}. Within the CDW language, the phase
slippage is required at junctions for the conversion from free to
condensed carriers \cite{Gor'kov,Ong,Gill}. When the CDW is
depinned between current contacts, CDW wave fronts are created
near one electrode and destroyed near the other, leading to CDW
compression at one end and to its stretching at the other end. In
a purely 1D channel, the order parameter can be driven to zero at
once \cite{Gor'kov} which allows for the macroscopic phase slip.
For samples of finite cross-section, phase slippages develop as
dislocation lines proliferate across the sample, each dislocation
line allowing the CDW to progress by one wavelength \cite{Ong}.
Proliferation of dislocation lines or expansion of dislocation
loops is called the climb. As for any motion not parallel to $\bf{b}$,
the climb is not conservative with respect to the number of atoms (the
charge in electronic crystals). As such, it is ultimately related
to the current conversion requiring for the phase slips.

Effects of phase slips and the current conversion are closely
related to macroscopic strains of the sliding and/or pinned state.
Recent years have brought a new understanding of the fact that the
sliding state is also essentially inhomogeneous
\cite{Itkis93,Cornel,Grenoble}. A freedom for deformations is
demonstrated by the dilemma associated with the choice of a
solution for the generic equation describing the sliding motion
and deformation of the CDW phase:
$\gamma^{-1}\partial_{t}\varphi+C\partial_{x}^{2}\varphi=f$ (see
the Section 4). Taken alone, this equation is satisfied by any
solution of the type: $\varphi=f(c\gamma t+(1-c)x^{2}/2C)$ with an
arbitrary value of the partition coefficient $c$. Then, at first
sight, the response to the driving force $f$ is optionally
distributed between the viscous $\sim t$ and the elastic $\sim
x^{2}$ reactions, leaving the collective current undetermined. It
is specifically the equilibrium with respect to phase slips which
selects the solution $c=1$ leaving only the viscous nondeformed
regime $\varphi\sim t$ in the bulk \cite{Grenoble,Braz-Kir-99}.
Phase gradients originating from the external force cannot grow
indefinitely because the associated strain is released through
$2\pi$-phase jumps repeating in time, with a rate dependent on the
magnitude of the remaining strain \footnote{The phase slip rate is
given by the space- time vorticity
$I=(\partial_{t}\partial_{x}\varphi-\partial_{x}\partial_{t}\varphi)/2\pi$.}.

\begin{figure}[hbt]
\centerline{\epsfxsize=9cm \epsfbox{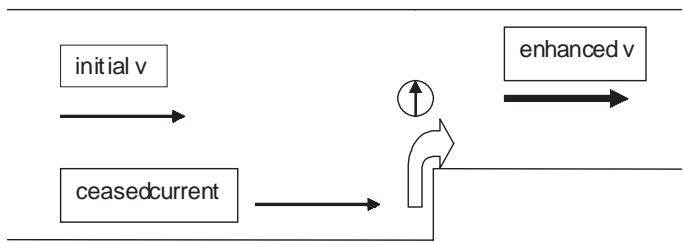}} \caption{Generation
of the perpendicular flow of dislocations by sliding through
narrows. The encircled up-arrow indicates the proliferation of a
dislocation line. The up-right double arrow indicates the material
flow; its conversion provides the climb of dislocations which
results in the velocity enhancement.} \label{fig:step}
\end{figure}

There is a particular case of the conversion with the help of
dislocations which brings us closer to the problem of strong
pinning. Namely, consider the sliding along a step-shaped host
sample (which is quite an important issue in reality) as shown on
the Fig. \ref{fig:step}. Coming to the threshold, the sliding in
cut layers is terminated, so that in the narrows the phase
velocity $v=-\dot{\varphi}$ must increase $v\rightarrow v+\Delta
v$. To keep the crystal connectivity, new periods must be
introduced with the phase slip rate $\Delta v/2\pi$ which are
provided by the flow of dislocation lines in the cross-section.
Understanding this macroscopic example leads us to expectations
for the role of plasticity and topological defects in the pinning
problem, even beyond the strong pinning limit.

Let us reduce the size and the sharpness of the obstacle (remind
soft macroscopic defects studied by the space resolved X-ray
diffraction \cite{rideau}). That will be then a local region of
the enhanced pinning force. It can be either mesoscopic,
originated by rare fluctuation of the collective pinning
strength\footnote{Interpretation of pinning in terms of large
scale fluctuations of the pinning force was suggested in
\cite{coppersmith}, where also the phase slips have been
discussed}
 (e.g. the concentration of impurities), or
microscopic: a single strong pinning center. Our only requirement
is that the local pinning enhancement is strong enough to reduce
the mean sliding velocity that is to provide -- at least from time
to time -- the retardation by the whole lattice period. Then the
retarded zone must be surrounded by dislocation loops to provide
the matching with the rest of the crystal.

\begin{figure}[hbt]
\centerline{\epsfxsize=6cm \epsfbox{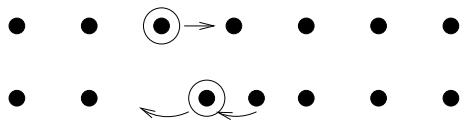}} \caption{Motion of
a sliding atomic array through the strong attractive pinning
center,  as viewed from the co-moving frame. The upper row: the
straight arrow indicates the relative displacement of the impurity
together with the trapped atom. The lower row: having been
displaced by more than half of the period, the trapped atom is
released to its, distant now, position (the left arc arrow), while
the more close now next atom is trapped instead (the right arc
arrow). } \label{fig:atoms}
\end{figure}

The microscopic case is illustrated on the Figure \ref{fig:atoms}
for an attractive  impurity moving across the array of ``atoms''
of e.g. the Wigner or the vortex crystal. Usually we shall assume
the co-moving frame where the pinning center moves through the
asymptotically immobile crystal.

\begin{figure}[hbt]
\centerline{\epsfxsize=6cm \epsfbox{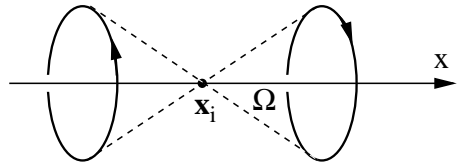}} \caption{A pair of
dislocation loops generated by a strong pinning center after the
nearly complete period of sliding. The cross-section (the figure's
plane) corresponds to the quaternion of dislocations in the r.h.s.
of the Figure \protect\ref{fig:dislocations}. The phase deficit at
the impurity point $\bf{x}_{i}$ is determined by the steric angle
$\Omega$ of the loop. } \label{fig:chain}
\end{figure}

There are three apparent regimes. A weak attractor will only
perturb atoms which will smoothly return to their equilibrium
position. A stronger attractor will draw the atom for more than a
half of the period, then it becomes apparently favorable to
release the overdrawn left atom and catch instead the next atom at
the right which is now closer, as it will release the energy of
deformations. Finally, let the attractor be strong enough to draw
the initial atom over the whole period to the next regular atomic
position. Then the crystal comes again to the local equilibrium
but in expense of creating zones of dilatation and compression
behind and ahead of the attractor. Being integers of the atomic
period, these deformations correspond just to dislocation loops of
the minimal size, the solitons of the CDW language, embracing the
path of the attractor motion, see the Figure \ref{fig:chain}.
Their energy $2E_{s}$ will be payed for the pinning preserved over
one period.

\begin{figure}[hbt]
\centerline{\epsfxsize=8cm \epsfbox{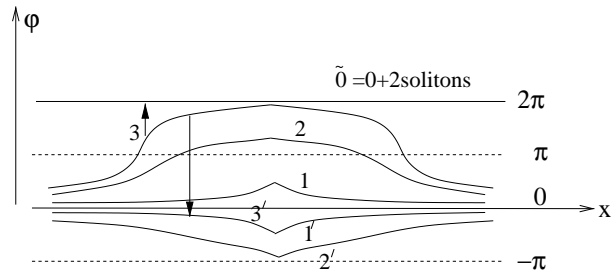}}
\caption{Evolution of the phase profile $\protect\varphi(x)$ (for
the chain passing through the impurity) in the course of the
relative motion. Starting from the equilibrium position $0$ when
$\protect\varphi(x)\equiv0$, the profile evolves gradually through
the shapes $1,2,3,\tilde0$ finally developing the bisolitonic
shape. These configurations correspond to the retarded branch
$E_{+}$ which becomes metastable after $\protect\varphi(0)$
crosses $\protect\pi$. Since then, the advanced profiles
$1^{\prime},2^{\prime},3^{\prime},0$ of the branch $E_{-}$ are
less deformed and hence cost a smaller energy $W$. If the
relaxation $E_{+}\rightarrow E_{-}$ does not happen, the new
circle starts with the profile $\tilde0=0+2\protect\pi$
corresponding to the infinitely divergent pair of solitons. For a
weak impurity the level $\protect\pi$ is never reached and only a
smooth reversible evolution is allowed following shapes
$0,1,0,1^{\prime},0$.} \label{fig:bisoliton}
\end{figure}

An opposite case to the atomic lattice is the CDW which density is
smoothly distributed over the whole period, see the Figure
\ref{fig:dislocations}. Now the development of elastic and plastic
deformations in the course of the impurity motion is illustrated
on the Figure \ref{fig:bisoliton}, with the same consequences for
metastable states.

In summary, we can generally anticipate several regimes which
existence can be verified for particular models considered below.
We shall refer to some parameter $V$ characterizing the pinning
center strength with respect to certain thresholds $V_{1},V_{2}$
for plastic deformations.

1. $V<V_{1}$. The pinning center is very week, local deformations
do not grow to the plastic threshold but they smoothly return to
the original unperturbed state after the whole period is passed in
the course of sliding.

\begin{figure}[hbt]
\centerline{\epsfxsize=8cm \epsfbox{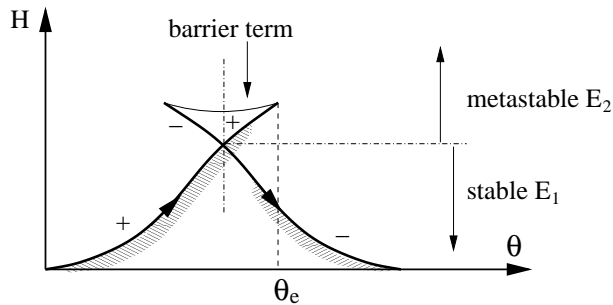}}
\caption{Energy branches for a restrictedly bistable impurity. The
uppermost thin line shows the barrier branch $E_{3}$. Thick lines
show the locally stable branches $E_{\pm}$, also classified as
$E_{2}>E_{1}$. The difference $\Delta E=E_{2}-E_{1}$ gives the
dissipated energy. The difference $U=E_{3}-E_{2}$ gives the
activation energy for a decay of the metastable state $E_{2}$.}
\label{fig:few-terms}
\end{figure}

2. $V_{1}<V<V_{2}$. The pinning center is strong enough to provide
a retardation for more than the half of the sliding period
($|\Delta\varphi|>\pi$ in the CDW language). Since then the branch
becomes metastable: it is favorite to switch the deformation from
the overdeformed retarded configuration
$-2\pi<\Delta\varphi<-\pi,\,\,|\Delta\varphi|>\pi$ to the weaker
deformed advanced one
$\,0<\Delta\varphi<\pi,\,|\Delta\varphi|<\pi$ which saves the
energy of elastic deformations around the pinning center, see the
Fig. \ref{fig:bisoliton}. A coexistence of {\it stable} $E_1$ and
{\it metastable} $E_2$ branches -- the absolute and the local
minima of the energy -- implies the existence of the third branch
$E_3$, the energy maximum, which is the {\it barrier} separating
the two minima. Since we postulated here that the metastable
branch cannot be maintained over the whole period, then there must
be a termination point (the ''end'' point) $\theta_{e}$ where the
metastable and the barrier branches merge together to disappear at
larger retardation, see the Figure \ref{fig:few-terms}.

3. $V_{2}<V$. An even stronger pinning center can sustain a
retardation by the whole period or more, then there is no
termination point and all branches always coexist. After the whole
period passes, the pinning center is again at the equilibrium with
the surrounding crystal, but it is the equilibrium modulo $2\pi$,
the difference of one period is accumulated between the pinning
center vicinity and the crystal at the infinity. And now this is
just the job of dislocation loops to compensate for this
difference. Their diverging pair forms a cylinder (containing the
pinning center) where this difference is just provided. In the
quasi-1d picture, see the Figure \ref{fig:bisoliton}, this process
is easily visualized as a creation of the soliton-antisoliton pair
(the bisoliton) which opens the $2\pi$ retarded allowed ground
state in between.

Notice finally that the ''mesoscopic'' case of a cluster of
impurities will have more degrees of freedom which can originate a
large number of close metastable states, thus merging gradually
with the collective pinning picture.

\section{Local metastable states.\label{sec:ms}}

All the following content will exploit the efficient language of phases
for Charge/Spin Density Waves (CDW/SDW), which is widely used since \cite{fukuyama76}.
The CDWs are characterized by the sinusoidal density profile $\sim\cos
(\bf{Q}\bf{x+}\varphi)$, see (\ref{eq:rho.cdw}), and have elastic
properties of uniaxial crystals, see the Table 1. The order
parameter can be taken as $\sim\exp[i\varphi]$ so that
dislocations are easily viewed as usual vortices. For a periodic
sliding media in general, the natural choice for the microscopic
length scale is the unperturbed lattice period $b$ along the
sliding direction. It corresponds to the CDW convention to use the
phase $\varphi$ for the description of the displacements
$u\Rightarrow -\varphi$: $\varphi=-2\pi u/b=-\bf{Qu}$. The
velocity becomes the \emph{phase velocity} for which we shall use
the same notation $v=\partial u/\partial
t\Rightarrow-\partial\varphi/\partial t$. This phase velocity is
accessed directly in experiments by measuring the so called Narrow
Band Noise (NBN) interpreted as the washboard or the phase slip
frequency \cite{Gruner88}. Correspondingly the force $f$ is
naturally defined as the work done via sliding by one period, that
is $f\Rightarrow f2\pi/b$. Particularly for Electronic Crystals
$f$ coincides with the electric field strength $f=e\mathcal{E}$
(for Wigner crystals, for $4k_{F}$ CDWs) or with $2e\mathcal{E}$
(for CDWs, SDWs  where one period caries the double electronic
charge $2e$).

\begin{table}[hbt]
\caption{\bf Relations between parameters of conventional crystals and
CDWs.}
\begin{tabular}{l@{\quad\quad}l}\\
\hline
\rule[-3mm]{0mm}{8mm}
displacements & ${\bf u}/b\rightarrow{\bf\hat{x}}\varphi/2\pi$,\,\,
$\bf{\hat{x}}=(1,0,0)$ \\
\rule[-3mm]{0mm}{3mm}
velocity, density & ${\bf v}=\partial{\bf u}/\partial t\Rightarrow
v=-\partial\varphi/\partial t$\thinspace ,$\;
-\bf{\nabla u}\Rightarrow\partial\varphi/\partial x$ \\
\rule[-3mm]{0mm}{3mm}
driving and pinning forces & $f\Rightarrow f2\pi/b=Qf$ \\
\rule[-3mm]{0mm}{3mm}
strain, stress & $\bf{\nabla}\varphi\,$, $ C\bf{\nabla}\varphi\;$\\
\hline
\end{tabular}
\label{tab:phase}
\end{table}

To quantify and prove the intuitive picture of the Sec.
\ref{sec:plastic}, we consider an isolated local pinning center
which can be described by a single degree of freedom $\psi_{i}$
and monitored by another single one $\theta_{i}$. They are the
local mismatches of phases
\begin{equation}\label{}
  \psi_{i}=\varphi(\bf{x}_{i})-\bar{\varphi}\;,\;
  \theta_{i}=-\bf{Qx}_{i}-\bar{\varphi}
\label{psi-the}
\end{equation}
 relative to the bulk value $\bar{\varphi}$. The latter can be taken to be homogeneous in space,
static $\bar{\varphi}\approx const$ or sliding
$\bar{\varphi}\approx -vt$, within the correlation domain of the
collective pinning which by definition contains many impurities,
see  more discussion in the Sec. \ref{sec:kin}).

Beyond a close vicinity (of a microscopic scale $a_0$) of the
pinning center and well within the collective pinning domain, we
can use the energy functional  (\ref{eq:Hamiltonian})

\begin{equation}
\mathcal{H}= \int_{\mathcal{D}}d^{d}x \left[
\frac{C}{2}({\bf\nabla}\varphi)^{2} +
\sum_{i}V_{i}(\varphi(\bf{x})
+\bf{Q}\bf{x}_{i})\delta(\bf{x}-\bf{x}_{i}) \right]\,,
\label{H-vph}
\end{equation}
and typically $V_{i}(\varphi)=V_{i}(1-\cos(\varphi))$.

This energy should be minimized over $\varphi(x)$ at the
asymptotic condition $\varphi\rightarrow\bar{\varphi}$. By
minimizing  the energy over $\varphi(x)$ we can get rid of  the
phase everywhere except at $\bf{x}=\bf{x}_{i}$. By analogy with
electrostatics, the ''potential'' $\varphi(\bf{x})$ originates
from the ''point charge''
$V_{i}^{\prime}=dV_{i}(\varphi)/{d\varphi}$, and the elastic
energy $W_{el}$ will be the one of a site charged at the potential
$\psi$ with respect to the infinity: $W_{el}=(K/2C)\psi^{2}$ where
$K\sim a_{0}$ (in $d=3$) is the ''capacitance''\footnote{
Certainly, for the ideal point impurity we face the divergency of
elastic deformations which would not allow to determine
$\varphi(\bf{x}_{i})$: the extremal solution of (\ref{H-vph}) is
divergent at $\bf{x}\rightarrow\bf{0}$ $\varphi(\bf{x})
-\bar{\varphi}\sim V_{i}^{\prime}/|\bf{x}|^{d-2}$. As usual, the
problem can be regularized at a cutoff length $a_{0}$ since
microscopically there is a finite width of the pinning site.
Actually, there is the microscopic coherence length $\xi_0\gg
a_0$, where the amplitude of the order parameter $A\sim\sqrt{C}$
can pass through zero if the phase gradients become too high. An
even larger length scale appears in quasi 1d CDWs: this is the
soliton width $l_{s}$ (Figures
\ref{fig:bisoliton},\ref{fig:soliton}, and the Section
\ref{ss:models}) below which the system cannot sustain the shear
deformations - see the Section \ref{ss:models}.}.
 The case $d=2$ is always problematic: now the inverse capacitance diverges at the
upper limit $K^{-1}\sim\ln L$ where $L$ is a limiting size. We
shall face this effect once again in the Section
\ref{sec:quantum}. In many respects, the local pinning scheme
needs revisions in $d=2$ which dimension is particular also for
the collective pinning, see the previous Sections.

The elastic regime $W_{el}\sim\psi^{2}$ is not valid at large
deviations where it must give rise to the more efficient plastic
regime. To see this more clearly, suppose that a very strong
pinning center allows for the retardation by many periods $N$,
$\psi\approx2\pi N$, then $W_{el}\sim N^{2}$. The plastic
alternative (to emit a pair of elementary dislocation loops, the
solitons) after each period of retardation -- would give the lower
energy $W_{pl} \approx 2E_{s}N$ which grows only as $\sim N$
rather than as $\sim N^{2}$. Actually, for large $N$ a further
drastic reduction of plastic deformations is possible: as much as
from $\sim N$ to $\sim N^{1/2}\ln N$ in $d=3$ and to $\sim\ln N$
in d=2 dimensions, respectively. To see it, remind that the
coalescence of dislocation loops is allowed provided the total
number of embraced chains, that is the total increment/deficiency
of the crystal periods, is preserved. Then it will be favorable to
aggregate the sequence of $N$ emitted elementary dislocation loops
into the growing single loop embracing $N$ chains which energy is
$\sim\ln N$ per unity of its length (the perimeter is $\sim
N^{1/2}$ in $d=3$). The expansion of the pair of wide dislocation
loops, see the Figure \ref{fig:chain}, at both sides of the strong
impurity will redistribute the retardation by multiple periods
along the\emph{\ }defected line to the retardation by the single
period over many lines embraced by the dislocation loop.

We shall return more systematically to the topic of dislocations
in the Sec. \ref{sec:loops}.

The above arguments do not tell us yet what is going on within one
period of sliding and for intermediate pinning strengths. Some
peculiarities of long range interactions between diverging
dislocation loops will require to consider carefully also the
marginal region between two successive periods. These questions
will be addressed in the next Section.

\subsection{Basics of metastability.\label{ss:ms}}

\subsubsection{Definitions and classification.}

Locally equilibrium states are determined by extremal, over
$\varphi(\bf{x)}$, values of the energy functional at presence of
only one impurity at the point $\bf{x}_{i}$ and for the asymptotic
condition $\varphi\rightarrow\bar{\varphi}$ at
$|\bf{x}-\bf{x}_{i}|\rightarrow\infty$. Here we keep in mind the
full scale, generally nonlinear (see examples below) model which
is regular at small distances and reduces to the elastic model
(\ref{H-vph}) at a sufficient distance from the impurity. Since
the ultimately nonlinear pinning energy depends only on the local
phase $\varphi(\bf{x}_{i})+\bf{Qx}_i=\psi-\theta_{i}$ (see Eq.
(\ref{psi-the}) for definitions), it is convenient to keep this
value fixed in preliminary stages and optimize for it only in the
end. Minimization over all other $\varphi(\bf{x})$, at given
$\varphi (\bf{x}_{i})$ and $\bar{\varphi}$, reduces the energy
functional to the function. (Below we shall omit the index $i$
addressing only one impurity positioned at $\bf{x}_{i}=0$.)
\begin{equation}
H(\psi,\theta)=V(\psi-\theta)+W(\psi)  \label{H=W+V}
\end{equation}

\begin{figure}[hbt]
\centerline{\epsfxsize=8cm \epsfbox{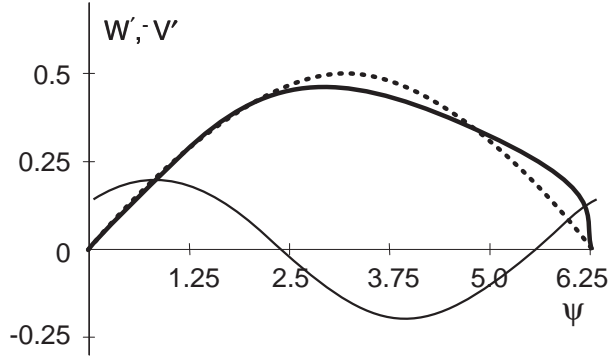}} \caption{
Solutions for all branches $\protect\psi_{a}(\protect\theta)$ are
obtained by crossing of a thick line ($dW/d\protect\psi$) with a
thin line ($-dV(\protect\psi -\protect\theta)/d\protect\psi$)
which are shown for $\protect\theta=3\pi/4$. The doted thick line is drawn
for the short range  model, i.e. without taking into account the
long range interaction of dislocation loops. The solid thick line
shows that these effects lead to the steep fall for the actual
$W^{\prime}$.} \label{fig:solution}
\end{figure}

\noindent This variational energy contains the pinning potential
$V(\psi-\theta)$ (its amplitude we shall call simply $V$) and the
energy of deformations $W(\psi)$. It is clear from the above
discussion that $W(0)=0$, $W(2\pi)=2E_{s}$ and the next circle
starts from this level (the configuration $0^{\prime}$ on the
figure \ref{fig:bisoliton}), so within each period $W(\psi)$ is
the same function with $\min W(\psi)=W(0)=0$ and $\max
W(\psi)=W(2\pi)=2E_{s}$. The function $W^{\prime}(\psi)$, together
with $-V^{\prime}(\psi-\theta)$ for several values of $\theta$, is
shown on the Figure \ref{fig:solution}. Their intersection
determines extremal values of (\ref{H=W+V})  over $\psi$.

The minima and maxima of the variational energy over $\psi$ at a
given $\theta$ determine the branches $\psi_{a}(\theta)$. There
are always locally stable states, which can be either absolutely
stable, $a=1$, or metastable, $a=2$,  the unstable barrier
branches are denoted by $a=3$, that is
\begin{equation}
\mathrm{any\,}a\,:\;\frac{\partial H}{\partial\psi}\equiv0\,;\;
H(\psi_{a}(\theta),\theta)=E_{a}(\theta)  \label{dH=0}
\end{equation}

\begin{equation}
a=1,2\,:\;\frac{\partial^{2}H}{\partial\psi^{2}}>0\ ;\;a=3:\;
\frac{\partial^{2}H}{\partial\psi^{2}}<0 . \label{d2H>0}
\end{equation}
Differentiating $H_{a}$ over $-\bar{\varphi}$, that is over
$\theta$ along the branch, we obtain the contribution $F_a$ of a
given impurity to the total pinning force $f_{pin}$. It depends on
the instantaneous value of $\theta$ and on the branch $a=1,2$
being currently occupied:
\begin{equation}
\frac{F_{a}(\theta)}{2\pi}=\frac{dH}{d\theta}=-\frac{\partial
V}{\partial\theta}=\left. \frac{\partial W}{\partial\psi}\right|
_{\psi=\psi_{a}(\theta)}  \label{F}
\end{equation}
\noindent Thus $F_a$ is positive/negative for ascending/descending
branches, see the Figure \ref{fig:few-terms}. In the Eq. (\ref{F})
and henceforth, we normalize the force in an invariant way, as a
work performed by displacing over one elementary period, $2\pi$ in
our case of density waves.

\begin{figure}[hbt]
   \centerline{\epsfxsize=6cm
   \epsfbox{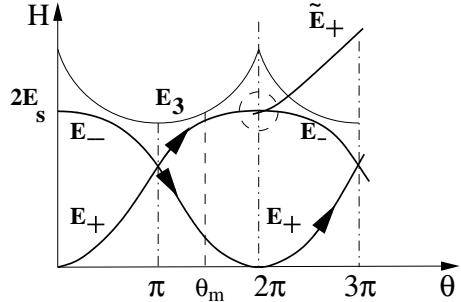}}
\caption{The complete structure of energy branches (thick lines)
for pinning centers of highest strengths $V>V_2$. Uppermost (thin)
lines show the barrier energies. Contrarily to the case of the
Fig. \ref{fig:terms1}, all branches pass continuously through the
whole period. Notice the point $\protect\theta_{m}$ where the
activation $U$ is minimal. The half of the second circle,
$\theta>2\pi$, is also shown. Here, the branches $E_+$,$E_-$ are
identical to those at $0<\theta<2\pi$, assuming that the system is
totally relaxed. The actual adiabatic continuation of the branch
$E_+$ is $\tilde E_+$ which differs by the presence of two
solitons at infinite separation which have been created over the
previous circle of the branch $E_+$. The details of the crossover
between $E_+$ and $E_-$ (dashed circle at the figure) are given on
the Figure \protect\ref{fig:terms3}.}
 \label{fig:terms2}
   \end{figure}

\begin{figure}[hbt]
 \centerline{\epsfxsize=6cm \epsfbox{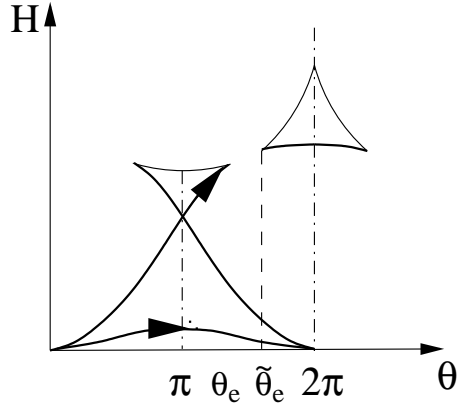}}
\caption{ The complete structure of energy branches
$E_{a}(\theta)$ for pinning centers of different strengths: weak
($V<V_{1}$, lowest thick curve) and intermediate ($V_{1}<V<V_{2}$,
other thick curves). Uppermost (thin) lines show the barrier
energies. For intermediate strengths, notice the disconnected
region of higher energy branches around
$\protect\theta=2\protect\pi$, in addition to continuously
accessible ones corresponding to the Figure \ref{fig:few-terms}.
The termination points of the two regions
$\tilde{\protect\theta}_{e}$ and $\protect\theta_{e}$ coalesce at
$V=V_{2}$ giving rise to the structure of the regime $V>V_{2}$
shown on the Figure \ref{fig:terms2}.} \label{fig:terms1}
   \end{figure}

Figure \ref{fig:few-terms} and Figures
\ref{fig:terms2},\ref{fig:terms1} illustrate a typical and more
complicated cases. The whole interval of $\theta$ or some parts of
it can be either \textit{monostable} or \textit{bistable}, the
last case corresponds to the coexistence of two locally stable
branches: the  \textit{absolutely stable} one with the lower
energy $E_{1}$ and the  \textit{metastable} one of a higher energy
$E_{2}$. The same pair of branches can be regrouped also as the
 \textit{ascending} branch $E_{+}$ for which $F_{+}(\theta)>0$ and the
 \textit{descending} one $E_{-}$ with $F_{-}(\theta)<0$; they correspond to
the  \textit{retarded} and the  \textit{advanced} states at the
impurity, respectively. Evidently,  $E_{1}=\min\{E_{+},E_{-}\}$ and
$E_{2}=\max\{E_{+},E_{-}\}$. There is a symmetry
$E_{-}(\theta)=E_{+}(2\pi-\theta)$
so that these branches cross at $\theta=\pi$,
$\psi_{\pm}(\pi)<\pi$; also $E_{B}(\theta)=E_{B}(2\pi-\theta)$.
 In the following we shall assume a certain sign of the overall
displacement or its velocity, such that branches evolve towards
increasing $\theta$ and consider only the important semi-interval
$\pi<\theta<2\pi$; here $E_{+}\equiv E_{2}$ and $E_{-}\equiv
E_{1}$.

Notice that the barrier branch has appeared via its absolute
energy $E_{3}$. But the necessary quantity is its increment $U$
with respect to the metastable branch $E_{2}$ which gives the
activation energy for its decay:
\begin{equation}
U(\theta)=E_{3}(\theta)-E_{2}(\theta)  \label{Ea}
\end{equation}
This quantity corresponds to the barrier height definition $E_{B}$
in the collective pinning.

For  \textit{strong pinning centers of unrestricted
metastability}, the two locally stable states coexist over the
whole period, hence $U(\theta)\ne 0$. Typical models show that
$U(\theta)$ is largest at $\theta=\pi$ and around $2\pi$ (while
not exactly at $2\pi$ because of special effects of long range
interactions, see also the Sec. \ref{sss:LLR} below). Then there must be the
minimum of the activation at some $\theta=\theta_{m}$ given by
$\min U(\theta)=U(\theta_{m})=U_m$ (see the Figure
\ref{fig:terms2}) which plays an important role.

For  \textit{moderate pinning centers of the restricted}
metastability, the coexistence resides over some intervals around
$\pi$ and $2\pi$: $\pi<\theta<\theta_{e}$ and
$\tilde{\theta}_{e}<\theta<2\pi$. At the end points $\theta_{e}\
$(or $\tilde{\theta}_{e}$) the metastable states terminate or
branch out of barriers, here the second derivative is zero:

\begin{equation}
\theta_{e},\ \psi_{e}=\psi(\theta_{e}): \quad\frac{\partial
H}{\partial\psi}=0 \quad\frac{\partial^{2}H}{\partial\psi^{2}}=0
\;;\; U\approx V_{e}(\theta_{e}-\theta)^{3/2} . \label{end}
\end{equation}
(see the Appendix \ref{app:ms}). The points
$\theta_{e},\tilde{\theta}_{e}$ appear by splitting off from
$\theta=\pi,2\pi$ at some $V>V_{1},V>\tilde{V}_{1}$. For  the
short range model (see below) $V_{1}=\tilde{V}_{1}$.

The points $\theta_{e},\ \tilde{\theta}_{e}$ coalesce and then
disappear at some higher $V=V_{2}>V_{1},\tilde{V}_{1}$ (Fig.
\ref{fig:terms2}) which gives rise to the point $\theta_{m}$ at
$V>V_{2}$ as described above. The full description for $V>V_{2}$
requires a detailed study of dislocations generated near the full
period $\theta\approx2\pi$. We shall postpone the analysis of all
special effects related to diverging dislocation loops or solitons
till the Sec. \ref{sec:loops}.

All together our qualitative arguments lead to the structure of
energy branches shown on the Figures
\ref{fig:terms2},\ref{fig:terms1}. There may be a more complicated
picture of termination points for less local pinning centers, e.g.
the Figure \ref{fig:terms3}, and even a more complex hierarchy for
the collective pinning regime. At present, this feature of the
energy landscape stays beyond the scaling theory of the collective
pinning.

\subsection{Models.\label{ss:models}}

The above analysis exploited only the most general properties of
the variational energy $H$: the periodicity of the pinning
component $V(\psi-\theta)$, the monotonous energy of deformations
$W(\psi)$ with the minimum at $\psi=0$, the maximum at $\psi=2\pi$
and with the inflection point $\psi_{m}$ in-between; and
complemented by universal long-range effects of distant
dislocation loops (Section \ref{sec:loops}). Now we shall illustrate these
features on model examples.

\subsubsection{ Elastic model}
This model takes into account the periodicity of the pinning
potential but neglects the topological character of plastic
deformations derived from the same lattice periodicity. As a model
of metastable states it was already considered in
\cite{larkin.ovch79}. For the bare pinning energy it is always
natural to choose the point impurity pinning potential
$V(\varphi)\approx V(1-\cos\varphi)$. The deformational part
$W(\psi)$ is the energy for the distribution $\varphi(\bf{x})$
optimized at the condition that $\varphi(0)=\bar\varphi +\psi$ and
$\varphi(\infty)=\bar\varphi$. To calculate this energy we
consider a small sphere $\mathcal{S)}$ of the radius $a_{0}$
around the position of the impurity on which we assume the phase
is $\varphi=\psi+\bar{\varphi}$. We have then to solve the Poisson
equation $\nabla^{2}\varphi=0$ with the boundary conditions on the
collective pinning domain $\mathcal{D}$ far away from the impurity
($L_{D}\gg a_{0}$) and on $\mathcal{S}$. The solution and the
energy are, in $d=3$,
\begin{equation}
\phi({\bf x})\approx\bar{\phi}+\frac{a_{0}\psi}{|{\bf x-x}_{i}|}
\,\,\,\,,\;W(\psi)=2\pi a_{0}C\psi^{2}\,\,\,\\,\kappa\equiv
V/(4\pi a_{0}C) \label{eq.phi.solution}
\end{equation}
Depending on the relative impurity strength there is only one
($\kappa<1$) or more ($\kappa>1$) solutions of Eq.(\ref{dH=0}) for
the energy branches
 \begin{equation}
 dH/d\psi=0\,,\;\kappa\sin(\psi-\theta)+\psi=0.\nonumber
 \end{equation}
The  condition $\kappa=1$ identifies $V_{1}\equiv4\pi a_{0}C$ and
the results of the above general treatment follow correspondingly,
except for the region $V>V_{2}$. The latter requires that the
periodicity to be fully taken into account: $W\thicksim\psi^{2}$
does not show the inflection point $\psi_{m}$.

Detailed calculations for this model can be found in
\cite{kobelkov}.

\subsubsection{ Solitons in quasi 1d system: the short range model.
\label{sss:srm}}

\begin{figure}[hbt]
\centerline{\epsfxsize=6cm \epsfbox{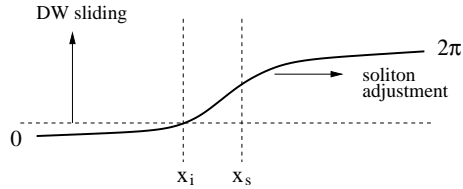}} \caption{
Extinction of a point impurity pinning at presence of the
$2\protect\pi$-soliton. The phase profile
$\protect\varphi(x-x_{s})$ can be adapted (the vertical arrow) to
the phase mismatch at the impurity position $x_{s}$ by the
adjustment (the horizontal arrow) of the soliton position
$x_{s}$.} \label{fig:soliton}
\end{figure}

We consider a quasi 1D system of interacting CDW-chains with an
impurity at the chain $n=0$ at the position $x=0$; the Hamiltonian
is

\begin{equation}
\mathcal{H}=\int dx\left\{ \sum_{n} \right[
\frac{1}{2}C_{\parallel}(\nabla_{\parallel}
\varphi_{n})^{2}-\sum_{m}C_{mn}^{\bot}\cos(\varphi_{n}-\varphi_{m})\left]
-V\cos(\varphi_{0}-\theta)\delta(x)\right\}.  \label{quasi}
\end{equation}
Here the first term in square brackets is the on-chain elasticity
and the second term is the interchain coupling which is reduced to
the shear elasticity $\sim C_{\bot}(\nabla_{\perp}\varphi)^{2}$
when perturbations are small, $C_{\bot}=\sum_{m}C_{mn}^{\bot}$.
The $2\pi$ periodicity of the pinning energy allows to skip the
$2\pi$ quanta in $\varphi_{0}$ to optimize the total energy which
already can originate local metastable states as we have discussed
above. Moreover, the $2\pi$ periodicity of the regular energy in
(\ref{quasi}) allows for topological defects, the solitons. For
the soliton centered at the position $X$ at the chain $n=0$, the
phase profile $\varphi_{s}(x-X)$ describes stretching/dilatation
by one period along the chain $n=0$ relative to surrounding
chains, see the Figure \ref{fig:soliton}. It is distributed over
the length $l_{s}\sim\sqrt{C_{\parallel}/C_{\bot}}$\ and costs the
energy $E_{s}\sim\sqrt{C_{\parallel}C_{\bot}}$, the last defines
their equilibrium concentration $n_{s}\sim\exp(-E_{s}/T)$.

\begin{figure}[hbt]
\centerline{\epsfxsize=6cm \epsfbox{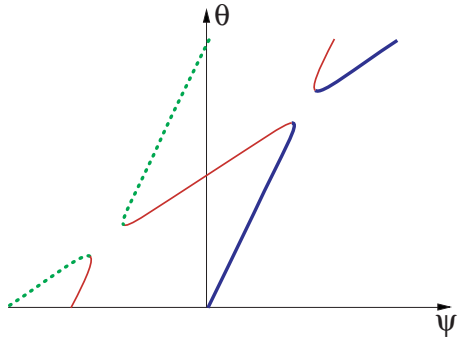}}
\caption{Solutions $\psi_{a}(\protect\theta)$ for the short range
model shown at $V_1<V<V_2$. They correspond to the plots of the
Fig. \protect\ref{fig:terms1}. The double period for
$-2\pi<\psi<2\protect\pi$ allows to see both retarded (medium
lines) and advanced (thick lines) configurations; thin lines
correspond to barriers.} \label{fig:psi-theta}
\end{figure}

While single solitons can be created only by phase slips, their
pairs are non topological configurations which can be continuously
developed by driving $\theta$. Hence we are looking for extremal
values of (\ref{quasi}) with trivial boundary conditions
$\varphi(\pm\infty)=0$. They can be visualized (see the Figure
\ref{fig:bisoliton}) as a combination of two pieces of $\pm2\pi$
solitons at positions $\pm X$:
$\varphi_{s}(-X)-\varphi_{s}(X)=\psi$ from which one concludes the
relation $X=X(\psi)$. We can specify the shapes $\varphi_{s}(x)$
within a short range model \cite{Larkin94} suggesting that in
(\ref{quasi}) only the central chain $n=0$ (passing through the
impurity) is perturbed while its $Z\gg1$ neighbors stay at
$\varphi_{n\ne 0}\equiv \bar{\varphi}$ homogeneously. Then the
energy functional is simplified as
\begin{equation}
\int dx\left[\frac{1}{2}C_{\parallel}(\varphi^{\prime})^{2}-
C_{\bot}\cos(\varphi)\right] -V\cos(\varphi-\theta)\delta(x)
\label{srm}
\end{equation}
Its extremum is the function $\varphi_{s}(x-X)-\varphi_{s}(x+X)$
where $\varphi_{s}(x)$ is the standard saddle point solution for
the Sine-Gordon soliton. The sequence of plots of $\varphi(x)$\
for different $\theta$\ is shown in the Figure
\ref{fig:bisoliton}. The energy is
$W(\psi)=E_{s}(1-\cos(\psi/2))$; in accordance with our general
expectations, over one period $W(\psi)$ changes monotonously
within $0=W(0)\leq W(\psi)\leq W(2\pi)=2E_{s}$,
$W(\psi)\sim\psi^{2}$ at $\psi\rightarrow0$. Now we can identify
the threshold values as $V_{1}=E_{s}/4$, $V_{2}=E_{s}/2$,
$\psi_{m}=\pi/2$. Typical solutions of the equation (\ref{dH=0})
for this model are shown on the Figure \ref{fig:psi-theta}.

The short range model (\ref{srm}) already contains most important
features necessary in applications of the local pinning. It fails
only for high velocity regimes of very strong impurities when the
two well formed solitons diverge at $\pm X\rightarrow\pm\infty$.
In this regime their interaction with each other and with the
impurity penetrates very efficiently, as a power law, via the
elastic deformations of the whole media. For the short range
model, where the surrounding chains $n\neq0$ are frozen, the
perturbations fall off exponentially as $\exp(-X/l_{s})$ which
gives incorrectly the analytic phase dependence of the energy
$W(\psi)-2E_{s}\approx E_{s}(\psi-2\pi)^{2}$ at
$\psi\rightarrow2\pi$. Contrarily, the true power law for
long-range elastic interactions results (for $d=3$) in
$W(\psi)-2E_{s}\sim E_{s}(\psi-2\pi)^{3/2}$ which leads to
particular instabilities. This effect relies upon the view of
solitons as nucleus dislocation loops, and we postpone its
analysis till the Sec. \ref{sss:LLR} which collects all
information on special contributions of dislocations.

\section{Kinetics and relaxation, v-f characteristics.\label{sec:kin}}
Results of the previous chapter provide the basis for the picture
of the local/strong pinning. In this chapter, we shall apply this
picture of local metastable states to kinetics, and finally we
shall describe nonlinear $v-f$ characteristics and to the linear
response function $\chi(\omega)$.

Before going on, it is appropriate here to rectify our definition
of the local or strong pinning. When these notions were introduced
\cite{larkin.ovch79,fukuyama.lee78}, the strong pinning case
implied primarily that the local adaptation of the elastic media
follows closely the minimum of the impurity pinning potential. Our
definition compiles with this tradition but generalizes it to the
multivalued case when the minimum is allowed also to be the
metastable one. But the same time we disagree with a rather common
view that for the strong pinning the correlation volume is of the
order of the one per impurity. At least at $d>2$, we see that
however strong the impurity potential is, the deformation falls
off with distance as a power law. At $d>4 $ the perturbation would
be levelled out completely while in $d=3$ it will contribute to
long range fluctuation of the collective pinning for which there
is no difference with respect to the impurity strength. The next
ambiguity is to identify the strong pinning via the linear
dependence of the critical field $f_c$ on the impurity
concentration $n_i$. This definition is also traced back to the
old epoch when the importance of time scales was not appreciated
yet. Today it is clear that, because of limited heights of
potential barriers, the local/strong pinning describes relatively
high velocities or frequencies, or respectively low temperatures,
where its contribution is indeed linear in $n_i$.

\subsection{Kinetic equation.\label{kin-eq}}

We consider now the kinetics in an ensemble of impurities
possessing local metastable states. With the exception of
particular regimes of strongly divergent dislocations, these
states are formed locally, at a distance shorter than the mean
impurity spacing $L_{i}$ which, in its turn, is much smaller than
the pinning length $L_{p}$. Hence we can define a reference phase
for a volume $D$ staying well within the collective pinning
correlation volume $L_{p}^{D}$ but still extending over the large
number of impurities:
\begin{equation}
\bar{\varphi}({\bf x},t)= \mathcal{D}^ {-1}\int_{\mathcal{D}}d{\bf
x}\varphi({\bf x},t) \,;\;L_{i}^{D}\ll\mathcal{D}\ll
L_{p}^{D}\nonumber
\end{equation}
Here we can neglect the dependence on $\bf{x}$, so that
$\bar{\varphi}\approx\bar{\varphi}(t)$. For the same reason we can
neglect the direct contribution $\sim fa_{0}$ of the driving force
to the energies $E_{a}$. Moreover, we can separate the local
pinning force $f_{pin}$ from impurities and the regular viscosity
$f_{visc}\sim v/\gamma$ coming from the phenomenological damping
(\ref{eq:eq.motion}): $f=f_{visc}+f_{pin}$  Here the time delay
related to viscosity $\delta t\sim a_{0}^{2}/\gamma C$ is small in
compare to $v^{-1}$. For the collective pinning this problem is
more complicated because the microscopic scale $a_{0}$ is enlarged
to an intermediate one $L_{p}$ (see Eq. (\ref{eq:v2}) and the
related text afterwards). In the following we shall speak only
about the pinning part $f_{pin}$ of $f$ implying that $f_{visc}$
can be added in the end. Actually even that is not necessary: we
shall see below that the linear damping $\sim v$ is generated by
the pinning itself which result is confirmed by experiments, see
the Sec. \ref{sec:applic}.

As in statics, each impurity can be characterized by the
positionally random phase
$\theta_{i}=-\bf{Qx}_{i}-\bar{\varphi}(t)$ which now evolves in
time following the moving reference phase $\bar{\varphi}(t)$. The
single monitoring parameter $\theta$ describes both the time
evolution and the distribution over impurity positions; the
average over randomness, $\langle ...\,\rangle_{R}$ becomes
\begin{equation}
\langle....\,\rangle_{R}\rightarrow\int\frac{d\theta}{2\pi}
\nonumber
\end{equation}
For each bistable impurity  (see the Sec. \ref{sec:ms} and the
Figures \ref{fig:few-terms},\ref{fig:terms2},\ref{fig:terms1}),
the state occupies instantaneously one of the two branches
''$\pm$'' with energies $E_{\pm}(\theta)$: ascending $F_{+}>0$ and
descending $F_{-}<0$ where $F_{\pm}=2\pi\partial
E_{\pm}/\partial\theta$ are the local pinning forces. The upper
and lower energy branches are $E_{2}=\max\{E_{-},E_{+}\}$ and
$E_{1}=\min\{E_{-},E_{+}\}$. There is also the barrier branch
$E_{3}(\theta),\ E_{3}\geq E_{2}\geq E_{1}$. The branches $E_{3},\
E_{2}$ can terminate at points $\theta_{e},\ 2\pi-\theta_{e}$,
where both the upper metastable branch $E_{2}$ and the barrier
$E_{3}$ split out: $E_{3}=E_{2}$ at $\theta_{e}$, hence at
$\theta_{e}$ there is no activation energy,
$U=E_{3}-E_{2}\rightarrow0$ at $\theta\rightarrow\theta_{e}$, for
the decay of the metastable state $E_{2}$ to the stable one
$E_{1}$. The barrier activation disappears at $\theta_{e}$ as
$U\sim V_{e}(\theta_{e}-\theta)^{\nu}$ with $\nu=3/2$. The
branches cross at $\theta=\pi$ and we neglect the repulsion
between $E_{1}$ and $E_{2}$ due to quantum tunnelling, see the
Section \ref{sec:quantum}. In the course of the density wave
motion $\theta=\theta(t)$, the distribution of occupation numbers
$n_{\pm}=\{n_+,n_-\}$ for branches $\pm$
\begin{equation}
n_{\pm}=\frac{1}{2}(1\mp n)\,;\;n=n(\theta,t) \label{n+-}
\end{equation}
obeys the kinetic equations (see more in the Appendix
\ref{app:kin})
\begin{equation}
{\frac{dn}{dt}}=\frac{n_{eq}-n}{\tau}\,; \quad{\frac{d}{dt}}=
{\frac{\partial}{\partial
t}}+\dot{\theta}{\frac{\partial}{\partial\theta}}
\quad\dot{\theta}= \frac{d\theta}{dt}=v \,; \; n_{eq}=
\tanh{\frac{\Delta E}{2T} \,,\;\Delta E=}E_{+}-E_{-} \label{df/dt}
\end{equation}
where $n_{eq}$ is the value of $n$ in thermal equilibrium.

Here and mostly below, we imply an internal relaxation which is
due to passing over the local barriers. Its rate $\tau(\theta)$
does not depend on the velocity $\dot{\theta}$ but is an essential
function of the position $\theta$: $\tau\sim\tau_{0}\exp(U/T)$
where $U=U(\theta)$ and $\tau_{0}^{-1}$ is an attempt rate. Later,
in the Sec. \ref{sec:loops}, we shall discuss also the external
relaxation which is due to the mean free path of diverging pairs
of dislocation loops or solitons.

Finally, the pinning force averaged over both the sliding period
and the initial conditions is given as
\begin{eqnarray}
f & = & n_{i}\int_{0}^{2\pi}{\frac{d\theta}{2\pi}}
\left[F_{+}n_{+}+F_{-}n_{}\right]
 =  n_{i}\int_{0}^{2\pi}{d\theta}{\frac{n}{2}}{\frac{d}{d\theta}}\Delta E
\nonumber \\
& = & \frac{n_i}{2}\left[\Delta
En\right]_{\theta_{left}}^{\theta_{right}}-
\frac{n_i}{2}\int_{\theta_{left}}^{\theta_{right}}{d\theta\Delta E
{\frac{dn}{d\theta}}} \label{pin-av}
\end{eqnarray}
where
${\theta_{left}},{\theta_{right}}=2\pi-\theta_{e},\theta_{e}$ are
the bistability limits. In the last form of $f$, the first term
$n_{i}\Delta E|_{\theta_{max}}=f_{max}$ gives the energy
dissipation by the ultimate falling from the termination point,
while the second term erases this value $f_{max}$ due to occasions
of earlier fallings down which are more frequent at lower $v$.

\subsection{Stationary motion.\label{ss:stationary}}
Consider the stationary process when the density wave moves with a
constant phase velocity
$v=-\dot{\bar{\varphi}}=\dot{\theta}=const$, then $\partial
n/\partial t=0$. The solution of the kinetic equation
(\ref{df/dt}) is trivial, see Eqs. (\ref{f(s0)}, \ref{s=}) in the
Appendix \ref{app:kin}, being even simpler at low $T\ll\Delta
E(\theta)$: For $\Delta E(\theta)\gg T$, the pinning force can be
written as a weighted distribution of instantaneous forces:
\begin{equation}
f=n_{i}\int_{\pi}^{\theta_{max}}d\theta{F(\theta)}
\exp\left({-\int_{\pi}^{\theta}
{\frac{d\theta_{1}}{v\tau(\theta_{1})}}} \right) \,\,,\;
\frac{F}{2\pi}={\frac{d}{d\theta}}\frac{\Delta
E}{2}=\frac{F_{+}-F_{-}}{2} \label{f-the-F}
\end{equation}
or of energies $\Delta E$ dissipated via falling from the
metastable to stable branches:
\begin{eqnarray}
f & = & f_{max}\exp\left({-\int_{\pi}^{\theta_{max}}
{\frac{d\theta_{1}}{v\tau(\theta_{1})}}}\right)-\nonumber\\
& & - 2\pi n_{i}\int_{\pi}^{\theta_{max}}
{\frac{d\theta}{v\tau(\theta)}}\Delta E(\theta)
\exp\left({-\int_{\pi}^{\theta}
{\frac{d\theta_{1}}{v\tau(\theta_{1})}}}\right) \,,\nonumber\\
f_{max} & = & 2\pi n_{i}\Delta E(\theta_{max})  \label{f-the-E}
\end{eqnarray}
Here $f_{max}$ is the largest value of the pinning force to which
it saturates at high $v$; $\theta_{max}=\theta_{e},2\pi$
(depending on the impurity strength) is the most distant point
reachable by the metastable branch.

\begin{figure}[hbt]
\centerline{\epsfxsize=6cm \epsfbox{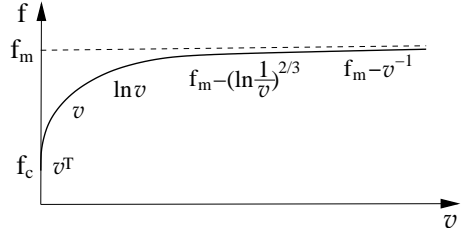}} \caption{ Schematic
plot of the $f(v)$ dependence showing several regimes of
relaxation, see the Sec.\protect\ref{sec:kin} for explanations,
here for shortness $f_{max}\rightarrow f_m$. The zoomed vicinity
of small $v=0$ and $f\approx f_{c}$ should recover the collective
pinning sliding regime with an opposite curvature of $f(v)$ and
finally, at very small $v$ of the collective creep regime, the
curvature will change once again (according to the Figure
\ref{fig:vf}). Note that the viscous force is not included in this
figure; it would simply give an inclination to dashed line of the
asymptotic regime.} \label{fig:f(v)}
\end{figure}

The Eqs. (\ref{f-the-F}) and (\ref{f-the-E})  are suitable for
calculations at small and large velocities correspondingly. After
simple calculations presented in the Appendix \ref{app:kin}, we
arrive at the results shown schematically at the $v-f$ plot of the
Fig. \ref{fig:f(v)}.

\bigskip

1. \emph{Small velocities} $v\ll\tau_{\pi}^{-1}$:\\
 where
$\tau_{\pi}=\tau(\pi)\sim\exp(U_{\pi}/T)$, $U_{\pi}=U(\pi)$ is the
maximal relaxation time in the region of the branch crossing point
$\theta=\pi$. The main contribution comes from the close vicinity
of $\pi$: $\theta\approx\pi +\delta\theta$ where $\delta\theta\sim
v\tau(\theta) $. We can distinguish between two subregimes.

1a. \emph{Very small velocities} $v\ll v_{\pi}=
(T/F_{\pi})\tau_{\pi}^{-1}\ll \tau_{\pi}^{-1}$, $F_{\pi}=F(\pi)$ :
\newline The decay happens as soon as the branch becomes
metastable in a vicinity of $\pi$, even before the $\theta$
dependence of $\tau$ is seen. The life time interval is
$\delta\theta\sim v\tau_{\pi}$, hence the law (\ref{f-the-F})
yields
\begin{equation}
f=\pi n_{i}v\tau_{\pi}\left. {\frac{d\Delta E}{d\theta}}\right|
_{\pi} =n_{i}vF_{\pi}\tau_{\pi}  \label{tau-pi}
\end{equation}
which emulates the phenomenological viscosity. This is the regime
of the \textit{\ linear collective conductivity }
$\sigma=v/f=const$. It shows an activated behavior via
$\tau_{\pi}^{-1}$ which can emulate the normal conductivity.

1b. \emph{Moderately small velocities} $v_{\pi} \ll v
\ll\tau_{\pi}^{-1}$:
\newline
Deviations $\delta\theta$ are still localized around $\pi$ but
they may be already large enough to see the decrease of the
barrier height: $ U\approx U_{\pi}-F_{\pi}\delta\theta/2\pi \,,\;
\tau=\tau_{\pi}\exp[-\delta\theta F_{\pi}/(2\pi
T)]\,,\;F_{\pi}/T\gg 1$
 We have exploited the fact that the branch $E_{b}(\theta)$ has
a minimum at $\pi$ so that the linear dispersion of $U(\theta)$ is
given by $dE_{2}(\theta)/d\theta=F_{\pi}$. The condition
$v\tau(\theta)\sim\delta\theta$ is fulfilled at $\delta\theta
F_{\pi}/T\sim\ln[v\tau_{\pi}]$ and finally the dependence $f(v)$
or $v(f)$ become
\begin{equation}
f\sim
n_{i}T\ln[v\tau_{\pi}]\,\;;\;v\sim\tau_{\pi}^{-1}\exp(f/n_{i}T)
\nonumber
\end{equation}
Convenient interpolation formulas for cases 1a,b are
\begin{equation}
f\approx 2\pi Tn_{i}\ln\left(1+v\frac{\tau_{\pi}F_{\pi}}{2\pi
T}\right) \;;\; v=\frac{2\pi T}{\tau_{\pi}F_{\pi}}
\left(\exp\frac{f}{2\pi Tn_{i}}-1\right) \nonumber
\end{equation}
but the integral representation (\ref{1ab}) is more precise.

The physics of $f\sim v$ regime is given by the high  probability
to stay on the metastable branch during a small displacement
$\delta\theta\sim$ $\tau_{\pi}v$. The physics of $f\sim\ln v$
regime is that at higher $v$ a wider region of $\delta\theta$ is
explored and the metastable branch starts to feel the decrease of
the barrier (long ahead there is either the termination point
$\theta_{e}$ or the minimal barrier point $\theta_{m}$, even if
unreachable yet at these moderate $v$).

\bigskip
2a. \emph{High velocities $v\gg{\tau_{\pi}}^{-1}$: restricted
metastability }$V_{1}<V<V_{2}$: \\
At higher velocities, $v\tau_{\pi}\gg 1$, the points distant from
$\theta=\pi$ can be explored, and especially
$\theta\approx\theta_{e}$ becomes important. The motion along the
branch $E_+$ starts to  reach a close vicinity of $\theta_e$ at
high enough $v_e\gg v\gg v_{\pi}$ where, see (\ref{s-e}),
\begin{equation}
v_{e}=\left(\frac{T}{V_{e}}\right)^{1/\nu}\frac{1}{\tau_{0}}
\;;\;\nu =\frac{3}{2} \;;\; V_e=cnst \label{v-e}
\end{equation}
and the coefficient $V_e$ was defined in (\ref{U-e}). Within the
limits $v_e\gg v\gg v_{\pi}$, the diminishing barrier $U\approx
V_e(\delta\theta)^{3/2}$ still stays at $U\gg T$ so that the decay
of the metastable branch $E_+$ is still activated. Finally we
obtain from (\ref{f-the-E})
\begin{equation}
f=f_{max}-C_en_{i}F_{e}\left( \frac{T}{V_{e}}
\ln{\frac{v_{e}}{v}}\right)^{1/\nu}\ ;\; v\sim v_{e}\exp\left[
-\frac{V_{e}}{T} \left(
{\frac{f_{max}-f}{C_en_{i}F_{e}}}\right)^{\nu}\right]\;;\; C_e\sim
1 \label{f-e}
\end{equation}
Realistically, this regime can be found only at very low $T$: the
crossover velocity $v_{e}$ must drop well below the
microscopically high values of the attempt rate $\tau_{0}^{-1}$.
 \bigskip

2b. \emph{High velocities $v\tau_{\pi}\gg1$: unrestricted
metastability } $V>V_{2}$.

For very strong impurities the metastability is maintained over
the whole period. Both the metastable branches and the barrier
branch stretch over all $\theta$ and $\tau\neq 0$ everywhere. The
important role is played now by the point $\theta_{m}$ where the
barrier activation is minimal: $\min U=U(\theta_{m})=U_{m}$. It
gives rise to the minimal (over the whole branch) relaxation time
$\tau _{m}\sim\exp(U_{m}/T)$, see the Figure \ref{fig:terms2}. In
the vicinity of $\theta_{m}$ we can write
$U(\theta)=U_{m}(1+(\theta-\theta_{m})^{2}B/2)$, $B\sim1$. There
are two different regimes described below.

\bigskip

3a. \emph{High velocity range} $v\gg
v_{m}=\tau_{m}^{-1}=\max\{\tau^{-1}\}$:

 Now the $1/v$ expansion works
in (\ref{f-the-E}) and we find from (\ref{1/v})
\begin{equation}
f={f}_{max}-const\frac{n_{i}}{v\tau_{m}}\sqrt{\frac{T}{BU_{m}}}\,
(\Delta E(2\pi)-\Delta E(\theta_{m})),\;{f}_{max}=2\pi n_{i}\Delta
E(2\pi) \label{fmx-1/v}
\end{equation}
The asymptotic force ${f}_{max}$ is the energy to create the pair
of dislocation loops (solitons), it is determined only by the
final point $\theta=2\pi$. It is approached by the law
$f-{f}_{max}\sim-1/v$ which reminds, at first sight, the collective
pinning corrections for high velocity, but the sign is opposite!

\bigskip

3b.\emph{ Moderately high velocities $v_{m}\gg v\gg
\tau_{\pi}^{-1}$}:

 Now $v$ is high enough to reach the point $\theta_{m}$, but not
that high yet as to bypass it easily - still $v\tau_{m}\ll 1$,
hence the point $\theta_{m}$ will provide the major relaxation.
This effects will be particularly pronounced near the threshold
$V_{2}$ when $U_{m}$, and consequently $\tau_{m}$, become much
smaller with respect to their typical values over the branch. We
can easily obtain (up to numerical coefficients)
\begin{equation}
f=2\pi n_{i}\left\{ \Delta E(\theta_{m})+(\Delta E(2\pi)-\Delta
E(\theta_{m}))P \right\} \,,\; P\sim\exp\left(
-\sqrt{\frac{T}{BU_{m}}}\frac{1}{v\tau_{m}} \right) \nonumber
\end{equation}
This formula tells us that the main force is provided by the part
of the branch between $0$ and $\theta_{m}$ or, equivalently, by
the energy released from the relaxation at $\theta_{m}$ - similar
to termination points for the case 1b. The second contribution
$\sim P$ comes from the remaining part of the branch, between
$\theta_{m}$ and $2\pi$, but the penetration probability $P$ to
this part through the ''hot point'' $\theta_{m}$ is exponentially
small in $1/v$. Nevertheless this small probability is responsible
for the irreversibility and memory effects. Indeed $P$ is a
probability to create dislocation loops (solitons) which are long
living plastic deformations.

{\em Summary:} The most important cases of these regimes are shown
on the Figure \ref{fig:f(v)} and will be discussed again in
comparison to experiments in the Sec. \ref{sec:applic}. We have
skipped from consideration the most limiting cases: highest $v$
for strongest $U$ will be considered in the Sec. \ref{sec:loops}
devoted to effects of dislocations; lowest $v$ (the law $v^T$ on
the Figure \ref{fig:f(v)}) will be considered in the Sec.
\ref{sec:ensemble} devoted to the ensemble averaging.

\subsection{Linear response. \label{response}}
The standard response function $\chi=\delta\bar{\varphi}/\delta f$
 is measured in CDWs as the dielectric susceptibility
$\varepsilon\sim\chi$. Within the collective pinning picture it
can be found with the help of the same kinetic equation as
described in the Appendix \ref{app:kin}. Here we shall follow a
more intuitive and transparent approach first considered in Ref.
\cite{Larkin95}. Consider the reaction of local bistable states to
a weak perturbation for low $\omega$ or at large $t$. In
equilibrium, the impurities occupy the lowest branch $E_{1}$ (for
a moment we neglect the effects of the thermal population). For
small variations $\delta\theta$ the main contribution comes from
the degeneracy point $\theta=\pi$ where the two stable branches
cross each other. Impose a small homogeneous shift of the relative
impurity positions $\delta\theta$ at $t\geq0$. In the whole
ensemble of impurities, those ones become metastable which have
been occupying the interval of stable positions
$\pi-\delta\theta<\theta<\pi$ while the stable positions at
$\pi<\theta<\pi+\delta\theta$ become empty, as shown on the Figure
\ref{fig:few-terms}. The subsequent evolution follows the
relaxation towards the thermal equilibrium. The imbalance of
forces gives rise to the inverse response $\chi^{-1}$. Going from
the real time $\chi(t)$ to the Fourier representation
$\chi_{\omega}$ we find
\begin{equation}
\delta f(t)=\delta\theta n_{i}F_{\pi}\exp(-t/\tau_{\pi}),\,\,\,\;
\theta_{\omega}=\frac{\delta\theta}{i\omega},\,\,\,\,\;
\chi_{\omega}^{-1}= \frac{\delta
f_{\omega}}{\delta\theta_{\omega}}=
\frac{2n_{i}F\pi}{1+1/i\omega\tau_{\pi}}. \label{F-omega}
\end{equation}
At high $\omega\gg\tau_{\pi}^{-1}$, $\chi_{\omega}$ saturates at
its maximal, real value $\max\chi\sim\left( n_{i}F_{\pi}\right)
^{-1}$. The small $\omega$ limit of $\chi(\omega)$ corresponds
completely to the small velocity limit of the $v-f$ law
(\ref{F-v}). At $\omega\tau_{\pi}\ll 1$, the system reaches the
thermal equilibrium and $\chi_{\omega}^{-1}$ disappears $\sim
i\omega$ giving only a contribution $n_{i}F_{\pi}\tau_{\pi}$ to
the damping parameter $\gamma^{-1}$ in full agreement with the
$f-v$ results.

\section{Generation of dislocations at high velocities.
\label{sec:loops}}

By now we exploited mostly the general properties of metastable
branches: the existence of points of levels degeneracy $\Delta
E=0$, of the barrier termination $U=0$ or of its minimum.
Topological defects were implied to exist in the background
providing peculiar reasons for the metastability. Quantitative
results were derived for a general position of $\theta$ when
dislocation loops have not emerged yet as distinct (and distant)
entities. A more elaborated analysis is required near the final
point $2\pi$ of unrestrictedly metastable branches accessible at
high $v$. Here the metastable configurations are formed explicitly
by divergent pairs of solitons, more generally dislocation loops,
centered around $\pm X$ -- see Figures \ref{fig:chain} and
\ref{fig:bisoliton}. Their interaction will modify both the
structure of energy branches and the related kinetics.

First of all, we add here a few technical notes necessary to work
with dislocations. For details see \cite{LLVol.7,friedel,nabarro}
in general and \cite{Feinb88,Brazov-Matv} specially for CDWs
\footnote{For more complicated techniques of working with
ensembles of dislocations see \cite{BK-ECRYS02}.}. In CDWs
particularly, dislocations have all properties of conventional
vortices in planar magnets or superfluids (with an exception for a
special conservation law for the total area embraced by the loops
which distinguishes the climb from the glide). Even more
pragmatically, we can invoke a common wisdom of the magnetostatics
considering dislocations as currents, the strain as the magnetic
field $\bf{\nabla}\varphi \Leftrightarrow\bf{H}$ and the stress as
the induction $C\bf{\nabla }\varphi\Leftrightarrow\bf{B}$. The
signs are however different: antiparallel dislocation lines
(dislocation lines with opposite polarities) attract each other.
For our typical case of dislocation loops lying within the  plane
$(y,z)$, i.e. perpendicular to the sliding axis $x$, we arrive at
the following prescriptions: In the inhomogeneous field
$\varphi(\bf{x)}$ created by other sources, e.g. another
dislocation loop or the impurity, the glide force  in $x$
direction applied to the unit length of the dislocation loop is
$\sim C\bf{\nabla}_{\bot}\varphi$. The energy per unit area of the
dislocation loop is $\sim C\partial_{x}\varphi$. The dislocation
loop self-energy has the standard vortex form: at large $R$,
$E_{DL}(R)$ $\sim CR\ln(R/a_{0})$ in $d=3$ or $\ln(R/a_{0})$ in
$d=2$, respectively. At the minimal $R\sim a_{0}$ the dislocation
loop is interpreted as the nucleus  dislocation loop embracing
just one chain - the soliton, $E_{DL}(R)\rightarrow E_{s}$. The
phase distortion by the dislocation loop itself at a given point
is the half of the steric angle $\Omega_{3}/2 $ (the angle
$\Omega_{2}$ in $d=2$) at which one views the dislocation loop
form this point. This angle evolves from $0$ to $2\pi$ along the
path crossing the dislocation loop, these asymptotic values are
approached as $\pm\delta \Omega_{d}\sim(R/X)^{d-1}$.

\subsection{Effects of dislocations upon metastable states.
\label{sss:LLR}}

\begin{figure}[hbtp]
   \centerline{\epsfxsize=6cm
   \epsfbox{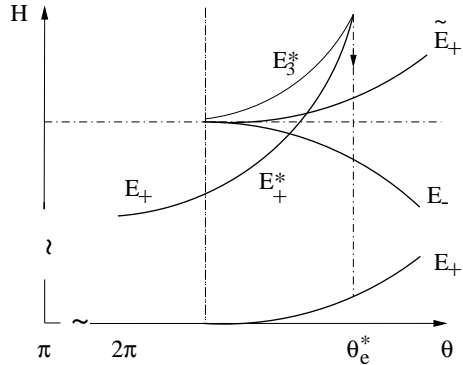}}
\caption{Special effects near the crossover between two successive
periods which are originated by interactions of distant
dislocation loops. $E_{+}^*$ is the overshooting part of the
branch $E_{+}$. The branch $\tilde{E}_+$ differs from the lowest
branch $E_+$ by presence of two solitons at $\pm\infty$.}
\label{fig:terms3}
\end{figure}

Consider firstly the long range instabilities near
$\psi,\theta=2\pi$ corresponding to the divergent pair of
solitons. From large distances, the dislocation loops interact
with each other and with the impurity via long range elastic
forces. By definition, the presence of dislocation loops at points
$\pm X$ displaces the phase in between, at $x=0$, by $2\pi$ with a
deficiency $-\delta\psi=2\pi-\psi(0)$. The latter is given by the
sum of (steric) angles $\Omega_{d}\sim(R/X)^{d-1}$ (at large
$X/R$) of their view from the point $x=0$, see the Figure
\ref{fig:chain}. The energy $W$ is equal to the energy of two
dislocation loops $2E_{DL}(R)$, also taking into account their
mutual interaction at finite $X$. The attractive potential $\delta
W $ of the loops is given by the longitudinal stress
$C\partial_{x}\varphi\sim-(C/R)(R/2X)^{d}$ produced by one loop
over the area $\sim R^{d-1}$ of another one at the distance $2X$:
$\delta W\sim-CR^{d-2}(R/2X)^{d}$. Finally, we exclude $X$ in
favor of $\psi$ to arrive at
\begin{equation}
\delta\psi\sim-(R/X)^{d-1}\, , \\
\delta W(\psi(X))\sim-CR^{d-2}(R/2X)^{d}
\sim-CR^{d-2}(-\delta\psi)^{d/(d-1)} \label{LLR}
\end{equation}
We notice that, while the force disappears as
$W^{\prime}\sim(-\delta\psi )^{1/(d-1)}\rightarrow0$, its
derivative diverges $W^{\prime\prime} \rightarrow\infty$ giving
rise to the branch instability. Namely, even for arbitrary large
$V\gg W$ we shall meet the condition
$W^{\prime\prime}=-V^{\prime\prime}$, hence there is always a
solution of the last equation in (\ref{end}) at some
$\psi_{e}^{\ast}<2\pi$.\footnote{ In CDW materials there are also
long range Coulomb forces which are not screened at low $T$. They
affect drastically the energetics of dislocations
\cite{Brazov-Matv} which leads to an even more singular law
$W(\psi)-2E_{s}\sim E_{s}|\psi-2\pi|$.}
 A more elaborated analysis given in  the Appendix \ref{app:ms}
 shows that this value of $\psi$ is reached at
 $\theta=\theta_{e}^{\ast}>2\pi$, that is
already in the next circle of the mean sliding. According to the
first relation in the Eq. (\ref{LLR}), there is a distance between
dislocation loops $X_{e}$ associated to the phase deficiency
$\psi_{e}-2\pi$. We see that instead of diverging
($X\rightarrow\infty$, which requires for $\psi\rightarrow2\pi$) at
the end of the period $\theta\rightarrow2\pi$, the pair of
dislocation loops looses its stability at some critical distance
$X_{e}^*$ corresponding to $\psi_{e}^{\ast}<2\pi$ of the branch
$E_+^*$. Then it falls to the higher distance
$X_e^*\Rightarrow\tilde{X}_e$ corresponding to
$\tilde\psi_{e}>2\pi$ of the branch $\tilde{E}_+$, and continues the
new circle along this branch.
 As a result, the region encircled by the dashed line on the Figure
 \ref{fig:terms2}, acquires the structure shown at the magnified picture
 of the Figure \ref{fig:terms3}.  We shall call $\theta_{e}^{\ast}$ the
\emph{overshooting } termination point of the overshooting part
$E_{+}^*$ of the branch $E_{+}$ penetrating into the next period
$\theta>2\pi$. The appearance of new termination points brinks
some similarity to the vicinity of the $\pi$ point at $V\gtrsim
V_{1}$.

The possibility for relaxation of $E_+^*$ to $\tilde{E}_+$ is not
unique. In principle, there is always an option to fall down
directly to the lowest branch $E_{+}$ (the same $\theta$ but
$\psi$ being close to $0$ rather than to $2\pi$), thus releasing
the energy $\approx2E_{s}$. But it requires for the annihilation
of the distant pair of dislocation loops which involves a large
barrier both in energy and configuration. There is one more
option: to fall to a close, in energy, descending branch $E_{-}$
(see the Figure \ref{fig:terms3}). But this is the strongly
advanced branch (all descending branches are advanced
configurations) corresponding to the pair of dislocation loops
with opposite polarities, so that this transition would require to
switch phases from $\approx2\pi$ to $\approx-2\pi$ along the whole
interval $(-X,X)$ which is hardly possible.

\subsection{Kinetic effects of diverging dislocation loops.}

The complete kinetics of these states is complicated for several
reasons, one of them is a larger number of branches involved. A
simplification comes from the high velocity condition to reach
this regime: $v\gg1/\tau(\theta)$ at all $\theta$. Here $\tau$ is
the relaxation time for dropping from the term $E_{+}$ to $E_{-}$
which we always kept in mind before. Neglecting this basic
relaxation in a small vicinity of $2\pi$, we can concentrate on
the short relaxation time $\tau^{\ast}\ll\tau$ to fall from
$E_+^*$ to $\tilde{E}_+$. Now on top of the law
$f-f_{max}\sim-n_{i}F/v\tau$ (the case 3a of the Sec.
\ref{ss:stationary}) we shall see the sequence of regimes similar
1a,1b,2a (also from the Sec. \ref{ss:stationary}) for the case of
the restricted bistability, but with much smaller $\tau^{\ast}$
instead of $\tau$. This new contribution to $f(v)$ falls off
slower (hence finally winning) than $\sim1/v$ but its emergence is
delayed because the force is reduced to the smaller value
$F^{\ast}$, as given by the inclination of the overshooting branch
$E_+^*$ (see the Figure \ref{fig:terms3}). We shall notice traces
of this regime in applications to CDWs in Sec. \ref{sec:applic}.

Actually, the relaxation time approach may not be applicable any
more. For well separated solitons, the extrinsic mechanisms of
relaxation enter the game: annihilation of solitons and
antisolitons produced by neighboring pinning centers along the
chain, aggregation of solitons into growing dislocation loops,
disappearance via phase slips. Now the final rate is determined by
the soliton distance $X$ in comparison to its collision mean free
path $\lambda$ rather than by the time in compare to $\tau $ as it
was for intrinsic processes. Phenomenologically, $\lambda $ is
included to the starting kinetic equation by the following
substitution which is noticeably  different from (\ref{df/dt}):
\begin{equation}
\frac{\partial }{\partial \theta }\rightarrow \frac{\partial }
{\partial\theta }+\frac{1}{\lambda
}\frac{dX(\psi(\theta))}{d\theta}\,. \nonumber
\end{equation}
where $X(\psi)$ is the distance associated to the retardation
$\psi (\theta )$ taken along the term $E_{+}$. It results in the
following modification of Eqs. (\ref{f-the-F},\ref{f-the-E}), as
well as of (\ref{s=}):
\begin{equation}
s=\int_{\pi }^{\theta }\frac{d\theta }{v\tau (\theta
_{1})}\Rightarrow s=\int \frac{d\theta }{\lambda
}\frac{dX}{d\theta }= \frac{X(\theta )}{\lambda } \label{X-lambda}
\end{equation}

Remind now(see Eq. (\ref{LLR}) and the text above it) that  the
diverging dislocation loops of the radius $R$ leave in between the
phase retardation approaching $2\pi $ as (at d=3) $\delta \psi
=\psi (X)-2\pi \sim -R^{2}/X^{2}$; and the same long range elastic
deformations provide their attraction with the potential $\sim
-CR^{4}/X^{3}$. Well before the overshooting instability develops,
that is at $\delta\theta<0$, we have (see the Appendix
\ref{app:ms}, \# 5) $\delta\psi\approx\delta\theta$ as it should
be for a very strong impurity. Then $\delta{\cal H}
=\mathcal{H}-2E_s\sim-E_s(-\delta\theta)^{3/2}$, hence $F\sim
E_{s}(-\delta\theta)^{1/2}$, and we arrive at
\begin{equation}
X\sim\frac{R}{\sqrt{-\delta\theta}}\,;\;
s\sim\frac{R}{\lambda\sqrt{-\delta\theta}} \nonumber
\end{equation}
The condition $s\sim 1$ in (\ref{X-lambda}) defines the
characteristic $\delta\theta\sim R/\lambda$ and finally we obtain
the force correction
\begin{equation}
\delta f_{max}  = -n_{i}\int 2F\frac{\delta\theta}{2\pi}
 \sim -n_i E_{s}(-\delta\theta)^{3/2}
\sim -n_i E_{s}\left(\frac{R}{\lambda}\right)^{3/2}
\end{equation}
The last formula relates the high $v$ asymptotics of the pinning
force and the mean free path $\lambda$ of dislocation loops.

\section{Quantum effects.\label{sec:quantum}}

The strong pinning picture gives also an access to quantum effects
which become important at very low temperatures, when the thermal
activation is not efficient. The quantum creep became the subject
of experimental studies since \cite{q-creep} which work has
attracted a substantial attention in theory, see e.g.
\cite{quantum-depinning}. The existing schemes concentrate upon
the quantum nucleation of CDW advances in regions free from
impurities. But this approach cannot tell us how the pinning is
released and the motion as a whole is initiated. For this goal, it
is necessary to consider the quantum decay of metastable states at
the pinning centers.

\begin{figure}[htb]
\centerline{\epsfxsize=6cm \epsfbox{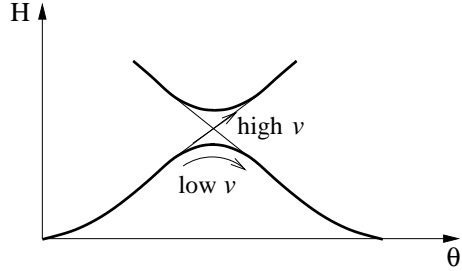}}
 \caption{
Pinning extinction by the quantum tunnelling between branches
$E_{\pm}$. The gap $\protect\delta _{q}$ opens between classical
branches $E_{1}$ and $E_{2}$ which were degenerate at
$\protect\theta=\pi$ (compare to the Figure \ref{fig:few-terms}).
Adiabatically, the system follows (the arc arrow) the lowest
branch $E_{1}^q$, thus giving the zero force in average over one
period. Only nonadiabatic transitions (the diagonal arrow)  from
$E_{1}^q$ to $E_{2}^q$ allow to reach the metastable branch to
gain a net pinning force.} \label{fig:quantum}
\end{figure}

Phenomenologically, the dynamics is introduced via the kinetic
energy ($I/2)\dot{\psi}^{2}$ with $I$ being the  ''moment of
inertia'' associated with the ''angle'' $\psi $. Typically, $I\sim
\omega _{0}^{-2}$ where $\omega _{0} $ is a microscopic scale for
frequency of local quantum vibrations, usually it is associated
with the attempt frequency $\omega _{0}\sim \tau _{0}^{-1}$. Then
the quantum Hamiltonian is
\[
\hat{H}=\hat{H}(\psi ,\theta )=P_{\psi }^{2}/2I+H(\psi ,\theta
)\;;\;P_{\psi }=-i\hbar \partial /\partial \psi \label{H-qu}
\]
where $P_{\psi }$ is the momentum conjugated to $\psi $ and
$\theta (t)$ is considered as the time dependent parameter. The
quantum interference   is efficient only near the branch crossing
point $\theta =\pi $. Then the degeneracy will be lifted by
splitting the levels $E_{1}$ and $E_{2}$ which opens the quantum
gap $\delta _{q}$ between them as shown on the Figure \ref
{fig:quantum}. Within the normal dynamics of Eq. (\ref{H-qu}), the
gap is $\delta _{q}\sim \exp (-cnst\sqrt{IU}/\hbar )$. But the
emittance of phase phonons to the bath will have a pronounced
dissipative effect upon the tunnelling as we shall describe in the
end.

Working within the normal dynamics, we arrive at the standard
Landau-Zener problem of tunnelling due to slow time dependence of
the Hamiltonian (via $\theta (t)$ in our case). The standard
notion  is that if $v,\ \omega $ are negligibly small, then the
system will follow adiabatically the exact quantum branch of the
lowest energy $E_{1}^{q}$ which is a mixture of classical branches
$E_{1,2}$
\begin{equation}
E_{1}^{q}\approx E_{2}+E_{1}-\sqrt{\left( E_{2}-E_{1}\right)
^{2}+\delta _{q}^{2}} \label{E-q}
\end{equation}
and the average force is zero within an exponential accuracy.
Namely, at low $v\ll\delta _{q}$ the force will be determined by
the small probability of the nonadiabatic transition to the upper
metastable branch $E_{2}$:
\begin{equation}
f\sim \exp \left( -\frac{1}{v\tau _{q}}\right) \,\,;\;\frac{\hbar
}{\tau _{q}}=\delta _{q}  \nonumber
\end{equation}

Oppositely, the tunnelling between bare branches $E_{1,2}$ is
suppressed for large velocities or frequencies of $\theta(t)$
($v,\omega\gg\delta_{q}$), and the average force is given by the
classical picture we have described above in terms of the thermal
relaxation time. In kinetics, the effects of quantum and thermal
fluctuations seem to be similar and in our simplified picture the
inverse times are additive $\tau ^{-1}=\tau _{\pi }^{-1}+\tau
_{q}^{-1}$.

The phenomenologically introduced inertial dynamics of the Eq.
(\ref{H-qu}) may not be valid. Actually, the dynamics of the
variable $\psi $ becomes retarded and dissipative because of the
nonlocal contribution from the whole field $\varphi({\bf x},t)$.
Its action (on the imaginary time axis) is given by
\begin{equation}
S_{bulk}[\varphi ({\bf x},t)]=\frac{C}{2}\int \int d^{d}{\bf
x}dt\left\{ u^{-2}\left( \partial _{t}\varphi \right) ^{2}+({\bf
\nabla }\varphi )^{2}\right\}   \nonumber
\end{equation}
where $u$ is the phason velocity and $C$ is the static elastic
modulus. Now we can repeat the prescription of the Sec.
\ref{ss:ms} to integrate out $\varphi $ at all ${\bf {x}\neq 0}$
keeping fixed the value $\varphi ({\bf 0},t)=\varphi (t)=\psi
+\theta $. In the Fourier representation we have
\begin{equation}
S_{kin}[\varphi (t)]=\frac{1}{2}\sum_{\omega }K_{\omega }\left|
\varphi _{\omega }\right| ^{2}\,,\;K_{\omega }=I_{\omega }\omega
^{2}  \label{L-t}
\end{equation}
where $I_{\omega }$ is the frequency dependent generalization of
the constant $I$, while the kernel $K_{\omega }$ is given by
\begin{equation}
\frac{C}{K_{\omega }}=\int \frac{d{\bf {k}}}{(2\pi )^{d}}
\frac{1}{{\bf {k}^{2}+\omega ^{2}/u^{2}}}=\int \frac{d{\bf
{k}}}{(2\pi )^{d}}\left\{ \frac{1}{{\bf {k}^{2}}}-\left(
\frac{1}{{\bf {k}^{2}}}-\frac{1}{{\bf {k}^{2}+\omega
^{2}/u^{2}}}\right) \right\}   \label{K}
\end{equation}

Mainly we shall address  the case $d=3$. Here we should use the
second form in the Eq. (\ref{K}) where the first term in brackets
$\left\{ {}\right\} $ gives, after the regularization at high
$k\sim a_{0}^{-1}$, just the elastic contribution
(\ref{eq.phi.solution}) we considered in the Sec. \ref
{ss:models}; its scale is $K_{0}\sim Ca_{0}$. The second term,
regular at high $k$ and hence model independent, gives the
contribution $\sim |\omega |/u$ and finally we obtain
$K_{\omega}\sim Ca_{0}^{2}|\omega |/u$, while the  regular
contribution $\sim \omega ^{2}$ appears only as the next order in
perturbation. The form (\ref{L-t}) with $K_{\omega }\sim |\omega|$
is typical for the dissipative quantum mechanics \cite{Leggett},
which route we shall follow below. In the time representation, the
$K_{\omega }$ gives rise to the kinetic action
\begin{equation}
S_{kin}[\varphi (t),t]\sim \hbar\Gamma
\int_{0}^{t}dt_{1}dt_{2}\left( \frac{\varphi (t_{1})-\varphi
(t_{1})}{t_{1}-t_{2}}\right) ^{2}\,\,,\; \Gamma
=\frac{Ca_{0}^{2}}{\hbar u}\gg 1 \label{K-d=3}
\end{equation}
It is logarithmically divergent in the tunnelling time $t$ if the
tunnelling trajectory acquires a final increment (between $\pi$
and $-\pi $ in our case). The total action can be written
schematically as
\begin{equation}
S(t)=\hbar \Gamma \ln (t\omega _{u})+\frac{I}{t}+Ut\;,\;\omega
_{u}=\frac{u}{a_{0}}  \nonumber
\end{equation}
Here we have included also the regular inertia moment $I$ for
which there are always some local sources. The tunnelling level
splitting is $\delta _{q}\sim \exp \left( -S_{min}/\hbar \right)
\,\,$where$\;S_{min}=\min S(t)$.

At intermediate  $1\ll \Gamma \ll \sqrt{IU}/\hbar $, we arrive at
the usual WKB result $S_{min}\sim \sqrt{IU}$, but with an
essential preexponential suppression of tunneling:
\[
\delta _{q}\sim \left( \frac{U}{I\omega _{u}^{2}}\right) ^{\Gamma
/2}\exp \left( -cnst\frac{\sqrt{IU}}{\hbar }\right)
\]

At higher $\Gamma \gg \sqrt{IU}/\hbar $, the tunneling suppression
is more drastic:
\begin{equation}
\;S_{min}\approx \Gamma \ln
\left(\frac{I\omega_{u}}{\hbar\Gamma}\right)\,\,,\; \delta
_{q}\sim\left( \frac{I\omega _{u}}{\hbar \Gamma }\right) ^{\Gamma
} \nonumber
\end{equation}
The last condition imposes the constraint upon the value of the
pinning potential which must be compatible with the metastability
condition. For typical models we find $\omega _{0}\gg \omega _{u}$
to be necessary.

In $d=2$, the first form of $K_{\omega}^{-1}$ in (\ref{K}) should
be used. Now the whole integral is diverging at small $k$ yielding
a universal result. We obtain an even slower frequency dependence
$K_{\omega}\sim C/\ln \left( u/|a_{0}\omega|\right) $ that is
$K(t)\sim 1/(t\ln ^{2}t)$ instead of $t^{-2}$ as in (\ref{K-d=3}).
The logarithmic divergence is the same  we have noticed for the
static problem. We see once again, remind the Section
\ref{ss:models}, that in $d=2$ short and long range effects cannot
be separated, whatever they concern: the interference of the
collective and the local pinning, or the local dynamics and the
one related to emittance of sound to the bulk.

Apparently further studies are necessary. Already now we can
understand the quantum creep as the tunnelling between retarded
and advanced configurations at the moment when they become almost
degenerate. The process is strongly affected by emitting sound
excitations which drive it to be dissipative even at $T=0$.

\section{Ensemble averaging of pinning forces. \label{sec:ensemble}}
Above, in studies of both $f-v$ and $\chi$, we have assumed the
simple exponential relaxation at identical pinning centers. In
reality, there may be either a broad distribution of impurities
strengths or a tail in addition to the peak at the value for a
typical pinning center. Effects of distributions can be important
in applications and they are particularly necessary to build a
bridge to the collective pinning regime where the broad
distribution is the basic ingredient. We shall concentrate on the
most pronounced effects at lowest $v$ and $\omega$ compatible with
the local pinning picture.

For a distribution of barriers $\mathcal{P}_{U}(U)$, the
distribution of $\tau$ is
$\mathcal{P}_{\tau}=\mathcal{P}_{U}dU/d\tau=\mathcal{P}_{U}T/\tau$
and we shall consider two limiting cases. Firstly, the model with
the exponential distribution of barriers corresponds to
microscopic fluctuations of e.g. a distance between the CDW chains
and impurities:
\begin{equation}
\mathcal{P}_{U}\sim U_{0}^{-1}\exp(-U/U_{0})\,,\quad
\mathcal{P}_{\tau}\sim(T/U_{0}\tau_{0})(\tau_{0}/\tau)^{1+T/U_{0}}.
\label{P_U}
\end{equation}
Similar effects appear for Poisson and Gaussian distributions.
Secondly, we can try also the scaling distributions which appear
intrinsically within the collective pinning regime, now
$\mathcal{P}_{\tau}\sim(\tau\ln^{\alpha}\tau)^{-1}$ where the
index $\alpha>1$ (this condition is necessary for convergence of
the normalizing integral) depends on the dimension $d$ and the
critical index $\chi$). Naturally, the distributions are
normalizable, but we also notice that in all cases their first
moment, which is the mean value of $\tau$, is divergent:

\begin{equation}
{\int{d\tau\mathcal{P}_{\tau}}(\tau)=1\;\mathrm{but}\;\left\langle
\tau\right\rangle
=\int{d\tau\mathcal{P}_{\tau}}(\tau)\tau=\infty}. \nonumber
\end{equation}
Remind now that the low $\omega,v$ asymptotic for both
$\chi_{\omega}^{-1}$ and $f(v)$ are linear in $\tau$
($\chi_{\omega}^{-1}\sim\omega\tau$ and $f(v)\sim v\tau$) and then
saturate or change to a slow growth at higher crossover values
$\omega\tau,v\tau\sim1$. Hence, their averages will be divergent
within the $\sim\tau$ regime and saturate at the crossovers. E.g.
for the response function (\ref{F-omega}) we obtain

\begin{eqnarray}
& & \chi_{\omega}^{-1}= {\int{d\tau\mathcal{P}_{\tau}}(\tau)
\frac{{\,2} n_{i}F_{\pi}} {1+1/i\omega\tau}\,}\,;\nonumber\\
& &
\Im\chi_{\omega}^{-1}\sim{\,}n_{i}F_{\pi}\omega^{1}{{\mathcal{P}_{\tau}}
(\omega}^{-1}{)\sim}n_{i}F_{\pi}
{{\mathcal{P}_{U}}(T\ln(\omega}^{-1}){)} \label{Ftotal}
\end{eqnarray}

\noindent Thus, the low $\omega$ tail of the imaginary part
$\Im\chi_{\omega}^{-1}$ gives the direct access to the
distribution of potential barriers. A similar result is obtained
for the real part $\Re\chi_{\omega}^{-1}$ which is given by the
integral of $\tau^{2}$, the second moment. The interpretation is
that at relatively low $\omega$ (still within the local pinning
domain) or high $T$, only those long living states contribute
which are due to rear occasions of large impurity potentials,
hence large barriers $U\sim T\ln\omega^{-1}$. Then
$\chi^{-1}\sim\mathcal{P}_{U}(T\ln1/\omega)$. E.g. for the
exponential distribution (\ref{P_U}) we find
$\chi_{\omega}^{-1}\sim\omega^{T/U_{0}}$.

The same procedure can be applied to the $f-v$ dependence. The law
(\ref{tau-pi}) was derived for a typical impurity. At the lowest
$v$, we may still find some rare regions - may be clusters of host
imperfections or particularly strong impurities - where barriers
$U$ are high enough so that $v\tau (U)\sim1\;$still\ holds, hence
we are looking for $U\sim T\ln\left(v_{0}/v\right) $ and the
pinning force is given just by their probability
$\mathcal{P}(T\ln(v_{0}/v))$. For the exponential distribution we
find that the ``current-voltage'' characteristics change from the
Ohm low $f\sim v$ at low $v$ to the nonlinear regime $f\sim
v^{T/U_{0}}$ with a diverging differential resistance at lowest
$v$, as shown on the Figure \ref{fig:f(v)}.

For the scaling distribution of the collective pinning we obtain a
very slow decrease $\chi_{\omega}^{-1}\sim|\ln\omega|^{-\alpha}$
and $f\sim|\ln v|^{-\alpha}$. These results are very encouraging
since they show the same functional forms as the formula
(\ref{eq:v_creep}) for the collective pinning creep with
$\alpha=1/\mu=(2-\zeta)/\chi$ and the formula
(\ref{eq:L-low.temp}) with $\alpha=2/\chi$. Actually there is no
discrepancy, even in powers, since Eq.(\ref{eq:v_creep}) was
derived already for the periodic media, where $\zeta=0$.

Certainly, the straightforward merging of results from collective
and local pinning theories is speculative and should be used only
as a suggestion for more rigorous studies. Nevertheless the
observed unification of different ($v(f)$ and $\chi(\omega)$) and
differently derived results of the collective pinning theory at
one side, and essentially different view of the local pinning
theory at another side, looks quite optimistic. We shall discuss
some other aspects of this correspondence in the next Section.

\section{Interference of local and collective pinnings.\label{sec:LP-CP}}
We have already seen that the simplified but explicit approach of
the local pinning provides clear interpretations for hypotheses of
the collective pinning, particularly on the origin of metastable
states. The same time, it raises challenges which have not been
met yet. E.g. the following items of the above analysis are
important for the theory of the $v-f$ dependencies:
\newline

1. A fraction of metastable branches terminate at the end points
or relax fast at minimal points. There are those points which
determine $f(v)$ at high enough $v$ but they are not accounted in
the collective pinning theory.
\newline

2. A fraction of stronger metastable branches do not posses this
instability which at first sight allows for the large $v$
perturbative approach of the collective pinning theories. But it
results in the even more obscure effect of generating  sequences
of dislocations or solitons. Now the $v-f$ dependencies are
determined by competing processes: the annihilation contrary to
the aggregation. These processes are not accessible yet to
existing theories except for the simple treatment of the local
pinning which also needs to be further elaborated. Particularly
demanding are studies of aggregation and annihilation of
dislocations, their own pinning, etc.

Consider more closely some other aspects of interference of
collective and local pinning centers. Pragmatically, we shall
concentrate on those which will be important in studies of the
response functions in applications to susceptibility of density
waves discussed in the next Section. The problem is more
fundamental, being related to interference of different scales
within the collective pinning picture. Here, a great
simplification comes from the  clear separation of both the length
and the time scales between the local pinning and the collective
pinning regimes. The slow evolution of the collective pinning
allows us to consider it within a given distribution of local
metastable states. Then the collective pinning is described by the
same Hamiltonian (\ref{H-vph}) where the generic pinning potential
$V(\varphi(\bf{x}))$ is substituted by the local pinning energy
$V(\varphi({\bf x}))\Rightarrow E_{a}(\varphi(\bf{x})+ {\bf
Qx}_{i})$. The collective pinning evolution will be described by
the same equation (\ref{eq:eq.motion}) where the generic random
force is substituted by $F_{a}=E_{a}^{\prime}$: within the phase
language it reads
\begin{equation}
\gamma^{-1}\partial_{t}\varphi-C{\bf\nabla}^{2}\varphi+
\sum_{i}F_{a}(\varphi({\bf x})+{\bf Q}{\bf x}_{i})=f \nonumber
\end{equation}
Without local effects, the sum over impurities would correspond to
the random force density $g(\bf{x},\varphi(\bf{x}))$ of the Eq.
(\ref{eq:eq.motion}). Here the major difference from the pure
collective pinning is that now $F=F_{a}$ is a two-valued function
where the degree of freedom $a=\pm$ or $a=1,2$ allows for the
thermodynamic or stochastic treatments, contrary to the frozen
disorder $\bf{x}_{i}$.

For the linear response, or for the motion with a small velocity,
most of impurities are locally relaxed, and we shall neglect
others for a moment. Then the random potential becomes single
valued, $a=1$ only, but its correlator $R$, defined by the
equation (\ref{eq:V-correlations}), changes in comparison with the
generic form $R\sim\cos(\varphi)$. To make things simple, consider
the limit of a very strong impurity potential $V$. Then the lowest
energy state is $\psi=\psi_{1}(\theta)=
\theta-2\pi\Theta(\theta-\pi)$ and the pinning energy becomes
$W(\psi_{1})$ which originates the discontinuous force. E.g. for
the short range model of the Sec. \ref{sss:srm} we have
$V(\theta)=V(1-\cos\theta)\Rightarrow V^{\ast}(\theta)$ with
\begin{equation}
V^{\ast}(\theta)=W(\psi_{1}(\theta))=
E_{s}\left(1-\sin\frac{\left|\theta-\pi\right| }{2}\right) \,,\;
F=\frac{E_{s}}{2}\sin\frac{\theta}{2}\mathrm{sgn}(\theta-\pi)
\nonumber
\end{equation}

 Remind that these energies and forces already take into account
the elastic adaptation, so their correlator should be compared  to
the renormalized correlator of pinning forces rather than with the
bare one (\ref{eq:g(u)}). We see that the force becomes
discontinuous at $\theta=\pi$
 which property is more general than our particular
choice of the model. The bare smooth correlator
(\ref{eq:periodicR(u)}) becomes singular, cusp-like, at
$\varphi=0$:
\begin{equation}
R=\int^{2\pi}_0\frac{d\theta}{2\pi}V^{\ast}(\theta)V^{\ast}
(\theta+\varphi))
\sim\cos(\varphi)\Rightarrow\left(\left|\varphi\right|-\pi\right)
\cos\frac{\varphi}{2}-2\sin\frac{|\varphi|}{2}
\nonumber
\end{equation}
Apparently it contains the nonanalytic therm
$\sim-|\varphi|^3$ which originates the kink
$\sim-|\varphi|$ in the correlator of forces,
\begin{equation}
R_{\varphi\varphi}\Delta(\varphi)\sim\frac{1}{4}\left(\left(
\pi-|\varphi|\right)
\cos\frac{\varphi}{2}-2\sin\frac{|\varphi|}{2}\right)
\label{eq:locpinap}
\end{equation}

We can already observe the apparent link to the cusp anomaly in
the force-force correlator discussed in the Sec.
\ref{subsec:const_driving_f}  as a clue to the threshold pinning
force. In this way, the local pinning picture suggests a
transparent view and straightforward interpretation for one of the
most important results of the collective pinning theory obtained
with the help of the FRG.
   \begin{figure}[hbt]
   \centerline{\epsfxsize=7cm
   \epsfbox{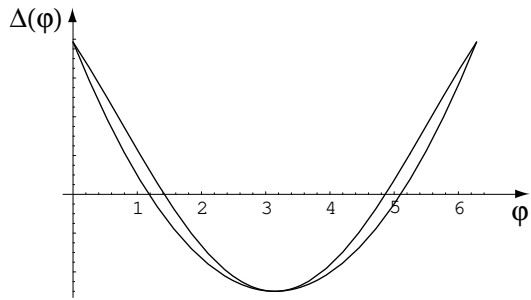}}
   \caption{Comparison of the force correlators
$\Delta(\varphi)\equiv -R_{\varphi\varphi}(\varphi)$ following
from local and collective pinning, respectively. The upper curve
shows the force correlator of the local pinning approach as given
by equation (\ref{eq:locpinap}). The lower curve shows the fixed
point function $\Delta^{\ast}(\varphi)\sim(\varphi-\pi)^2-\pi^2/3$
found from collective pinning \cite{narayan.fisher92}. Note that
$\Delta(\varphi)$ is periodic:
$\Delta(\varphi)=\Delta(\varphi+2\pi n)$, $n$ integer. The scales
on the vertical axis has been chosen differently for the two
curves to allow for a better comparison.}
\label{fig:force_correlator}
   \end{figure}

 The bare kink in the random force correlator,
 originated by the local pinning at $L\ll L_{p}$,
 provides the necessary boundary condition \cite{narayan.fisher92}
 for the kink formation,
and hence the development of the threshold field, within the
collective pinning domain $L\gg L_{p}$. Also we can get an
interpretation that the kink is formed by choosing the lowest
state every time when the retarded and advanced terms cross each
other changing their character from stable to metastable and vice
versa.

As well as in the case of the collective pinning, the cusp is
rounded, if $T\neq 0$, for small $\varphi$; here it happens at
$\Delta E=F_{\pi}\varphi<T$ when both levels $a=1,2$ become
comparably populated.

Consider now those pinning centers which are not in equilibrium;
each of them provides a point pinning force which is not random:
it is directed against the applied force $f$. These  states are
close to the degeneracy,
$\theta=\varphi(\bf{x})+\bf{Q}\bf{x}_{i}\approx\pm\pi+\delta\theta$
and their fraction $\nu=\delta\theta/2\pi$ is small (as $T/F_{\pi}$
for the linear response problem and as $v\tau_{\pi}$ for the slow
sliding). Their concentration $n_{ne}=\nu n_{i}=L_{ne}^{-d}$
determines the distance $L_{ne}$ which is large in compare to the
mean distance between impurities. Still it can be either larger or
smaller than the pinning length $L_{p}$. If $\nu$ is not too
small, such that $L_{i}\ll L_{ne}\ll L_{p}$, there are many
nonequilibrium impurities within the pinning volume, and their
point forces add to the total restoring force $f_{loc}$ of the
local pinning which we have been studying before. The collective
pinning will react to the difference $f^{\ast }=f-f_{loc}$
developing its own reaction $f_{col}(f^{\ast})$. Then for the
linear responses to both forces,
$f_{b}=\chi_{b}^{-1}\delta\theta$, we find the additivity of the
inverse susceptibilities which will be an important element of
applications (e.g. Eq. (\ref{eps-tot})). In case of very low
$\nu$, when $L_{p}\ll L_{ne}$, we face the picture of very distant
point sources of nonrandom forces. The reaction of the pinned
elastic media to the isolated point force is not quite known and
we can only guess that they will still contribute additively to
the average pinning force.

\section{Some applications to Density Waves.\label{sec:applic}}

Experimental observations on sliding charge and spin density
waves, are very rich and clean; most general effects are very
stable and observed similarly in different materials
\cite{ECRYS93,ECRYS99,ECRYS02,NATO96}. At high enough $T$, the
collective pinning picture is well confirmed in general. A typical
observation is the inverse relation between the critical field for
the onset of sliding and the real part
$\Re\varepsilon=\varepsilon^\prime$ of the dielectric
susceptibility $\varepsilon\sim\chi$: $f_{c}\propto
1/\varepsilon^\prime\sim n_{i}^{2}$ \cite{Monc-rev,Gruner88}. The
collective pinning is always affected by the $T$ dependence of the
elastic moduli (e.g. via the order parameter vanishing near the
transition temperature $T_{c}$ of the CDW/SDW formation, or via
screening of the long-range Coulomb interactions at low $T$), and
these are readily identified experimentally \cite{Maki}. The
critical dependencies of parameters $C$ and $V$ on $T_{c}-T$ are
known microscopically, interestingly different for CDWs and SDWs,
and their combination confirms in all cases the $T$ dependence of
the critical field $f_{c}$. The sliding also demonstrates the
expected saturation of the $v-f$ dependence to the linear law at high $v$.
The local pinning does not show up at these high $T$ as it should
be: the barriers $U$ cannot be higher than $T_{c}$, which is the
scale of $E_{s}$, so that the relaxation is too fast for any
observations.

The picture changes drastically (see \cite{Thorne02} for the
modern review) at $T$ low in comparison to $T_{c}$ and to the
activation energy $\Delta$ of normal carriers, the last one is
important because of the Coulomb hardening of the elasticity
$C_{\shortparallel}\sim\exp(\Delta/T)$. In addition to the usual
threshold $f_{c}=f_{t1}$, the $v-f$ curve shows a sharp upturn at
the ''second threshold field'' $f_{t2}$ (for reviews on earlier
observations see \cite{Gruner88,Maeda90,Itkis90}). Even beyond
details, the overall $v-f$ curvature becomes opposite to the high
$T$ one and hence to expectations of the collective pinning theory. The
dielectric susceptibility $\varepsilon\sim\chi$ starts to show
$\omega$ and $T$ dependencies, with $\Im\varepsilon$ showing the
maximum as a function of $\omega$ and $\Re\varepsilon$ showing a
surprising sharp peak as a function of $T$ \cite{Nad93}. These
changes can be related to the opposite $T$ dependencies for
strengths of collective and local pinnings, with the last one
playing the major role at low $T$ and not very low $\omega$. (Even
at low $T$ the collective pinning reemerges at ultra low $\omega$
which appear in measurements of the time delayed heat response
\cite{heat}.)

Below, we shall apply the picture of metastable plastic
deformations to interpret these observations. The same model will
allow to describe the two remarkable features which became
commonly observed in Charge and Spin Density Waves. There are both
the anomalous peak of $\varepsilon^\prime (T,\omega=const)$ and
the nonlinear current-voltage $I-\mathcal{E}$ (that is $v-f$)
curve with the second threshold field in the sliding regime.
Namely, the features of $\varepsilon$ result from a competition of
the local relaxation with the collective pinning affected by the
freezing of the Coulomb screening. The upper critical field in
$I-\mathcal{E}$ curves is reached when the shortest life time
configurations are accessed by the fast moving density wave.

\subsection{Nonlinear f(v).}\label{ss.I-V}

Apparently, the first critical field $\mathcal{E}_{t1}$ can be
only the threshold due to the collective pinning
$\mathcal{E}_{t1}\sim f_{c}$. It seems to be almost
time-independent which requires for high barriers available only
within the collective pinning regime. The slow creep, observed as
a "broad band noise" at finite $T$, corresponds to a distribution
of high barriers in accordance with the collective pinning
picture.

Contrarily, $\mathcal{E}_{t2}$ appears to be the high velocity
limit of the pinning force via the energy dissipated by the moving
density wave which we identify with the maximal force derived
above for the local pinning: $\mathcal{E}_{t2}\sim f_{max}$. With
increasing $v$, we shall inevitably reach the local pinning regime
with its lower barrier heights necessary to provide the condition
$v\tau\sim1$.  The qualitative curve of the Figure \ref{fig:f(v)}
shows the necessary positive curvature and the approach
to the almost vertical $I-V$ as it is observed sometimes.

\begin{figure}[hbt]
   \centerline{\epsfxsize=8cm
   \epsfbox{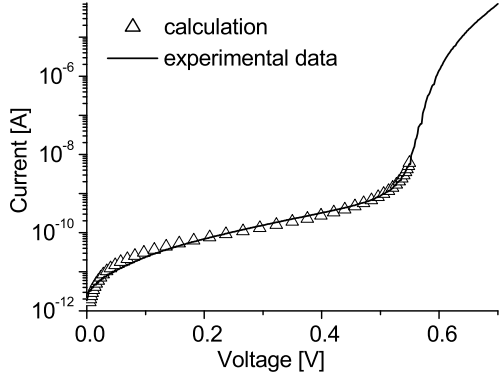}}
   \caption{The single model fits (after Miyano and Ogawa)
for the experimental $v(f)$ (current-voltage) curve through three
different regimes, 1a,1b,3a of the Sec.
\protect\ref{ss:stationary}. The middle part of the semi-log plot
clearly demonstrates the $f\sim\ln v$ regime. The left part, if
plotted in the normal scale, shows the linear $f\sim v$ law. These
two regimes indicate the presence of bistable pinning centers of
either intermediate or high strengths. The sharp upturn at higher
$f$ discriminates in favor of the unrestricted bistability
($f\approx f_{max}-const/v$). The curvature sign changing at low
$v$ is an artifact of the logarithmic rescaling of the current
axis. The decreasing growth rate at highest $v$ agrees with
expected effects of dislocations, Sec. \ref{sec:loops}.}
\label{fig:fit}
   \end{figure}

In recent experiments \cite{Thorne02,Lemay99}, two distinct regimes have
been established: the linear one $I\sim\mathcal{E}$ at small
$\mathcal{E}$ followed by the exponential growth
$I\sim\exp[cnst\,\mathcal{E}]$. Next, very recent experiments
\cite{Miyano} have shown that $\mathcal{E}_{t2}$ is a steep
crossover, rather than a kink as it was supposed for a long time,
and it is closer to our picture. Now it is possible to fit
quantitatively, within the same set of parameters, the $f-v$
dependence over several orders of magnitude of $v$ encompassing
three different regimes of the theory described in the Sec.
\ref{sec:kin} as 1a, 1b and 2, that is $f\sim v$, $\ln v$ and
$f-f_{max}\sim-1/v$. The Figure \ref{fig:fit} shows such a fit
\cite{Miyano} done by the general formula (\ref{f-the-F}) with
functions $U(\theta)$ and $F(\theta)$ specified for the
short-range model (\#2 of the Section \ref{ss:models}). Even the
slowing down at high currents corresponds qualitatively to high
velocity effects related to generation of dislocations, Sec.
\ref{sec:loops}.

\subsection{Low $T$, low $\omega$ susceptibility peak.
\label{ss:epsilon}}

\begin{figure}[hbt]
\includegraphics*[width=8cm]{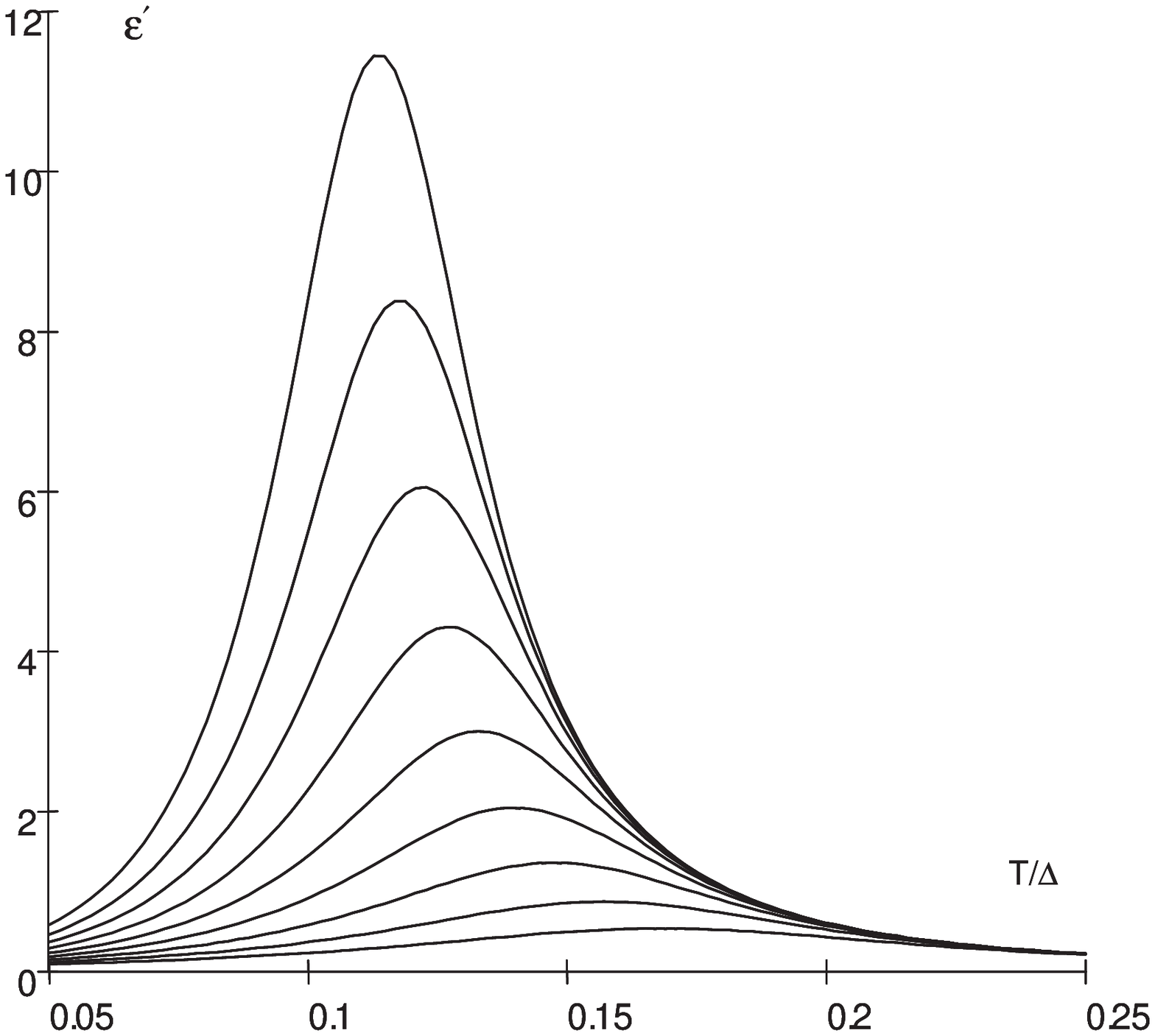}
\caption{The dielectric susceptibility of density waves from
interference of the local and the collective pinnings. The plots
show the $T$ dependencies of $\protect\epsilon^{\prime}(\omega,T)$
at various $\protect\omega$. Calculations have been done for the
formula (\protect\ref{Pexp}) with $A/B=9$ and  $U_0=0.5$ (both
$U_0$ and $T$ are in units of $\Delta$). While units of
$\varepsilon^\prime$ and $\omega$ arbitrary, their changes
correspond to the experimental plots below: at each step, $\omega$
was rescaled by one order of magnitude resulting in the overall
change of $\varepsilon^\prime$ by one order.}
\label{fig:re-eps-the}
\end{figure}

\begin{figure}[hbt]
\includegraphics*[width=10cm]{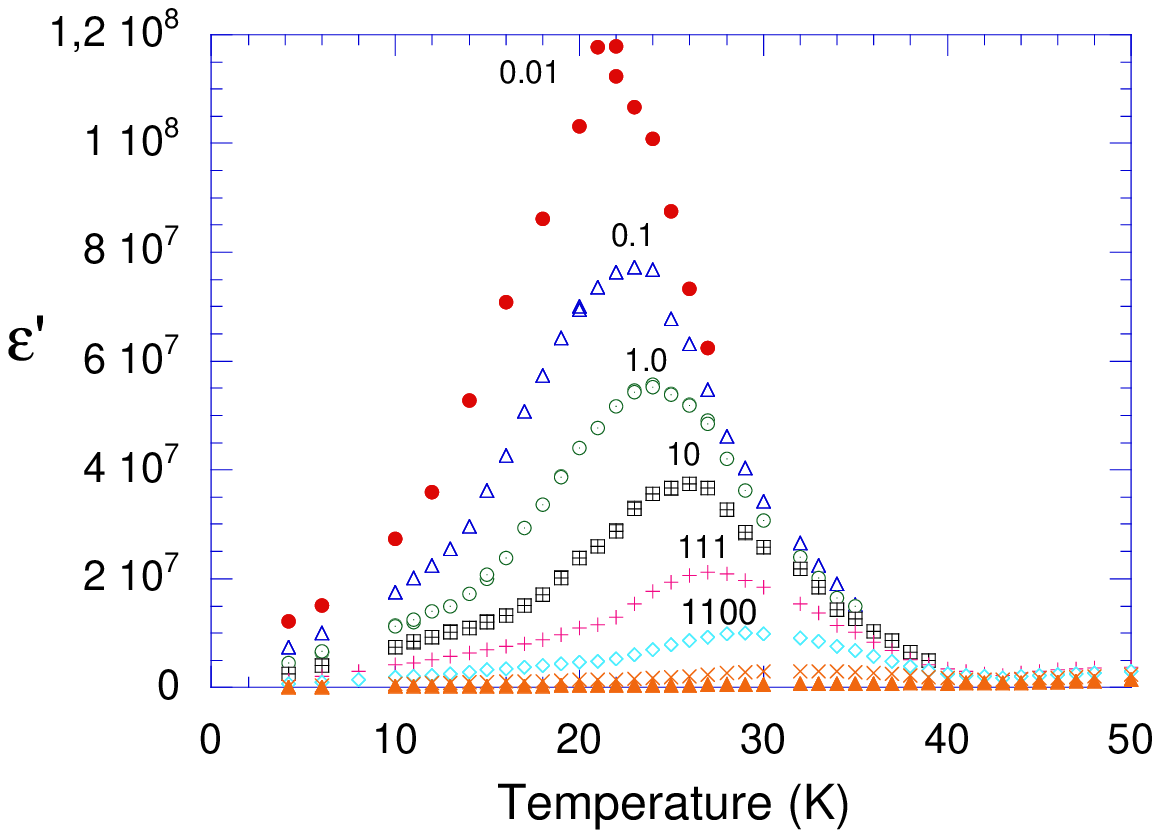}
\caption{Experimental data by Nad \textit{et al} \cite{Nad93}.
Labels at the top of each curve show the frequency, in
Hz.}\label{fig:re-eps-exp}
\end{figure}

Sliding density waves are principally characterized by their giant
dielectric susceptibility,
$\varepsilon^\prime=\Re\varepsilon\sim10^{6}-10^{9}$,
corresponding to the low threshold field $\mathcal{E}_{t}\sim
f_{c}$. Remarkably, a sharp maximum of
$\varepsilon^\prime(\omega,T)$ as a function of $T$ has been
observed in a wide range of density waves  materials at low $T$
and for very low frequencies $\omega$ \cite{Nad93,Lasj94}. With
$\omega$ decreasing from,typically, $10^{5}Hz$ to $10^{-2}Hz$, the
maximum height is growing while its position $T_{max}(\omega)$ is
shifting towards low $T$ as shown on the Figure
\ref{fig:re-eps-the}. Importantly, the uprising parts $T>T_{max}$
of all plots for $\varepsilon^\prime(\omega=const,T)$ follow the
same master curve $\varepsilon^\prime(T)$ and differ only by the
cutoff $T_{max}(\omega)$ below which $\varepsilon^\prime$ drops
sharply, see the Figure \ref{fig:re-eps-exp}.

While some features of the $T$ dependence (the rising high $T$
slope) are quite specific to CDWs and SDWs, we suggest this
example as showing simultaneously a combination of several
important ingredients: sensitivity of the collective pinning to
elastic parameters; separation of time scales between the two
types of pinning, as well as their interference in observations.

The electric polarization, being proportional to the average phase
displacement $\delta\bar{\varphi}$, creates the restoring force
$f_{pin}$, which may decay in time. The external electric field
$\mathcal{E}$ opposes $f_{pin}$ which has two well separated
contributions $f_{pin}=f_{col}+f_{loc}$. The inverse
susceptibility can be defined as:
\begin{equation}
\varepsilon^{-1}\sim-\frac{\partial
f_{pin}}{\partial\bar{\varphi}}
\Rightarrow\varepsilon_{col}^{-1}+\varepsilon_{loc}^{-1}
\label{eps-tot}
\end{equation}

Keeping track only of the major dependencies on $\omega,T,n_{i}$,
we can write with the help of (\ref{F-omega})
\begin{equation}
\varepsilon^{-1}=const\,n_{i}^{2}e^{-\Delta/T}+\frac{const\,\,n_{i}}
{1+(i\omega\tau_{0})^{-1}\exp(-U_{\pi}/T)} \label{1/eps}
\end{equation}
where the first and the second terms come from the collective and
the local pinnings correspondingly. The  activation law
$e^{-\Delta/T}$ in the first term comes from the effect of the
Coulomb hardening of the longitudinal elastic modulus which is
special to quasi one-dimensional density waves.

For the model with the exponential distribution of barriers of the
Sec. \ref{sec:ensemble}, we obtain

\begin{equation}
\varepsilon^{-1}=A\exp(-\frac{\Delta}{T}) + B\omega^{T/U_{0}}
\,,\;
  A\sim n_{i}^{2} \,,\,
  B\sim n_{i}
  \label{Pexp}
\end{equation}
where the second coefficient $B$ is a complex weak function of
$T,\omega$. In both cases, the function
$\varepsilon^\prime(T,\omega=const)$ is not monotonic, it
demonstrates a peak at $T_{max}(\omega)$ defined by the equation
\begin{equation}
\omega\sim{\frac{1}{\tau_{0}}}
\exp(-{\frac{{E_{a}U_{0}}}{T_{max}^{2}}}+
{\frac{V_{0}}{T_{max}}}\ln n_{i}) \label{T-max}
\end{equation}
Numerical calculations for (\ref{Pexp}), shown on the Figure
\ref{fig:re-eps-the}, are in agreement with the results of
\cite{Nad93} shown on the Figure \ref{fig:re-eps-exp}. Concerning
the frequency dependence, the function
$\Im\varepsilon_{DW}^{-1}(T=const,\omega)$ in (\ref{1/eps}) is
monotonous, but the measured function
$\Im\varepsilon_{DW}(T=const,\omega)$ shows a maximum which is
also in agreement with experiments.

The interpretation of these results \cite{Larkin95} is that
different types of pinning compete to contribute to
$\varepsilon^{-1}$, so that the lowest $\varepsilon$ dominates and
near $T_{max}$ the pinning force is minimal. Namely, at the higher
 $T$ slope the $T$ dependence arises from the dispersionless
(very high barriers for collective metastable states) collective
pinning affected by statically screened Coulomb interactions. At
the lower $T$ slope, the dispersion of $\varepsilon$ comes from
the relaxation of local metastable states. With increasing $T$ at
given $\omega$, the local states approach the thermal equilibrium;
then $f_{loc}\rightarrow0$, hence $\varepsilon (T,\omega)$ grows
with increasing $T$ until the collective pinning force becomes
dominating. Recall another, more phenomenological interpretation
\cite{Nad93} which suggested approaching a kind of a dipole glass
transition\footnote{ Indeed, there is an interesting, while
indirect link to physics of two-level systems in conventional
polar glasses \cite{anderson.halperin.varma72}. Thus,
$\varepsilon_{loc}^{-1}$ may be interpreted as
$\varepsilon_{loc}^{-1}=\sigma^{\ast}$, where
$\sigma^{\ast}=-i\omega\varepsilon^{\ast}$ are the effective
complex ''conductivity'' $\sigma^{\ast}$ and the ''dielectric
susceptibility'' $\varepsilon^{\ast}$ per one effective two-level
''dipole'' with ''polarization'' $\pm f_{\pi}$ under the ''random
field'' $\theta-\pi$ and the ''external field'' $\delta\theta$.}.

We conclude that the crossover picture of the local and the
collective types of pinning describes altogether the
susceptibility anomaly, long time relaxation and nonlinear $v-f$
characteristics. The peak of $\varepsilon(T)$ results from the
competition of the local metastable plasticity with the collective
pinning affected by freezing out of the Coulomb screening. In $f-v
$ curves, the upper threshold field $f_{2}$ is reached when the
metastable plastic deformations with shortest life time are
accessed by the fast moving density wave.

\section{Conclusions  \label{sec:conclusions-II}}

In this article we have suggested an overview of the present
status of our understanding of pinning phenomena in various
systems, stressing the unifying aspects. Both languages of the
collective and the local pinning exploit intensively the concept
of metastable states.

The picture of collective pinning discussed in the first part of
this review is to some extend worked out. The theory has been
successfully applied already for the explanation of equilibrium
phases, depinning and creep phenomena in vortex lattices, charge
density waves and  magnetic domain walls. The decisive step here
were the understanding of thermodynamic and dynamic scaling
behavior, as well as development of renormalization group methods
which are capable to consider the effects of metastability
emerging on large length scales even from weak pinning centers.
The domain of application of this theory are large length and time
scales, the critical field vicinity and the creep below it. The
theoretical description of plasticity in disordered systems is
still incomplete.

The picture of the local pinning, within its domain of low
temperatures and not too low frequencies or velocities, can be
effectively used to explain experimental data: qualitatively and
even quantitatively. The advantages come from the explicit
treatment of metastable states, their creation and relaxation,
their relation to plasticity and topological defects. It provides
also a clue to the quantum creep showing that the tunneling
repulsion of crossing branches destroys the pinning.
\bigskip

{\bf Acknowledgments}

\bigskip

 T.N. acknowledges the hospitality of the Laboratoire de Physique
Th\'{e}orique et des Mod\`{e}les Statistiques  of the
Universit\'{e} Paris-Sud and of the LPMMH group of the  Ecole
Superieure de Physique et de Chimie Industrielles de  Paris were
most of this work was done. He is also very grateful for the
support of the Volkswagen Foundation which made this stay possible
and to S.Malinin for a critical reading of the manuscript. S.B
acknowledges the hospitality of the ISSP, Tokyo University and
YITP, Kyoto University, the support from the INTAS grant 2212, and
the assistance of N.Kirova and S.Teber.

\newpage
\setlongtables
\begin{longtable}[c]{p{9cm}cl}
\caption{\bf Frequently appearing quantities}\\
\hline
\rule[-3mm]{0mm}{8mm}
{\bf Quantity} & {\bf Symbol} & {\bf see eq.,fig.,sec.}\\ \hline
\\
field coupling to $\bnabla u$ & $\bf A$
& (\ref{eq:A-perturbation}) \\
cutoff & $a_0$ & 
\\
Burgers vector & $\bf b$ & Sec.\ref{sec:plastic}
\\
stiffness constant &  $C=(C_{\parallel}C_{\perp}^2)^{1/3}$
& (\ref{eq:Hamiltonian}), (\ref{eq:elastic.dispersion})\\
elastic moduli & $C_{\parallel},\; C_{\perp}$ &
(\ref{eq:elastic.dispersion})\\
curvature & $\tilde{\mathcal{C}}$ & (\ref{eq:ansatz_u})\\
dimension of the elastic object & $D$ & (\ref{eq:Hamiltonian})\\
space dimension &  $d$ & (\ref{eq:random_potential}) \\
energy barrier on scale $L$ &  $E_B(L)$ & (\ref{eq:total_E_B})\\
energy of the dislocation loop of the radius $R$ & $E_{DL}(R)$ &
(\ref{LLR})
\\
soliton energy & $E_s$ &  Figure\ref{fig:terms2},
Sec.\ref{sss:srm},
\\
energy difference between low lying states &  $\Delta E$ & (\ref{eq:Delta-E}) \\
frequency dependent barrier &
$E_{\omega}$ & (\ref{eq:max_height_barriers})\\
energy branches (1: stable, 2: metastable, +: accending, -:
descending) & $E_{1,2}$, $E_{\pm}$ &  (\ref{dH=0},\ref{d2H>0})
\\
free energy & $\mathcal{F}(L,u)$ & (\ref{eq:KPZ})\\
disorder dependent free energy & $\mathcal{F}_{R}$ & (\ref{F-fluctuations1})
\\
forces from the single impurity $F_a$, $a=+,-,1,2$ &
$F_{\pm}(\theta)=E_{\pm}^{\prime}$ &  (\ref{F})
\\
driving force density &  $f$  & (\ref{eq:L_f}), (\ref{eq:motion})\\
pinning threshold & $f_c\approx f_p=C l/L_p^2$  & (\ref{f_c}),
(\ref{eq:threshold})
\\
pinning forces: total, collective, local and its maximum &
$f_{pin}$, $f_{col}$, $f_{loc}$, $f_{max}$ &
Fig.\ref{fig:f(v)}, Sec.\ref{ss:epsilon}\\
temperature dependent force &
$f_T$ & (\ref{eq:temp_dependent_force})\\
frequency dependent characteristic force & $f_{\omega}$ &
(\ref{eq:frequency_force})\\
free enthalpy & $G$ & (\ref{Eq:enthalpy}) \\
random pinning force &  $g({\bf x},u)$ & (\ref{eq:motion})\\
variational energy & $H=W(\psi)+V(\psi-\theta)$ & (\ref{H=W+V})
\\
Hamiltonian & $\cal H$ & (\ref{eq:Hamiltonian})\\
tunneling rate & $I_{\omega}$ & (\ref{L-t})
\\
dynamic parameters & $K_{\omega}$ & (\ref{K})
\\
length scale (variable)  & $L$ & (\ref{eq:w(L)}) \\
system size  & $L_0$  & 
\\
scale on which tilted potential  vanishes & $L_f$ & (\ref{eq:L_f})\\
Larkin length & $L_p$ & (\ref{eq:L_p})\\
frequency dependent scale & $L_{\omega}$ & (\ref{eq:max_height_barriers}) \\
frequency dependent diffusion length & $\tilde L_{\omega}$& (\ref{eq:diffusion  length}) \\

correlation length of the random potential & $l$ & (\ref{eq:w(L)}) \\
soliton length & $l_s$ & Figure\ref{fig:soliton},
Sec.\ref{sss:srm}
\\
concentration of impurities & $n_{imp},n_i$ & (\ref{eq:n_imp})
\\
occupation numbers of terms $\pm$, there difference and its
equilibrium value & $n_{\pm}$, $n=n_{+}-n_{-}$, $n_{eq}$ &
(\ref{n+-},\ref{df/dt})
\\
probability distribution of excited states & $P(\Delta E,L)$ &
(\ref{eq:Delta-E})\\
distribution functions for $U$ and $\tau$ & $P_{U}$, $P_{\tau}$ &
(\ref{P_U})
 \\
wave vector of charge density wave & $\bf Q$ & (\ref{eq:rho.cdw})\\
disorder average & $\langle...\rangle_R$ & (\ref{eq:disav}) \\
correlator of the random potential & $R$ & (\ref{eq:V-correlations}),
(\ref{eq:periodicR(u)})\\
correlation function of the random potential & $R(u)$ &
(\ref{eq:V-correlations})
 \\
running effective $t/\tau$ & $s(\theta(t))$ & (\ref{s=})
\\
thermal average & $\langle...\rangle_T$ & (\ref{eq:w_R})\\
temperature & $T$ &  (\ref{eq:w_T})\\
energy scale due to pinning & $T_p=C l^2L_p^{D-2}$ & (\ref{F-fluctuations1}) \\
position vector (D-dimensional)& $\bf x$ & (\ref{eq:Hamiltonian}) \\
barrier branch $E_{3}$ and metastable sate decay activation $U$ &
$U(\theta )=E_{3}-E_{2}$, $U_{\pi}=U(\pi)$ &
(\ref{d2H>0},\ref{Ea})
\\
displacement field (N-dimensional) &  $u$  &  (\ref{eq:Hamiltonian})\\
pinning potential, its magnitude and threshold values &
$V(\varphi)$; $V$, $V_{1}$,$V_{2}$ & (\ref{srm}),\ref{app:ms}
\\
random potential & $V_R({\bf x},u)$ & (\ref{eq:Hamiltonian}),
(\ref{eq:random_potential}) \\
velocity & $v$ & (\ref{eq:v_creep})\\
phase velocity  & $v=-d\bar{\varphi}/dt$ & Tab.\ref{tab:phase},
App.\ref{app:ss:stationary}
\\
impurity potential & $v_R(\mathbf{x},\mathbf{z})$ &
(\ref{eq:random_potential}), (\ref{eq:n_imp})\\
strength of individual pinning center $i$ & $V_i$ & (\ref{eq:n_imp}) \\
deformation energy & $W(\psi)$ & (\ref{H=W+V})
\\
roughness &  $w_R(L)$ & (\ref{eq:w(L)})  \\
dynamical critical exponent & $z$ & (\ref{eq:x,t,u})\\
exponents at the $T=0$ depinning transition &
$\tilde{\alpha},\tilde{\beta},\tilde{\zeta},\tilde{z},\tilde{\nu}$
& (\ref{eq:v})-(\ref{exponent.relations})\\
mobility & $\gamma$ & (\ref{eq:eq.motion})\\
correlator of random pinning forces &  $\Delta(u)$ & (\ref{g_correlator})\\
$\delta$-function of width $l$ & $\delta_l({\bf x})$ &
(\ref{eq:V-correlations1}) \\
quantum splitting of branches & $\delta_{q}=\hbar/\tau_{q}$ & \ref{E-q}\\
roughness exponent & $\zeta$  & (\ref{eq:x,t,u})\\
thermal noise & $\eta({\bf x},t)$  & (\ref{eq:eq.motion})\\
pinning phase mismatch and its special values &
$\theta=\theta_{i}(t)= \bf\textbf{}{Qx}_{i}-\bar{\varphi}(t)$;
$\theta_{e}$, $\tilde{\theta}_{e}$, $\theta_{m}$,
$\theta_{e}^{\ast}$ & (\ref{psi-the}),
Figs.\ref{fig:few-terms},\ref{fig:terms2}
\\
barrier exponent & $\mu=\chi/(2-\zeta)$ & (\ref{eq:tilde_E_B}) \\
size distribution of excited states & $\nu(L)$ & (\ref{eq:specific.heat})\\
correlation length & $\xi$ & (\ref{eq:xi})\\
relaxation time & $\tau=\tau_{0}\exp[U/T]$ & (\ref{df/dt})
\\
scaling function&  $\Phi(y)$ &  (\ref{eq:v(f,T)})
\\
phase field of the charge density wave &
$\varphi({\bf x},t)$  & (\ref{eq:u-phi-relation}) \\
mean value of phase & $\bar{\varphi}(t)$ &  (\ref{psi-the})
\\
soliton profile & $\varphi_{s}(x-X)$ &
Figs.\ref{fig:soliton},\ref{fig:bisoliton}
\\
exponent describing the free energy fluctuations & $\chi$ &
(\ref{F-fluctuations1})
\\
response function and the susceptibility of CDWs &
$\chi\sim\varepsilon$ & (\ref{eps-tot})
\\
local phase mismatch and its special values &
$\psi=\varphi(\bf{x}_{i},t)-\bar{\varphi}(t)$, $\psi_{e}$,
$\psi_{e}^{\ast}$ & (\ref{psi-the},\ref{end})
\\
frequency &  $\omega$  & (\ref{eq:ac_driving_force})\\
pinning frequency &  $\omega_p=\gamma C/L_p^2=C f_p/l=v_p/l$ &
(\ref{eq:max_height_barriers}) \\ \\[-0.5ex]
\hline
\end{longtable}

\newpage
\centerline{APPENDICES.}
\appendix

\section{Free energy fluctuations in $D=1$ dimensions}\label{sec:app1}

An illustrative example is given by a linear $D=1$-dimensional
object such as a
magnetic flux line with the boundary
conditions $u(L)\equiv u$ and $u(0)= 0$. Changing $u$ enforces the
object to see another disorder environment. Using the transfer
matrix technique, it can be shown that $F(L,u)$ obeys the equation
\cite{Huse+85}
\begin{equation}
\frac{\partial \mathcal{F}_R}{\partial
L}=\frac{T}{2C}\frac{\partial^2\mathcal{F}_R}{\partial u^2}-
\frac{1}{C}\left(\frac{\partial \mathcal{F}_R}{\partial
u}\right)^2+V_R(L,u). \label{eq:KPZ}
\end{equation}
As a side remark we mention that if we read $L$, $u$ and
$\mathcal{F}_R$ as time, space and height coordinates,
respectively, eq. (\ref{eq:KPZ}) becomes the
Kardar-Parisi-Zhang-equation, which describes the height profile
of a growing surface under the random influx $V_R(L,u)$ of
particles \cite{KPZ}. The correlations of the restricted free
energy show the following scaling behavior:
\begin{equation}
\left\langle\big(
\mathcal{F}_R(L,u)-\mathcal{F}_R(L,u')\big)^2\right\rangle_R^{1/2}=
T_p\left(
\frac{L}{L_p}\right)^{\chi}\Phi\left(\frac{u-u'}{w_R(L)}\right),
\label{eq:F-fluctuations}
\end{equation}
For small values of the argument $y=({u-u'})/{w_R(L)}$ of $\Phi$
the difference of the free energies should not depend on $L$ since
the configurations dominating the free energy will be the same for
most parts of the elastic object (apart from $x$ close to $L$).
This gives $\Phi(y)\sim y^{\chi/\zeta}$. For large arguments $y$
the elastic stiffness dominates over the disorder and
$\left\langle\big(\mathcal{F}_R(L,u)-\mathcal{F}_R(L,u')\big)^2\right\rangle_R^{1/2}
\sim\frac{|u^{2}-{u^{\prime}}^{2}|}{L}$
as in pure systems.
A numerical solution of (\ref{eq:KPZ}) shows that for large $L$ and
intermediate values of $y$ the free energy $\mathcal{F}_R(L,u)$
forms a rugged landscape as a function of $u$ with typical valleys
of width $w(L)$ separated by hills of height $T_p(L/L_p)^{\chi}$
\cite{Kardar8787}. The general picture of a rugged energy
landscape as concluded from eq. (\ref{eq:F-fluctuations}) is
believed to hold also for higher dimensional elastic objects.

\section{Strong pinning in $D=1$--dimensional CDWs}
\label{sec:app2}

To give a specific example for the case of strong pinning, we
consider a lattice model for a charge density wave with  the
Hamiltonian given by
\begin{equation}
H=\sum_{<i,j>}\frac{1}{2}C_{i,j}(u_i-u_j)^2- \sum_i
V_i\cos\left(2\pi (u_i-\kappa_i))\right) \label{eq:lattice.cdw1}
\end{equation}
where $2\pi\kappa_i=-{\bf Qx}_i$ and ${\bf x}_i$ is a random
impurity positions at which $V_i\neq0$. If we assume for
simplicity $V_i\equiv V$ for all $i$ and consider the limit
$V\rightarrow \infty$, then $u_i=\kappa_i+n_i$ with  $n_i$ integer
and the Hamiltonian can be rewritten as
\begin{equation}
H=\sum_{<i,j>}\frac{1}{2}C_{i,j}(n_i-n_j+\kappa_i-\kappa_j)^2
\label{eq:lattice.cdw2}
\end{equation}
The minimization of this Hamiltonian leads to a set of integers
$\{n_{i,0}\}$ from which a well defined result for ground state
and hence the roughness $w_R$ follows.  The ground state consists
of regions of constant $n_i$ separated by oriented domain walls at
which $n_i$ changes by $\pm 1$. A very simple situation exists in
$d=1$ dimensions, where $<i,j>=i,i+1$ and the ground state follows
trivially as $n_{i+1}=n_i+[\kappa_{i+1}-\kappa_i]$. Here $[...]$
denotes the Gauss bracket which replaces its argument by the
closest integer. Thus the $u_i$ undergo a random walk and hence
$\zeta=1/2$. For a more detailed discussion of the one-dimensional
case see e.g. \cite{Glatz+01}.

The specific Transfer Matrix technics (Appendix \ref{sec:app1}) in
the dimension $D=1$  allows for a more detailed description of
interference between the pinning and the thermal motion
\cite{brazov79}. Thus for temperatures high in compare to the
characteristic elastic energy $T\gg T^{\ast}\sim Cn_{imp}$, and
arbitrary with respect to $V$,  the heat capacitance $c(T)$ is
\[
c\approx n_{imp}/2~,~T\ll V\ \ ;\ \ c\sim n_{imp}(V/T)^{2}~,~T\gg V\
\]
In the same regime, the correlations of the order parameter
$\cos(2\pi u_i)$ decay exponentially with the correlation length
$\xi(T)$ such that $\xi ^{-1}=\xi_{T}^{-1}+\xi_{R}^{-1}$. Here
$\xi_{T}\sim C/T$ is the correlation length of a pure system while
the randomness contribution to $\xi^{-1}$ is
\[
\xi_{R}\sim n_{imp}^{-1}\left(I_{0}(V/T)/I_{1}(V/T)\right)^{2}
\]
where $I_{m}$ are the modified Bessel functions.

For low temperatures with a constraint to the strong pinning regime $T\ll
T^{\ast}\ll V$ the correlation function of displacements behaves as a kind of
the Mott law
\[
\left\langle \left\langle \left\vert u_{i}-u_{j}\right\vert
\right\rangle _{T}\right\rangle _{R}\sim\left\vert i-j\right\vert
n_{imp}\left( \frac{T^{\ast}}{T}\right)
^{1/4}\exp\left(-cnst\left(\frac{T^{\ast}}{T}\right)^{1/4}\right)
\]
This nontrivial $T$ dependence appears because thermal jumps take
place primarily within segments of an optimal spacing $\left\vert
i-j\right\vert_{opt}\sim n_{imp}^{-1}(T^{\ast}/T)^{1/2}$ which is
much larger than the typical one $n_{imp}^{-1}$.

\section{Details on metastable branches.\label{app:ms}}

Here we give details to the results of Sec. \ref{sec:ms} on the
energy branches and their special points in the language of the
CDWs and its phase.

Consider first the termination points defined in (\ref{end}); all
quantities at this point will be labelled by the index $e$.
Expanding at the vicinity of the end point
\begin{equation}
\psi=\psi_{e}+\delta\psi;\quad\theta=\theta_{e}+\delta\theta,\quad\delta
\theta<0  \label{dpsi,dtheta}
\end{equation}
we find from (\ref{dH=0}), (\ref{end}) and (\ref{Ea}) the
solutions
\begin{equation}
\delta\psi_{3,2}=\pm\left({2{
\frac{V_{e}^{\prime\prime}}{H_{e}^{\prime\prime}}}|\delta\theta|}\right)^{1/2}
\,,\; U=V_{e}|\delta\theta|^{3/2} \,,\;V_{e}=-{\frac{2}{3} }
{\frac{(2V_{e}^{\prime\prime})^{3/2}}{(H_{e}^{\prime\prime})^{1/2}}
\,,\;}
H_{e}^{\prime\prime}=V_{e}^{\prime\prime}+W_{e}^{\prime\prime}
\label{U-e}
\end{equation}
 As a function of $V$, the coefficient $V_{e}$ is singular at $V=V_{1}$
 when the end point emerges and at $V=V_{2}$ when it annihilates with the next one,
$\theta\rightleftharpoons \tilde{\theta}_{e}$:

1. Consider the emergence of metastable branches when the points
 $\theta_{e},-\theta_{e}+2\pi$ split from the point $\pi$. We find

\begin{equation}
V=V_{1}+\delta V\geq V_{1}\ :\;\theta_{e}-\pi\sim(\delta V)^{3/2}
\,,\;U\sim(\delta
V)^{-1/4}(-\delta\theta)^{3/2}\,,\;F_{e}\sim(\delta V)^{1/2}\qquad
\label{dV1}
\end{equation}

2. Consider the crossover to the unrestricted bistability:
$V\rightarrow V_{2}$ when the two sets of end points $\theta _{e},
\tilde{\theta}_{e}$ join together and with the point $\theta_{m}$
of the minimal barrier.  At $V=V_{2}+\delta V$ $<V_{2}$ the
degeneracy is lifted and the branch crossing point
$(\theta_{m},\psi_{m})$ splits into two end points

\begin{equation}
\theta_{e},\,\tilde{\theta}_{e}=\theta_{m}\mp\delta\theta_{e}
\;\;\delta\theta_{e}\sim\sqrt{-\delta V}\,;\;U\sim(\delta
V)^{3/4}(-\delta\theta)^{3/2} \label{the-m}
\end{equation}

At $V>V_{2}$, $U(\theta)$ passes through the minimum
$U_{min}\sim(\delta V)^{3/2}$ at $\theta=\theta_{m}$. For both
signs of $\delta V$ we can write an interpolation
\begin{equation}
V\lessgtr V_{2}:\;U=\left( B_{1}\delta\theta^{2}+B_{2}\right)
^{3/2} \,\,;\;B_{1}\sim\delta V^{1/2}\,,\;\ B_{2}\sim\delta V
\nonumber
\end{equation}

3. Consider the limit of very strong impurities which allows for
an explicit treatment. The equation
$W^{\prime}(\psi)+V\sin(\psi-\theta)=0$ at $V\gg\max W^{\prime}$
has the following solutions:
\begin{eqnarray*}
E_{+} &:&\psi \approx \theta -V^{-1}W^{\prime }(\theta
)\,,\;E_{+}\approx
W(\theta )-(2V)^{-1}W^{\prime 2}(\theta ) \\
E_{-} &:&\psi \approx \theta -2\pi -V^{-1}W^{\prime }(\theta -2\pi
)\,,\;E_{-}\approx W(\theta -2\pi )-(2V)^{-1}W^{\prime 2}(\theta -2\pi ) \\
E_{3} &:&\psi \approx \theta -\pi +V^{-1}W^{\prime }(\theta -\pi
)\,,\;E_{3}\approx 2V+W(\theta -\pi )-V^{-1}W^{\prime 2}(\theta ) \\
U &=&E_{3}-E_{+}\approx 2V+W(\theta -\pi )-W(\theta
)\,,\;W^{\prime }(\theta _{m}-\pi )=W(\theta _{m})
\end{eqnarray*}

4. Fortunately, for a point impurity we can order the branches and
simplify the energy a priori even at arbitrary $V$. In what follows, $\pi
<\theta <2\pi$, while $W,W^{\prime }$ are functions of $\psi$ at
$-2\pi <\psi <2\pi$. For each term, we determine its own function
$\psi =\psi _{a}(\theta )$ with $a=\{-,+,3\}\equiv \{1,2,3\}$
(within the selected semiperiod of $\theta $).

\begin{eqnarray*}
\psi & = & \psi _{+}:\theta =\psi +\arcsin \frac{W^{\prime
}}{V}\,,
\;E_{+}=W-V\sqrt{1-\frac{W^{\prime 2}}{V^{2}}}\,+V \\
\psi & = &\psi _{-}:\theta =\psi +\arcsin \frac{W^{\prime
}}{V}+2\pi
\,,\;E_{-}=W-V\sqrt{1-\frac{W^{\prime 2}}{V^{2}}+V} \\
\psi & = &\psi _{3}:\theta =\pi +\psi -\arcsin \frac{W^{\prime
}}{V}
\,,\;E_{3}=W+V\sqrt{1-\frac{W^{\prime 2}}{V^{2}}}+V \\
U(\theta )& = & E_{b}(\psi _{b}(\theta ))-E_{+}(\psi _{+}(\theta
))\,
\end{eqnarray*}

These expressions were the bases for our plots on Figures
\ref{fig:solution},\ref{fig:psi-theta}.

\medskip

5. Consider in more details the overshooting branches which appear
due to special long range effects of dislocations  (recall the end
of the Section \ref{ss:models} and the Section \ref{sss:LLR}).
This is the regime of small $\delta \psi \approx \psi -2\pi $ and
$\delta \theta =\theta -2\pi $. Here, $V$ is close to its minimum
$\delta V \approx b/2(\delta\psi-\delta\theta )^{2}$,
$b=V^{\prime\prime}(0)$; $W$ is close to its maximum
$W_{2\pi}=2E_{s}$, but the expansion is not analytical. We shall
write it, according to (\ref{LLR}) taken for $d=3$, as $\delta
W=-4/3a(-\delta \psi )^{3/2}$ where $a\sim CR$ for the dislocation
loop of the radius $R$. The minimization of $H$ over $\psi$ gives
$\delta\theta=\delta\psi+2a/b(-\delta \psi )^{1/2}=0$.

At $\delta\theta<0$, there is one solution
$\delta\psi=-\left(a/b+((a/b)^{2}-\delta\theta)^{1/2}\right)^{2}$;
it gives the branch $E_+$ approaching the end of the period,
$\theta\rightarrow 2\pi - 0$ with some deficiency $E_+(2\pi)<0$:
$\delta E_+=-(8/3)a^{4}/b^{3}$ corresponding to the retardation
$\delta \psi=-(2a/b)^{2}$.

At $\delta\theta>0$, there are two solutions
$\delta\psi=-\left(a/b\pm((a/b)^{2}-\delta\theta)^{1/2}\right)^{2}$.
Here the sign $-$ corresponds to the barrier branch $E_3^*$, the
sign $+$ corresponds to the overshooting part $E_+^*$ of the
branch $E_+$.

Entering the next circle $\delta\theta>0$, the energy $E_+$ keeps
increasing, passing through the energy $2E_s$ at
$\delta\psi=-\left(3a/2b\right)^{2}\,,\,\delta\theta=3/4(a/b)^{2}$.
Further on it crosses with the branch $\tilde E_+$ to become
metastable. Since then, the difference $E_3^*-E_+^*=U$ gives the
relaxation barrier. Finally, the two solutions $E_+$ and $E_3$
collapse at the termination point
\begin{equation}
\delta \theta _{e}=(a/b)^{2}\,,\;\, \delta
\psi_{e}=-(a/b)^{2},\;\delta H_{e}=20a^{4}/3b^{3} \nonumber
\end{equation}

 The above  results give rise to the picture of the Figure \ref{fig:terms3}
 and related conclusions.

\section{Details on the kinetic equation.\label{app:kin}}
The kinetic equation is derived from the balance law for
occupation numbers of branches $+,-$, see Eq. (\ref{n+-}):
\begin{equation}
{\frac{dn_{+}}{dt}}=W_{\mp}n_{+}+W_{\pm}n_{-}\ ,\quad
{\frac{dn_{-}}{dt}}= W_{\pm}n_{-}+W_{\pm}n_{+}  \label{df}
\end{equation}
Here $W_{\pm},W_{\pm}$ are transition rates between the branches
and the full time derivative is
\begin{equation}
{\frac{d}{dt}}= {\frac{\partial}{\partial t}}+
\dot{\theta}{\frac{\partial}{\partial\theta}}; \quad\dot{\theta}=
{\frac{d\theta(t)}{dt}=v}  \label{dt}
\end{equation}

\begin{equation}
W_{\mp}/W_{\pm}=\exp{(\Delta E/T)\,},\quad W_{\pm},\
W_{\mp}\sim\exp{(-E_{b}/T)},\quad \Delta E=E_{+}-E_{-}  \label{W}
\end{equation}
The relaxation rate is
\begin{equation}
{\frac{1}{\tau(\theta)}}= W_{\pm}+W_{\mp}\sim{\frac{1}{\tau_{0}}}
\cosh{\frac{\Delta E}{2T}} \exp\left(
\frac{E_{+}+E_{-}-2E_{3}}{2T}\right) \label{tau}
\end{equation}
where $\tau_{0}^{-1}$ is an attempt rate. For $\Delta E>>T$,
$\quad\tau\sim\exp(U/T)$ with the activation energy
$U=E_{3}-E_{2}$. Notice that at the end points the metastable
branch disappears and there must be $\tau(\theta
\rightarrow\theta_{e})\rightarrow0$. Still the expression
(\ref{df/dt}) leaves us with small but finite value of
$\tau\sim\tau_{0}$ even at $U(\theta _{e})\rightarrow0$. Hence it
should be corrected to provide $\tau\rightarrow0$ at $U\ll T$; it
happens via a dependence $\tau_{0}(\theta)$ which plausible form
is $\tau_{0}\sim(\theta-\theta_{e})^{k},\ k>1$.

\subsection{Stationary motion.\label{app:ss:stationary}}

Consider a stationary process when the density wave moves at a
constant phase velocity
$v=-\dot{\bar{\varphi}}=\dot{\theta}=const$, then $\partial
n/\partial t=0$. Now the solution of the Eq. (\ref{df/dt}) is
trivial, but the boundary conditions must be properly specified.
Suppose first that there are no end points which is the case of
very strong impurities $V>V_{2}$, see the Figure \ref{fig:terms2}.
Then for $\theta$ approaching $2\pi$, both branches $+$ and $-$
contribute to initial conditions for the branch $+$ at $\theta=0$
adding the pair of solitons at infinity, see the Figure
\ref{fig:bisoliton}. Contrarily, there is no source for the branch
$-$ at $\theta=0$. This conditions read
\begin{equation}
\left. {\frac{\partial n_{+}}{\partial\theta}}\right| _{0}=
{\frac{\partial}{\partial\theta}}(n_{+}+n_{-})_{2\pi}=0\,,\;
\mathrm{hence}\quad n(0)=n_{eq} \label{f(0)}
\end{equation}
The solution of (\ref{df/dt}), (\ref{f(0)}) is
\begin{equation}
n=n_{eq}(s_{0})e^{s_{0}-s}+\int_{s_{0}}^{s}n_{eq}(s_{1})e^{s_{1}-s}ds_{1}
\label{f(s0)}
\end{equation}
where $s(\theta)$ is an effective $t/\tau$ over the branch:
\begin{equation}
s=s(\theta)=\int_{\pi}^{\theta}{\frac{d\theta}{v\tau(\theta_{1})}};\quad
s_{0}=s(0)=-s(2\pi)  \label{s=}
\end{equation}
(We shall keep the same notations for functions of $\theta$ and of
$s=s(\theta)$.) At presence of end points ( right $\theta_{e}$ and
left $2\pi-\theta_{e}$) there is only one branch of lowest energy
$E_{1}$ which survives beyond $(2\pi-\theta_{e},\ \theta_{e})$,
that is
\begin{equation}
0<\theta<2\pi-\theta_{e}\ :\quad n=1\,;\;
\theta_{e}\leq\theta<2\pi\ :\quad n=-1  \label{f=+1,-1}
\end{equation}
The contributions of these monostable regions to the pinning force
are exactly compensated as it should be. Within the bistability
region $(2\pi-\theta_{e},\ \theta_{e})$ the solution is
\begin{equation}
n=\int_{0}^{\infty}{n_{eq}(s-s_{2})\exp(-s_{2})ds_{2}}
\label{f(s,infty)}
\end{equation}

The substitution to the general expression for the force
(\ref{pin-av}) yields (at  presence of end points)

\begin{equation}
f=n_{i}\int_{-\infty}^{\infty}{ds\Delta E}
\int_{0}^{\infty}{ds_{2}e^{-s_{2}} \left[
n_{0}(s-s_{2})-n_{0}(s)\right] } \label{f-theta}
\end{equation}

Mostly we shall consider the case of low $T$ when $\Delta E\gg T$
in essential regions, then
\begin{equation}
n_{eq}(s)\approx-\mathrm{sgn}s,\quad n-n_{eq}=\Theta(s)e^{-s}
\label{12}
\end{equation}
($\Theta$ is the unit step function). The point of symmetrical
population, $n=0$, is shifted to $\ s=\ln2$. At $0<s\leq\ln2$
there is an inverted population, $n_{2}>n_{1}$ as shown on the
Fig. \ref{fig:few-terms}. Finally the expression (\ref{f-theta})
is simplified to the form (\ref{f-the-F}).

\subsection{Various regimes for $f(v)$.\label{app:ss:kin-results}}

\bigskip\textbf{Small velocities for all cases.}

At $v\tau_{\pi}\ll1$, $s(\theta)$ is large almost everywhere,
except for a vicinity of $\pi$ where the barrier activation energy
takes its largest value $U(\pi)=U_{\pi}$; then we should use
(\ref{f-theta}). Namely, if for largest $\tau(\pi)=\tau_{\pi}$ we
have $v\tau_{\pi}\ll1$, then $s\sim1$ already at $\theta-\pi\sim
v\tau_{\pi}\ll1$ so that at $(\theta-\pi)\sim1$ we have $s\gg1$,
then the series in $v$ is well convergent. In lowest order of
$v\tau$ we find
\begin{equation}
f\approx\pi n_{i}\int_{0}^{\infty}ds\Delta
E{\frac{d^{2}n_{eq}}{ds^{2}}} = \pi
n_{i}\int{\frac{v\tau/T}{\cosh^{2}(\Delta E/2T)}} \left(
{\frac{d\Delta E}{d\theta}}\right)^{2} d\theta \approx
v\tau_{\pi}F_{\pi} \label{F-v}
\end{equation}
At low $T$ the dependence of the expression under the last
integral in (\ref{F-v}) is governed by the factor, see
(\ref{tau}), $\exp\left[-\left((E_{3}-E_{2})-(E_{2}-E_{1})\right)
/T\right]$. It has a maximum $\exp(-U_{\pi}/T))$ at $\theta=\pi$
and we arrive at the result (\ref{tau-pi}).

At higher velocities, still only the vicinities of the crossing
point $\theta\approx\pi$ is important, but we must take into
account the reduction of the barrier with increasing $\theta$:
$U=U_{\pi}-F_{\pi}(\theta-\pi)/2\pi$. We obtain
\begin{equation}
f=n_{i}T\int\frac{ds\exp(-s)}{s+v\tau_{\pi}F_{\pi}/(2\pi T)}
\approx\left\{
\begin{array}{cc}
n_{i}\tau_{\pi}F_{\pi}v & \mathrm{at\;}v\tau_{\pi}F_{\pi}T\ll1 \\
n_{i}T\ln(v\tau_{\pi}F_{\pi}/T) &
\mathrm{at}\;v\tau_{\pi}F_{\pi}/T\gg1
\end{array}
\right.  \label{1ab}
\end{equation}
\bigskip

\textbf{ High velocities: restricted metastability.} Let
$v\tau_{\pi}>>1$, then $s\sim1$ only at $\theta\approx\theta_{e}$.
The form (\ref{f-the-E}) is more appropriate for calculations.
Since $e^{-s}-1$ is small at $s\ll1$ i.e. at almost all $\theta$,
then only a vicinity of $\theta_{e}$ contribute, hence we can take
$F=F_{e}$ at $\theta=\theta_{e}$. We obtain
\begin{equation}
f=f_{max}-2\pi n_{i}F_{e} \int_{\pi}^{\theta_{e}}d\theta
\left(1-e^{-s(\theta)}\right) \,\,;\; f_{max}=2\pi n_{i}\Delta
E_{e}
\end{equation}
Remind that at $\theta_{e}$ the activation
$U=E_{3}-E_{2}\rightarrow0$ vanishes while the force is finite
$F_{e}=\partial\Delta E/\partial\theta \neq0$. We find
\begin{equation}
{f_{max}-f}\sim F_{e}n_{i}\delta\theta_{v} \nonumber
\end{equation}
where $\delta\theta_{v}$ is defined by the condition

\begin{equation}
s(\delta\theta_{v})\sim\frac{v_{e}}{v} \delta\theta_{v}^{-\nu+1}
{\exp(-\frac{V_{e}}{T} \delta\theta_{v}^{\nu})}\sim{1};\quad
v_{e}= \left( \frac{T}{V_{e}}\right) ^{1/\nu}{\frac{1}{\tau_{0}}}
\label{s-e}
\end{equation}
 and we arrive at the result (\ref{f-e}), valid in this simple form at $v_{e}\exp(-V_{e}/T)\ll v\ll v_{e}$.
 \bigskip

\textbf{ High velocities: unrestricted metastability.}

The calculations are similar to the above case of the restricted
metastability and we shall skip equivalent steps. The difference
is that now there is a high velocity range $v\gg
v_{m}=\max\tau^{-1}$ where the $1/v$ expansion is valid:
\begin{equation}
f=2n_{i}\left\{ \Delta E(2\pi)-\frac{2}{v} \int_{\pi}^{2\pi}
{{\frac{d\theta}{\tau(\theta)}(}\Delta E(2\pi)- \Delta
E(\theta))}\right\} \label{1/v}
\end{equation}

\subsection{ Linear response.\label{app:ss:response}}

Consider $\dot\theta$ as a perturbation in the kinetic equation
and expand as $n=n_{eq}(\theta)+\delta n(\theta,t)$:
\begin{equation}
 \left({\frac{\partial}{\partial t}+
\frac{1}{\tau}}\right) {\delta n}+\dot{\theta}
\frac{\partial}{\partial\theta}n_{eq}=0\,;\;
f=n_{i}\int_{\pi}^{\theta_{max}}{d\theta F(\theta)\delta
n(\theta)} \label{the-ome}
\end{equation}
In the Fourier representation we have
\begin{equation}
{\delta n}_{\omega}  = \frac{i\omega\theta_{\omega}} {\left(
-i\omega{+\tau}^{-1}\right)} \frac{d}{d\theta}{n}_{eq}
\end{equation}

\begin{equation}
f_{\omega}= \frac{\theta_{\omega}n_{i}} {\left(
-1{+(i\omega\tau)}^{-1}\right)}\int_{\pi}^{\theta_{e}} {d\theta
{\frac{d\Delta E}{d\theta} \frac{d}{d\theta}{n}_{eq}}}
\approx\frac{F_{\pi}\theta_{\omega}n_{i}}
{\left(-1{+(i\omega\tau)}^{-1}\right) }
\end{equation}

\begin{equation}
\chi_{\omega}^{-1}= \frac{\delta
f_{\omega}}{\delta\theta_{\omega}}=
\frac{n_{i}F_\pi}{1+1/i\omega\tau},
\end{equation}
which confirms (\ref{F-omega}).

\end{document}